%% file: Ivanov.tex
\documentclass[12pt]{elsarticle}
\usepackage{refmerge}
\usepackage{epsfig}
\usepackage{multicol}
\usepackage[percent]{overpic}
\usepackage{textcomp}
\usepackage{float}
\floatstyle{plaintop}
\restylefloat{table}

\usepackage{booktabs}
\usepackage{amssymb,bm,mathrsfs,amscd}
\usepackage[tbtags]{amsmath}
\usepackage{natbib}

\begin{document}

\date{\today}

\title{\bf{CHARGED PARTICLE IDENTIFICATION WITH THE LIQUID XENON CALORIMETER OF THE \mbox{CMD-3} DETECTOR}}

\input{authors.tex}

\vspace{0.7cm}
\begin{abstract}
  \hspace*{\parindent}
  The paper describes a method of the charged particle identification, 
  developed for the \mbox{CMD-3} detector, installed at the VEPP-2000 
$e^{+}e^{-}$ collider. 
  The method is based on the application of the boosted decision trees  
  classifiers, trained for the optimal separation of electrons, muons, pions 
and kaons in 
  the momentum range from 100 to $1200~{\rm MeV}/c$. 
  The input variables for the classifiers are linear combinations 
  of the energy depositions of charged particles in 12 layers 
  of the liquid xenon calorimeter of the \mbox{CMD-3}.
  The event samples for training of the classifiers are taken from 
  the simulation. Various issues of the detector response tuning in 
simulation and  calibration of the calorimeter strip channels are considered.
  Application of the method is illustrated by the examples of
  separation of the $e^+e^-(\gamma)$ and $\pi^+\pi^-(\gamma)$ final states 
  and of selection of the $K^+K^-$ final state at high energies.
\end{abstract}

\maketitle

\baselineskip=17pt


\section{\boldmath Introduction}
\hspace*{\parindent}

The VEPP-2000 $e^+e^-$ collider~\cite{vepp1,vepp2,vepp3,vepp4} at the Budker 
Institute of Nuclear Physics (Novosibirsk, Russia) covers the 
center-of-mass (c.m.) energy ($E_{\rm c.m.}$) range from 0.32 to 
2.01~GeV and uses a technique of round beams to reach an instantaneous
luminosity of 10$^{32}$\,cm$^{-2}$s$^{-1}$ at $E_{\rm c.m.}=2.0$~GeV. The 
Cryogenic Magnetic Detector (\mbox{CMD-3}) described in~\cite{cmd3} is 
installed in 
one of the two interaction regions of the collider. 
The main goal of the \mbox{CMD-3} is the precise measurement of the 
exclusive cross sections of $e^{+}e^{-}$ annihilation into hadrons, which 
provides, for instance, a necessary input for the theoretical calculation of 
the hadronic contribution to the muon anomalous magnetic moment $(g-2)_\mu$ 
and the running fine structure constant~\cite{jegerlehner,davier,teubner,hagiwara}.

The precise measurement of any hadronic cross section requires  
selection of a clean sample of signal events. 
The latter often requires the effective procedure of particle 
identification (PID), {\it i.e.} separation of electrons, muons, pions, 
kaons {\it etc}.
In particular in the CMD-3: 
\begin{itemize}
\item $e^{\pm}$ identification can be performed on the base of the total energy deposition in the calorimeter~\cite{cal};
\item identification of muons can be carried out with the muon veto system;
\item identification of antineutrons can be done with the time-of-flight system~\cite{TOF};
\item separation of charged kaons and pions at the momenta less than 550~MeV/$c$
is performed using specific energy losses of particles in the drift chamber ($dE/dx_{\rm DC}$)~\cite{dc}. 
\end{itemize}

In this paper we describe a new technique of the charged PID
based on the multiple measurements of the energy depositions of a particle
in the layers of liquid xenon (LXe) calorimeter of the \mbox{CMD-3}~\cite{lxe}.
The LXe-based PID method is developed mainly for the task of $K^{\pm}/\pi^{\pm}$ separation
at momenta larger than 550~MeV/$c$, where the $dE/dx_{\rm DC}$-based PID is 
inefficient.
For the hadronic final states like $K^{+}K^{-}$, $K^{+}K^{-}\pi^{0}$, $K^{+}K^{-}2\pi^{0}$, $K_{S}K^{\pm}\pi^{\mp}$
the LXe-based PID turns out to be an efficient tool for background suppression 
at high c.m. energies.


\section{\boldmath The liquid xenon calorimeter of the \mbox{CMD-3} detector}
\hspace*{\parindent}

\begin{figure}[hbtp]
  \begin{minipage}[t]{0.52\textwidth}
    \centerline{\includegraphics[width=0.98\textwidth]{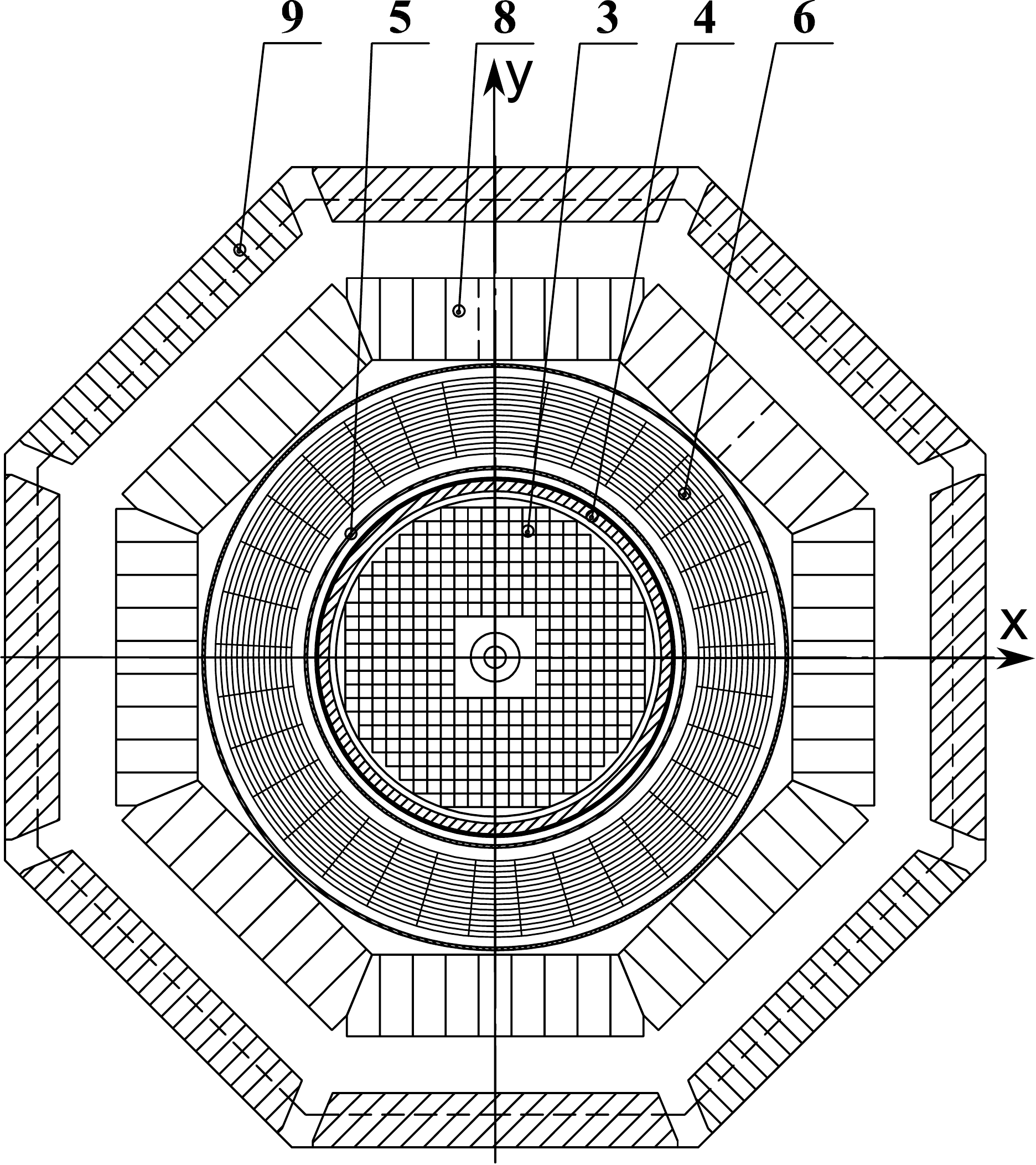}}
  \end{minipage}\hfill\hfill
  \begin{minipage}[t]{0.47\textwidth}
    \centerline{\includegraphics[width=0.98\textwidth]{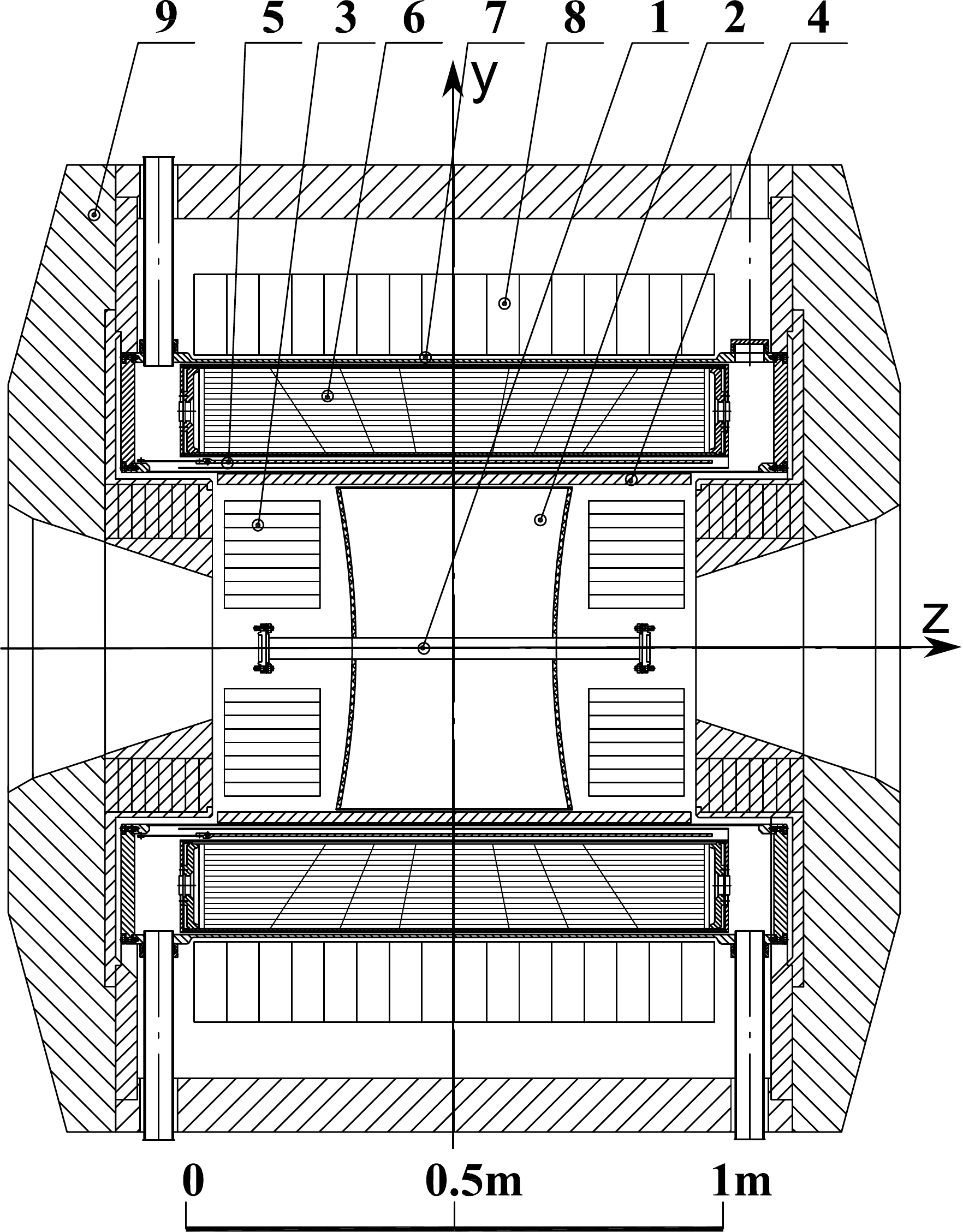}}
  \end{minipage}\hfill\hfill  
  \caption{\label{fig:cmd3} The \mbox{CMD-3} detector layout: 
1 --- beam pipe, 2 --- drift chamber, 
3 --- BGO endcap calorimeter,   4 --- Z-chamber (ZC), 
5 --- superconducting solenoid, 6 --- LXe calorimeter, 
7 --- time-of-flight system (TOF), 8 --- CsI calorimeter, 9 --- yoke.}
\end{figure}

The \mbox{CMD-3} detector layout is shown in Fig.~\ref{fig:cmd3}. 
The major tracking system is the 
cylindrical drift chamber (DC)~\cite{dc}, installed inside a 
thin (0.085 $\rm X_{0}$) superconducting solenoid with 1.3 T magnetic 
field. Amplitude information from the DC signal wires is used to measure 
$z$-coordinates of tracks and ionization losses of charged 
particles. The endcap calorimeter is made of 
BGO crystals 13.4~$\rm X_{0}$ thick~\cite{cal}.
The barrel calorimeter consists of the inner LXe-based (5.4~$\rm X_{0}$) 
ionization and outer 
CsI-based (8.1~$\rm X_{0}$) scintillation calorimeters~\cite{lxe_csi_bgo}. 
The total amount of material in front of the barrel calorimeter is~0.35$\rm X_{0}$.

\begin{figure}[hbtp]
  \begin{minipage}[t]{0.49\textwidth}
    \centerline{\includegraphics[width=0.98\textwidth]{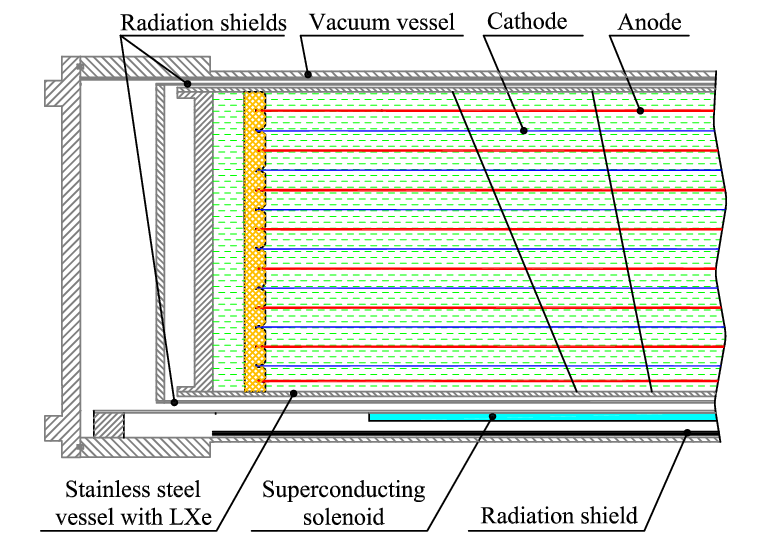}}
    \caption{Structure of LXe calorimeter electrodes.          
          \label{fig:LXe_layers}}
  \end{minipage}\hfill\hfill
  \begin{minipage}[t]{0.49\textwidth}
    \centerline{\includegraphics[width=0.98\textwidth]{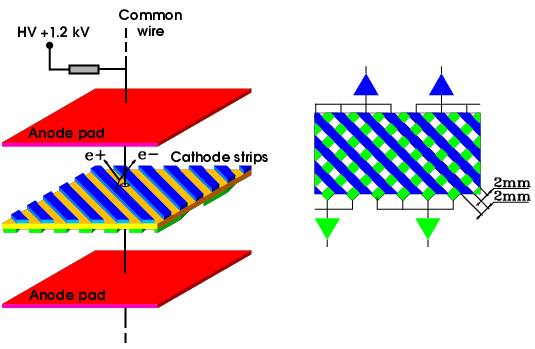}}
    \caption{Anode-cathode-anode double layer of the LXe calorimeter and the 
strip structure of the cathode.      
          \label{fig:LXe_strips}}
  \end{minipage}\hfill\hfill
\end{figure}

The LXe calorimeter consists of a set of ionization chambers 
with seven cylindrical cathodes and eight anodes with a 10.2~mm gap between 
them, see Fig.~\ref{fig:LXe_layers}. The electrodes are made of 
copper-plated G-10. 
The typical electric field in the gap is 1.1~kV/{\rm c}m. 
The conductive surfaces of the anode electrodes are divided into 
rectangular pads, 
electrically connected by the wire going through the cathode layers. 
These 264 sets of pads form the towers oriented to the interaction 
point and are used to measure the energy deposition of the particles. 

Both conductive surfaces of the cathode electrodes are divided into 
strips with a 1.2--1.8~mm width (depending on the layer), 
separated by the 1.5--2.0~mm gaps. A set of four consecutive strips is 
electrically connected to one superstrip making up one channel of electronics, 
see Fig.~\ref{fig:LXe_strips}. 
The number of supestrips on one side of the cathode is 147--156. 
In what follows we refer to superstrips as just strips.
The strips on the opposite sides of the cathode are mutually perpendicular that 
allows one to measure $z$ and $\varphi$ coordinates of the clusters.

A current induced on the strip during the ionization flow is 
integrated for 4.5 $\mu$s, where the integration time corresponds 
to the maximum drift time of electrons in the gap.
The strip channels are used for the measurements of the photon 
conversion point coordinates and the $dE/dx$ of the particle in each of 
the 14 layers. 
Due to the gaps between the strips the cathodes 
are semitransparent, {\it i.e.} the ionization in one layer of 
the anode-cathode-anode double layer induces the charge on the strips 
of both sides of the cathode. This allows one to measure the coordinates of
the point of photon conversion on the base of ionization that happened 
in one anode-cathode gap only.

\section{\boldmath The idea of the PID procedure \label{sec:PID_idea}}
\hspace*{\parindent}

\begin{figure}[hbtp]
  \begin{minipage}[t]{0.49\textwidth}
   \centerline{\includegraphics[width=0.98\textwidth]{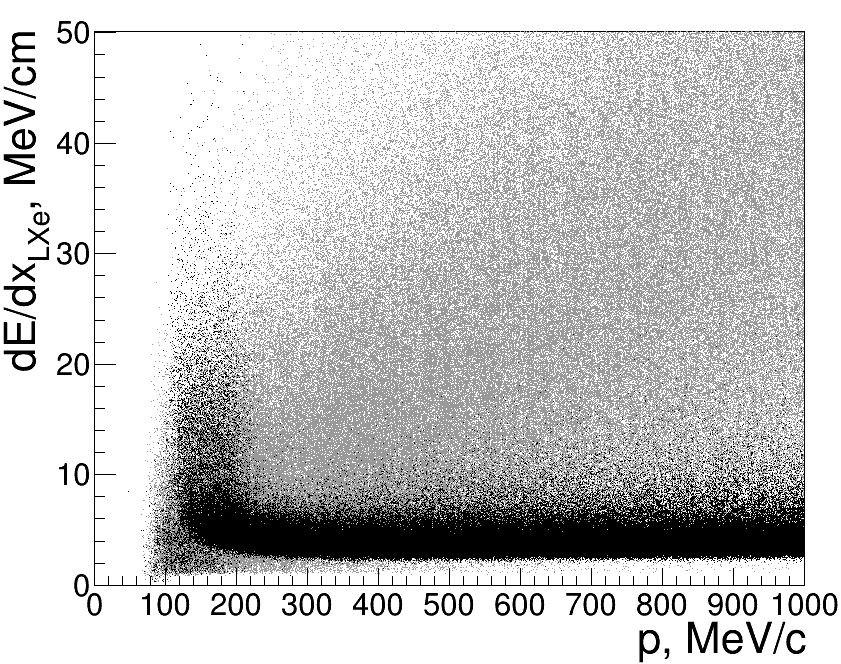}}
    \caption{$dE/dx_{\rm LXe}$ in all LXe layers vs. particle momentum 
    for $e^{\pm}$ (gray) and $\mu^{\pm}$ (black) in simulation. 
          \label{fig:dedx_e_mu}}
  \end{minipage}\hfill\hfill
  \begin{minipage}[t]{0.49\textwidth}
    \centerline{\includegraphics[width=0.98\textwidth]{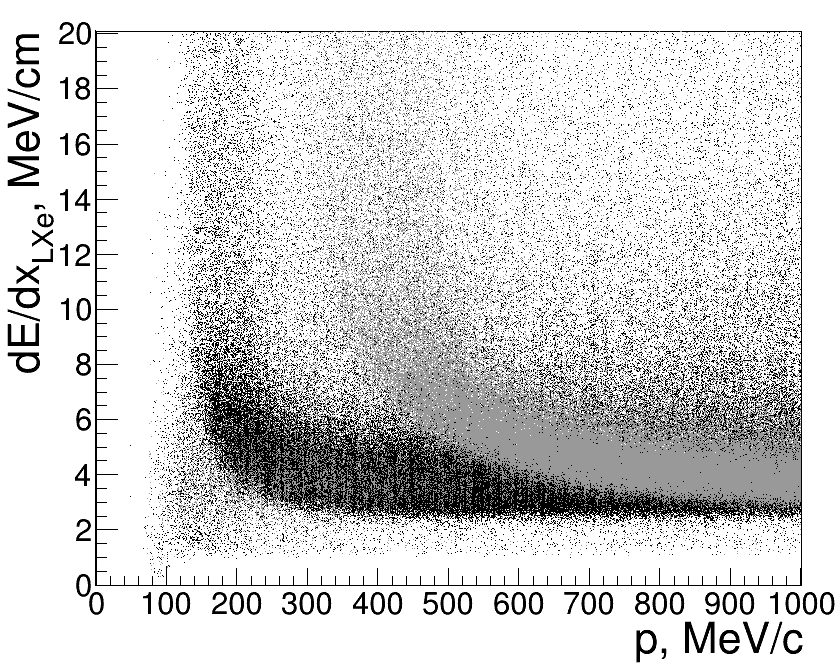}}
    \caption{$dE/dx_{\rm LXe}$ in all LXe layers vs. particle momentum 
    for $K^{\pm}$ (gray) and $\pi^{\pm}$ (black) in simulation.        
          \label{fig:dedx_k_pi}}
  \end{minipage}\hfill\hfill
\end{figure}

In what follows we denote by $dE/dx_{\rm LXe}$ 
the energy deposition produced by a particle in each LXe layer, 
normalized to the {\em expected path length} 
$d_{\rm LXe}$ of the particle in the layer, estimated 
via the DC-track extrapolation. $dE/dx_{\rm LXe}$
is the single designation for the minimum ionizing and 
nuclear interacting particles as well as for the electromagnetic showers.
The distributions of $dE/dx_{\rm LXe}$ in 14 LXe layers depending on
the particle momentum for the simulated single 
$e^{\pm}$ and $\mu^{\pm}$, $\pi^{\pm}$ and $K^{\pm}$ are shown in
Figs.~\ref{fig:dedx_e_mu} and \ref{fig:dedx_k_pi}, respectively.
One should note the following features of $dE/dx_{\rm LXe}$:

\begin{itemize}
\item In Figs.~\ref{fig:dedx_e_mu} and \ref{fig:dedx_k_pi} 
a certain momentum threshold $p_{\rm thr}$ is seen for each particle type that 
corresponds to a minimum energy necessary to pass through the material in front 
of the LXe. Below this threshold only the products of the particle decay
or nuclear interaction can reach the calorimeter. For kaons 
$p^{K}_{\rm thr}$ is about 300~MeV/$c$ for the normal incident;
\item The $dE/dx_{\rm LXe}$ spectra and the values of $p_{\rm thr}$ 
depend on $d_{\rm LXe}$. 
This dependence is caused by the dependences of the shower development rate, 
the nuclear interaction probability, the particle deceleration rate {\it etc.} 
on $d_{\rm LXe}$.
\end{itemize}

The LXe-based PID method uses the difference in the $dE/dx_{\rm LXe}$
in the LXe-layers for different particle types. 
Namely, for each track in the DC reaching LXe, 
we calculate six values of the responses of the boosted
decision trees (BDT) classifiers provided by the TMVA package \cite{tmva},
trained for optimal separation of particular pairs of particle types in 
certain ranges of the momentum $p$ and $d_{\rm LXe}$, {\it i.e.} in 
the ${\Delta}p_{i} \times {\Delta}d_{{\rm LXe}, \, j}$ cell.
In what follows we denote these six values as 
${\rm BDT}(e^{\pm},\mu^{\pm})$, ${\rm BDT}(e^{\pm},\pi^{\pm})$, 
${\rm BDT}(e^{\pm},K^{\pm})$,
${\rm BDT}(\mu^{\pm},\pi^{\pm})$, ${\rm BDT}(\mu^{\pm},K^{\pm})$, 
${\rm BDT}(\pi^{\pm},K^{\pm})$. 

For training of each classifier we use samples of ${\sim}10^{5}$ 
simulated events with single $e^{\pm}$, $\mu^{\pm}$, $\pi^{\pm}$, $K^{\pm}$, 
having the momentum and $d_{\rm LXe}$ uniformly
distributed in the ${\Delta}p_{i} \times {\Delta}d_{{\rm LXe}, \, j}$ cell. 
In total, we have 55 ${\Delta}p_{i}$ cells 
of 20~MeV/$c$ width in the momentum range from 100 to 1200~MeV/$c$ and 
eight ${\Delta}d_{{\rm LXe}, \, j}$ cells (from 1.0 to 1.5~cm at large 
momenta).
Thus, there is $2{\times}6{\times}55{\times}8=5280$ classifiers to be trained, 
where the factor of 2 stands for the two particle charges.
The input variables for the classifiers are linear combinations of the 
$dE/dx_{\rm LXe}$ values in LXe layers described later in Section~\ref{sec:sim_tuning}.

Since this PID method is based on the precise measurement of the energy
deposition for different type of particles we need in the accurate
energy calibration of the strip channels of the LXe calorimeter.
Issues of the calibration of strip channels and the detector response tuning
in MC are considered in the next sections.


\section{\boldmath Calibration of LXe calorimeter strip channels}
\hspace*{\parindent}

In what follows we call {\it cluster} the group of 
the neighboring triggered strips (on one side of the cathode) 
with at least one strip having an amplitude above the {\it cluster reconstruction threshold}. 
This threshold is set to 1.5 MeV (in terms of the calibrated amplitude), 
which corresponds to the minimum amplitude induced by the minimum ionizing 
particles (MIPs) on the strip
and is well above the level of electronics noise which energy equivalent is ${\sim}0.1$~MeV.
The cluster amplitude is equal to the sum of the amplitudes of its strips.
The typical number of strips in the cluster for MIPs is 2--3.

The calibration of the strips is performed using events with  
cosmic muons having the momentum larger than 1~GeV/$c$. 
There are three stages of the calibration:

\begin{enumerate}
\item Equalization of the strip amplitudes normalized to the 
particle path length, within each of seven cathodes separately;
\item Equalization of the cluster amplitudes normalized to the 
particle path length in all seven cathodes by bringing them to a 
common average;
\item Calculation of the MeV to ADC channel~\cite{LXe_electroincs} transition 
coefficient.
\end{enumerate}

The calibrated strip amplitude is calculated as 
$A_{\rm calib} = A_{\rm raw}K_{3}/(K_{1}K_{2})$, where $K_{1...3}$ are 
the calibration coefficients of the corresponding stages, 
$A_{\rm raw}$ is the raw amplitude with the pedestal subtracted. 
To achieve the convergence of the $K_{1...3}$ the calibration is carried out iteratively: 
reconstruction of events at the current iteration is performed with the 
application of the 
calibration coefficients calculated at the previous iteration.
To obtain the calibration precision of about 1\%, three iterations are 
sufficient. The $K_{2}$ and $K_{3}$ are not calculated at the first iteration, 
since it does not make sense to clusterize non-equalized strips.

\subsection{Equalization of the strip amplitudes within each cathode} 

\begin{figure}[hbtp]
  \begin{minipage}[t]{0.49\textwidth}
   \centerline{\includegraphics[width=0.98\textwidth]{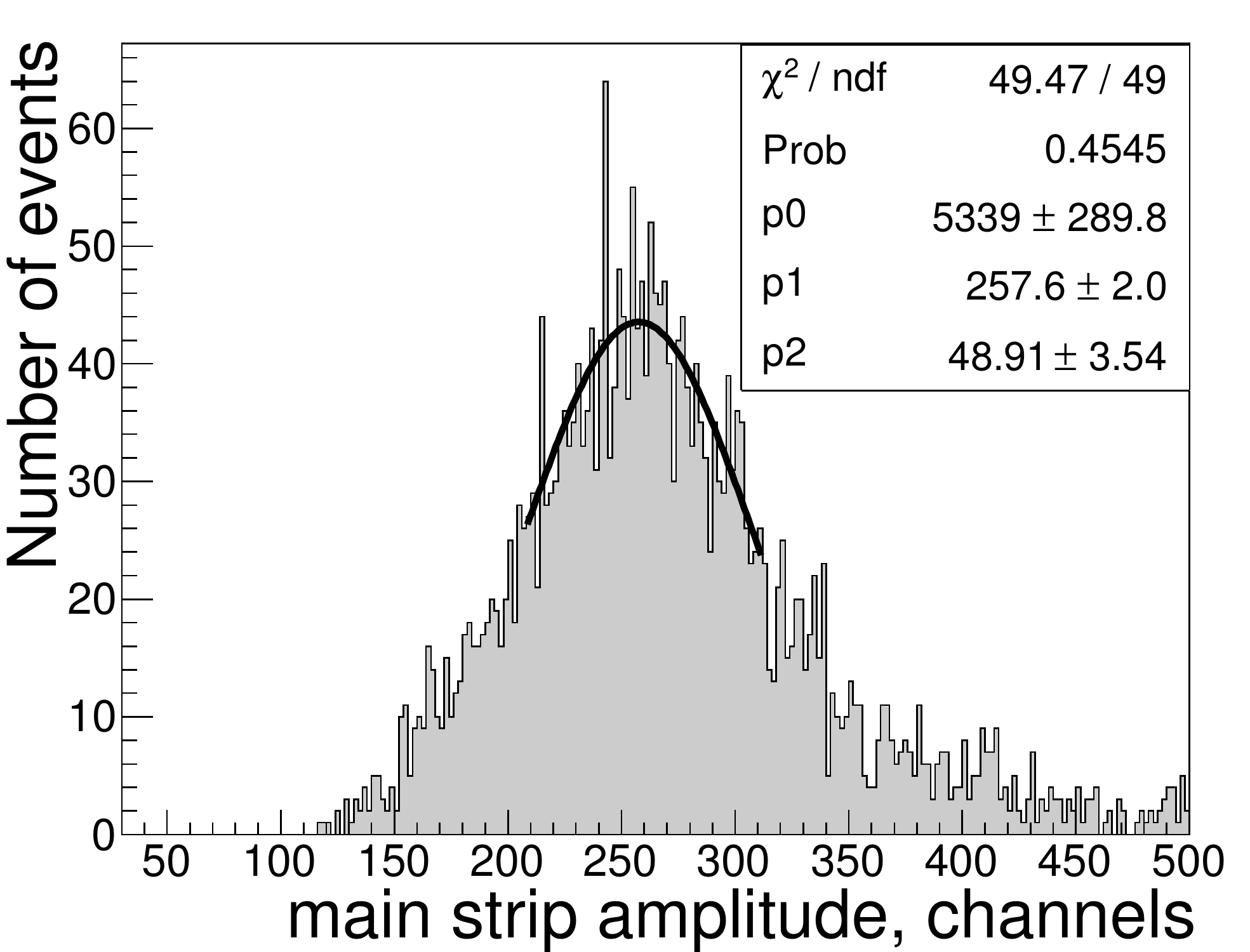}}
    \caption{Typical amplitude spectrum of the cluster main strip and 
    its Gaussian approximation near the maximum. The amplitude is normalized to
 the particle path length.
          \label{fig:main_s_spectrum}}
  \end{minipage}\hfill\hfill
  \begin{minipage}[t]{0.49\textwidth}
    \centerline{\includegraphics[width=0.98\textwidth]{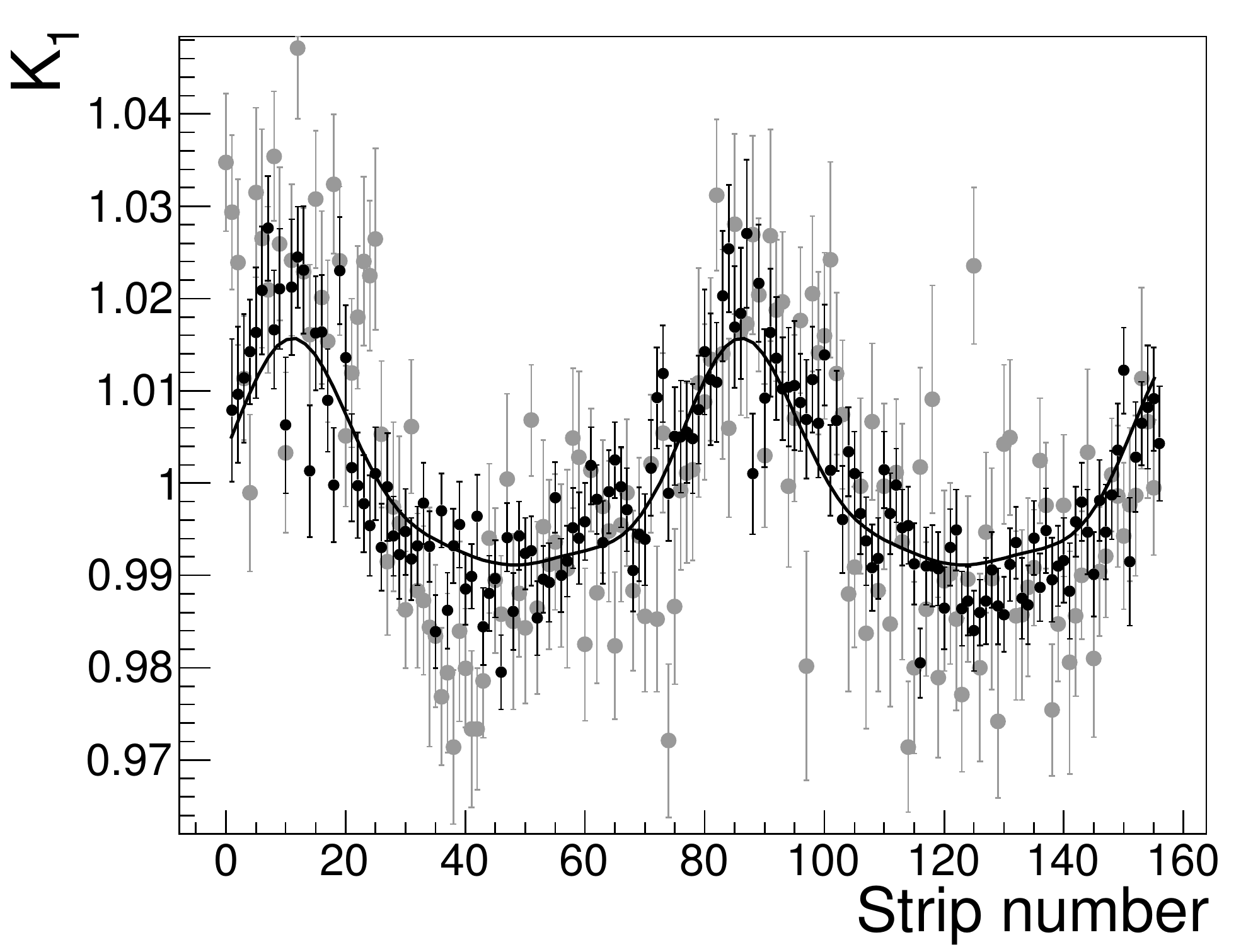}}
    \caption{$K_{1,\,\rm MC}$ dependence on the strip number 
    for simulated cosmic muons (black markers) and its approximation 
(black curve).
The $K_{1,\,\rm cosmic}-K_{1,\,\mu^{+}\mu^{-}}$ difference in the experiment 
as a function of the strip number is shown by the gray markers.        
          \label{fig:K1_modulation}}
  \end{minipage}\hfill\hfill
\end{figure}

The equalization of the strip amplitudes is performed by fitting 
the spectra of amplitudes of the {\em main} strips in the clusters, 
{\it i.e.} of the strips with the maximum amplitude, see Fig.~\ref{fig:main_s_spectrum}. 
These amplitudes are normalized to the particle's path length to 
suppress the dependence on the inclination of the track in the 
anode-cathode gap. Let us denote by $A^{\rm max}_{\rm main~strip}$ the position 
of the maximum of the normalized strip amplitude spectrum 
obtained from the Gaussian fit (Fig.~\ref{fig:main_s_spectrum}).
The $K_{1}$ for a given strip is calculated as the ratio

\begin{equation}
K_{1}=A^{\rm max}_{\rm main~strip}/\overline{A^{\rm max}_{\rm main~strip}},
\end{equation}
 where $\overline{A^{\rm max}_{\rm main~strip}}$ is the average maximum position for the strips on both sides of the given cathode.

Simulation of cosmic muons reveals the residual angular dependence 
of the $K_{1,\,\rm MC}$ manifested as the systematic $\pm$1\% modulation of the 
$K_{1,\,\rm MC}$ for different strips, see Fig.~\ref{fig:K1_modulation}.
The same modulation is seen in the difference between the $K_{1}$ coefficients, 
calculated in the experiment on the base of events with cosmic muons ($K_{1,\,\rm cosmic}$)
and using muons from the process $e^{+}e^{-}{\to}\mu^{+}\mu^{-}$  
($K_{1,\,\mu^{+}\mu^{-}}$).
Since muons from the process $e^{+}e^{-}{\to}\mu^{+}\mu^{-}$ have the uniform 
azimuthal angle distribution, there is no azimuthal modulation in $K_{1,\,\mu^{+}\mu^{-}}$.
To account for observed modulation the experimental $K_{1}$ is multiplied 
by the approximated angular dependence 
of the $K_{1,\,\rm MC}$ (Fig.~\ref{fig:K1_modulation}). 

Figure~\ref{fig:K1_trends} shows the $K_{1}$ trends for the first and second 
calibration iterations in the runs of 2020 year.

\subsection{Equalization of average cluster amplitudes between cathodes}

At the second stage we equalize the average cluster amplitudes 
normalized to the particle's path length, 
$\overline{dE/dx}^{l}_{\rm clust}$, $l=1...7$, between cathodes. The average 
is calculated near the maximum of spectra in the limits containing
 $\sim$90\% of events. The calibration coefficients $K^{l}_{2}$ 
 bringing the $\overline{dE/dx}^{l}_{\rm clust}$ on each cathode to the common 
interlayer average are calculated as 

\begin{equation}
K^{l}_{2}=\frac{\overline{dE/dx}^{l}_{\rm clust}}{\sum\limits_{l=1}^{7}\overline{dE/dx}^{l}_{\rm clust}/7}.
\end{equation}

\begin{figure}[hbtp]
  \begin{minipage}[t]{0.49\textwidth}
   \centerline{\includegraphics[width=0.98\textwidth]{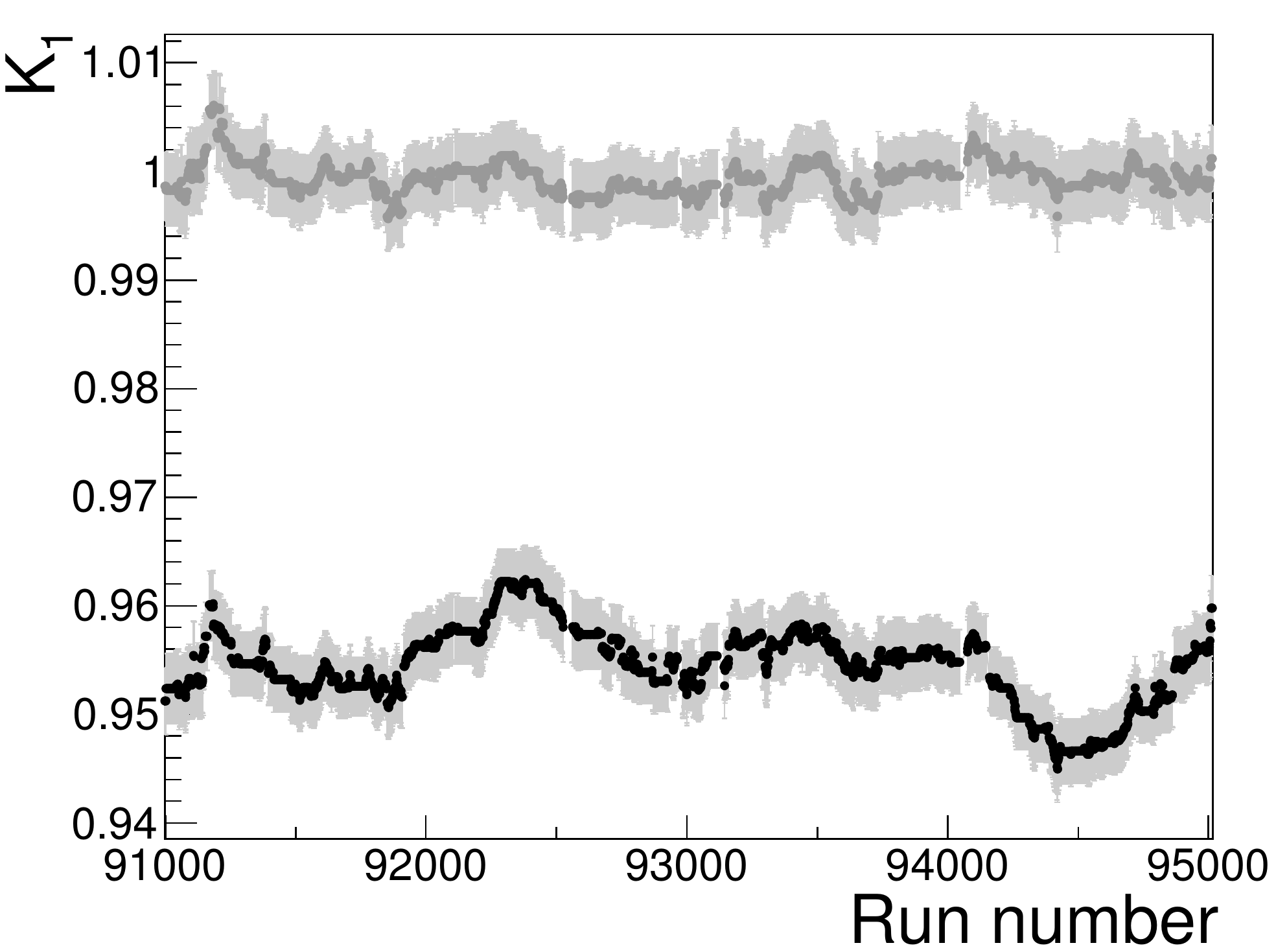}}
    \caption{The typical $K_1$ trend for one strip at the first 
    (black) and second (gray) calibration iterations in the 2020 runs. 
    The gray bands show the statistical uncertainties.
          \label{fig:K1_trends}}
  \end{minipage}\hfill\hfill
  \begin{minipage}[t]{0.49\textwidth}    
    \centerline{\includegraphics[width=0.98\textwidth]{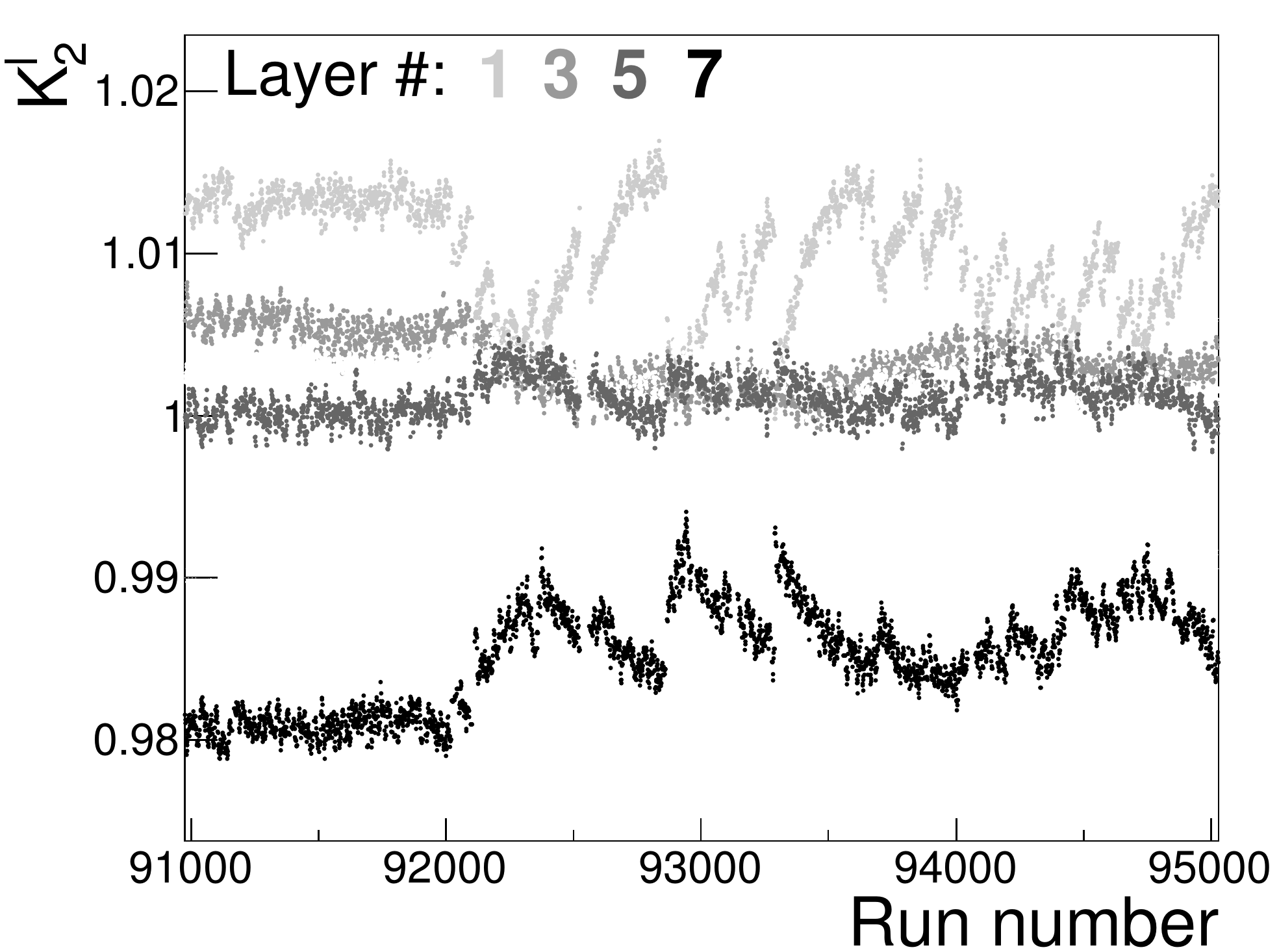}}
    \caption{The $K^{l}_{2}$ trends for different cathodes at the second 
    calibration iteration in the 2020 runs, 
    see color/double layer correspondence in the legend.   
          \label{fig:K2_trends_iter_2}}
  \end{minipage}\hfill\hfill
\end{figure}

The same equalization procedure is performed for simulation. 
Figure~\ref{fig:K2_trends_iter_2} shows the $K^{l}_{2}$ trends for different 
cathodes at the second calibration iteration in the 2020 runs.

\subsection{Calculation of the MeV/{\rm c}hannel transition coefficient}

\begin{figure}
  \begin{center}
    \includegraphics[width=0.49\textwidth]{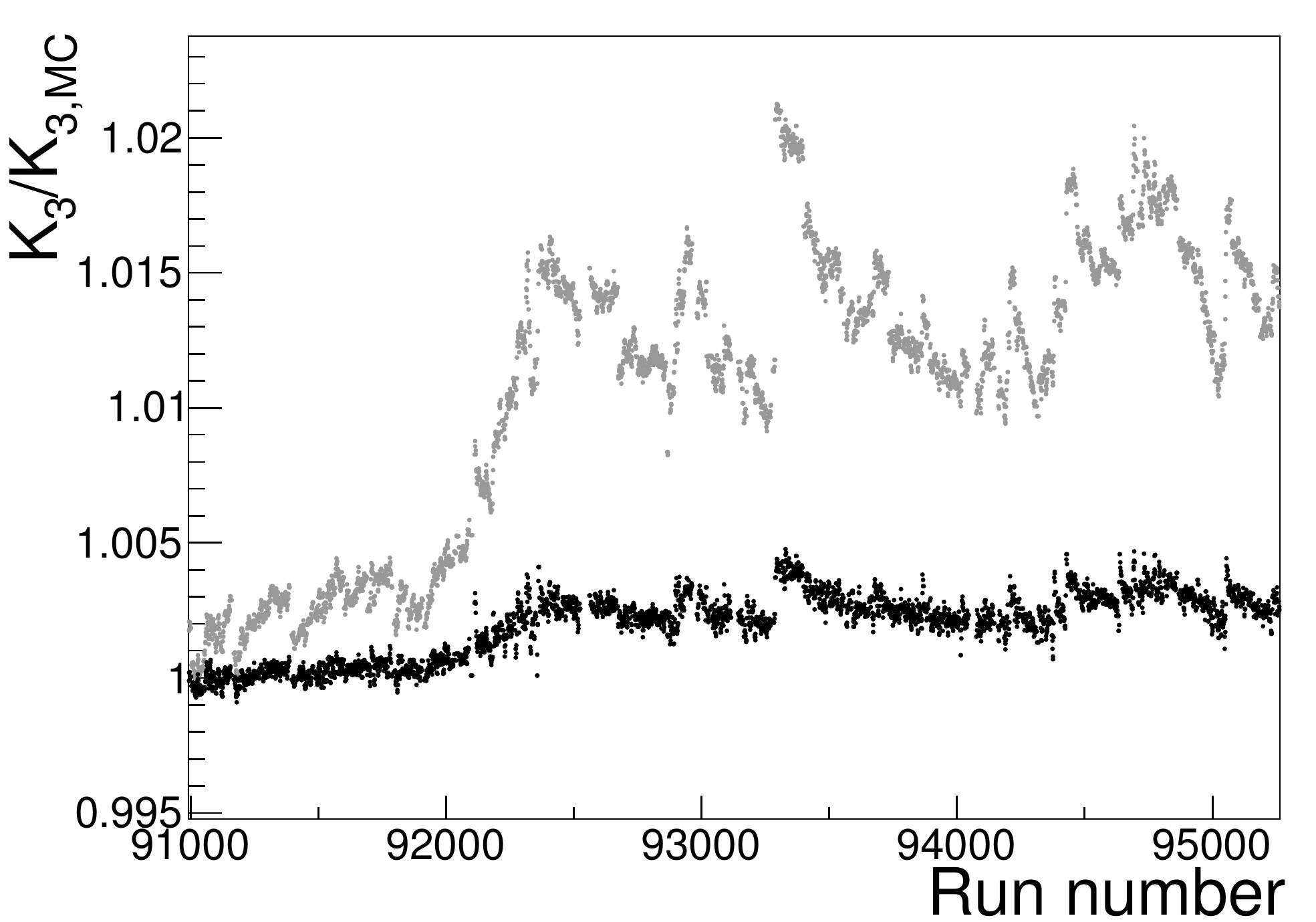}
    \caption{The $K_{3}/K_{\rm{3,MC}}$ trends at the second (gray) and third 
    (black) calibration iterations in the 2020 runs.
        \label{fig:K3_trends}}
  \end{center}
\end{figure}

We calculate the MeV/{\rm c}hannel transition coefficient $K_{3}$ via 
the relation

\begin{equation}
K_{3}=\frac{\sum\limits_{l=1}^{7}\overline{dE/dx}^{l, {\rm MC}}_{\rm clust}}{\sum\limits_{l=1}^{7}\overline{dE/dx}^{l,{\rm data}}_{\rm clust}}{\cdot}K_{3, \rm MC} [\rm MeV/{\rm c}hannel],
\end{equation}
where $K_{3,\rm MC}$ is the transition coefficient tabulated in the MC. 
Figure~\ref{fig:K3_trends} 
shows the $K_{3}$ trends for the second and third calibration iterations 
in the 2020 runs.

\section{\boldmath Detector response tuning in simulation \label{sec:sim_tuning}}
\hspace*{\parindent}

\subsection{Tuning for minimum ionizing particles}

\begin{figure}[hbtp]
  \begin{minipage}[t]{0.49\textwidth}
   \centerline{\includegraphics[width=0.98\textwidth]{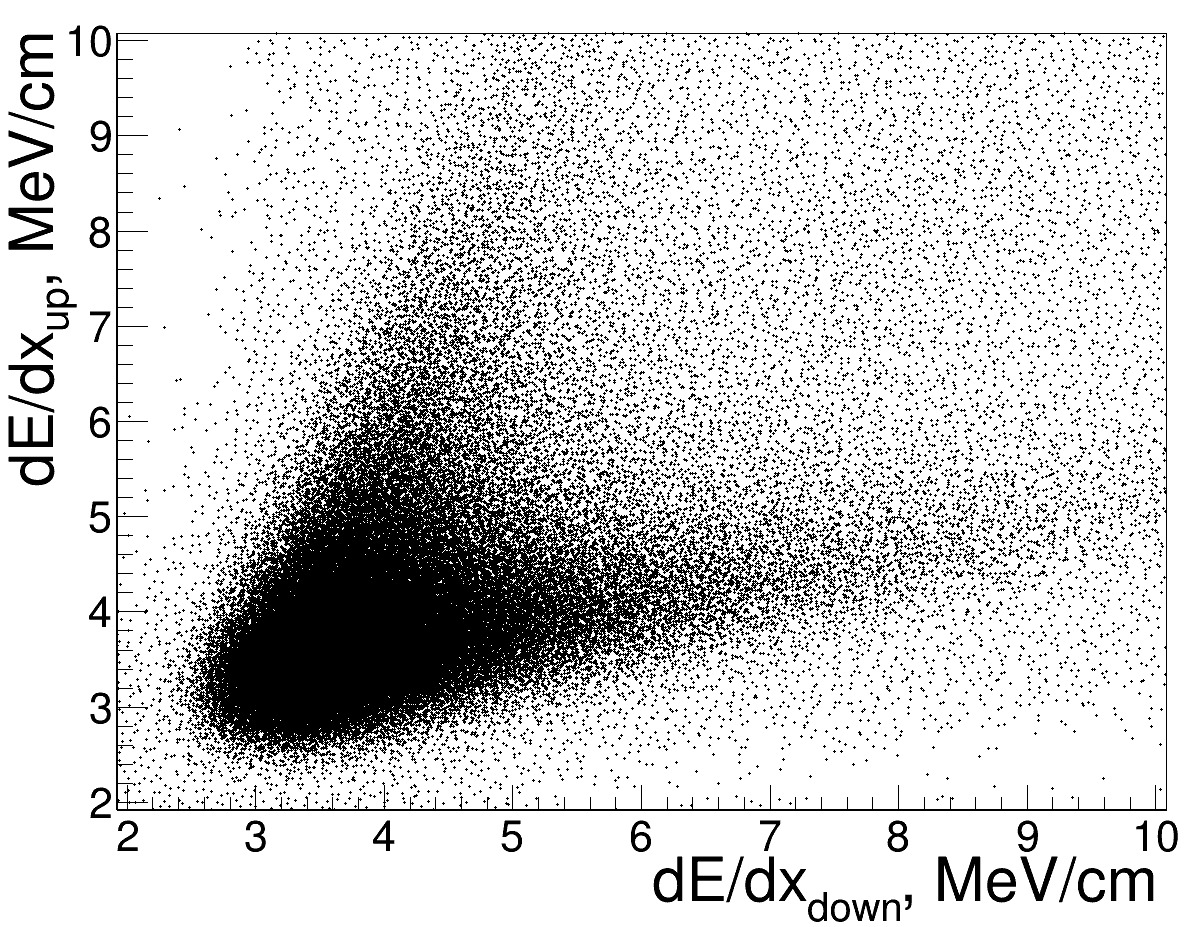}}
    \caption{$dE/dx_{\rm LXe}$ for cosmic muons measured in the 
innermost double layer by the upper strips vs. that measured by the lower 
strips in the experiment.
          \label{fig:dedx_up_vs_down}}
  \end{minipage}\hfill\hfill
  \begin{minipage}[t]{0.49\textwidth}
    \centerline{\includegraphics[width=0.98\textwidth]{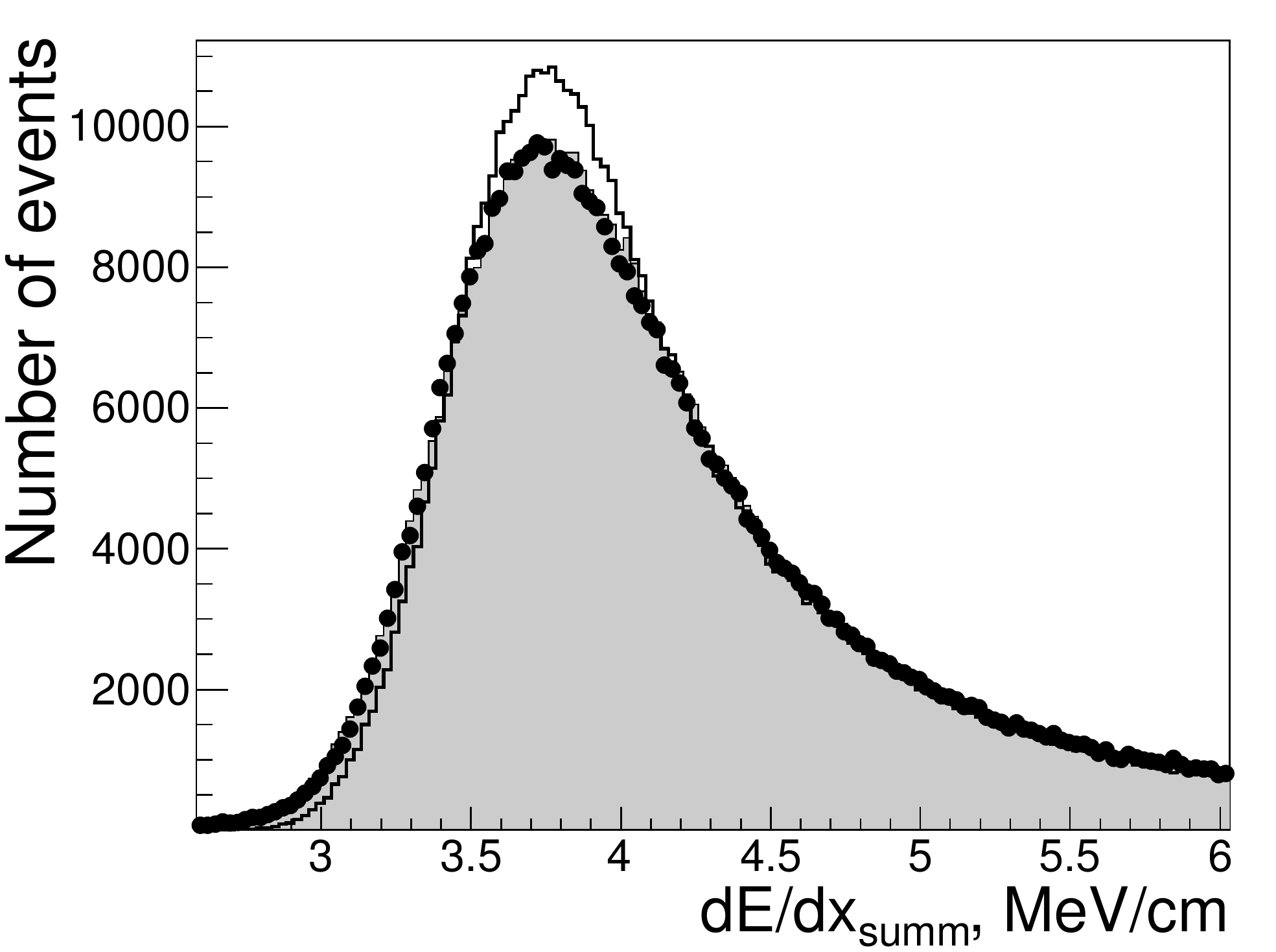}}
    \caption{The $dE/dx_{\rm summ}$ spectra in the innermost double layer for 
cosmic muons 
    in the experiment (markers) and MC before (open histogram) 
    and after (gray histogram) tuning.  
          \label{fig:dedx_summ_cosmics_tuning}}
  \end{minipage}\hfill\hfill
\end{figure}

Figure~\ref{fig:dedx_up_vs_down} shows $dE/dx_{\rm LXe}$ of the cosmic muons
 measured in the innermost double layer by the upper strips ($dE/dx_{\rm up}$) 
 vs. that measured by the lower strips ($dE/dx_{\rm low}$).
The events in the pair of inclined bands correspond to the cases, when 
the large ionization in one anode-cathode gap induces the large 
amplitude on the strips on the opposite side due to the cathode transparency. 
This cross-layer induction, compared with the normal interlayer induction,
 occurs with some suppression factor, which depends on the 
position of the ionization in the gap. The average of this factor over 
all possible ionization positions is called the {\em transparency 
coefficient} $T_{l},$ $l=1...7$. The $T_{l}$ depends on the geometry of the 
cathode, namely on the widths of the strips and the gaps and on the thickness 
of dielectric. Initially we fix the $T_{l}$ values to the {\it a priori} 
value of 0.17 for all double layers.

The transparency mixes up the real energy depositions 
$dE/dx^{\rm real}_{\rm low,\, up}$ into the amplitudes measured by the lower 
and upper strips $dE/dx^{\rm meas}_{\rm low,\, up}$:

\begin{equation}
\begin{bmatrix}
dE/dx^{\rm meas}_{\rm up}\\
dE/dx^{\rm meas}_{\rm low}
\end{bmatrix}
=
\frac{1}{1+T_{l}}
\begin{bmatrix}
1&T_{l}\\
T_{l}&1
\end{bmatrix}{\cdot}
\begin{bmatrix}
dE/dx^{\rm real}_{\rm up}\\
dE/dx^{\rm real}_{\rm low}
\end{bmatrix}.
\end{equation}

These relations should be understood as correct on average, or as the 
definitions of $dE/dx^{\rm real}_{\rm low,\, up}$.
For convenience in what follows we operate with the half sum and the
half difference of the $dE/dx^{\rm real}_{\rm low,\, up}$:

\begin{equation}
\begin{bmatrix}
dE/dx_{\rm summ}\\
dE/dx_{\rm diff}
\end{bmatrix}
=
\frac{1}{2(1-T_{l})}
\begin{bmatrix}
1& 1\\
1&-1
\end{bmatrix}{\cdot}
\begin{bmatrix}
1&-T_{l}\\
-T_{l}&1
\end{bmatrix}{\cdot}
\begin{bmatrix}
dE/dx^{\rm meas}_{\rm up}\\
dE/dx^{\rm meas}_{\rm low}
\end{bmatrix}.
\label{decorrelation}
\end{equation}

We use $dE/dx_{\rm summ}$ and $dE/dx_{\rm diff}$ in six inner double layers 
as the input variables of the BDT classifiers, described in 
Section~\ref{sec:PID_idea}. The outer seventh double layer suffers from the 
incomplete xenon fill and is not used in PID.
The data/MC comparison of the $dE/dx_{\rm summ}$ spectra for cosmic muons reveals 
the relative broadening of the experimental spectra, see 
Fig.~\ref{fig:dedx_summ_cosmics_tuning} (in what follows the simulated 
histograms are normalized to the number of events in the experimental one 
unless otherwise stated). 
The alleged reason of the broadening is the complicated cathode structure, 
not taken into account in the MC, where the 
cathode is supposed to be just a solid plane. To account for 
this broadening, we add in simulation
the random Gaussian noise to the amplitudes induced on the strips on 
both sides of the cathode. 
The width of the Gaussian noise is taken the same for all double layers,
its energy equivalent is ${\sim}0.8$~MeV. 
The resulting simulation agrees well with the data, see Fig.~\ref{fig:dedx_summ_cosmics_tuning}.

\begin{figure}
  \begin{center}
    \includegraphics[width=0.6\textwidth]{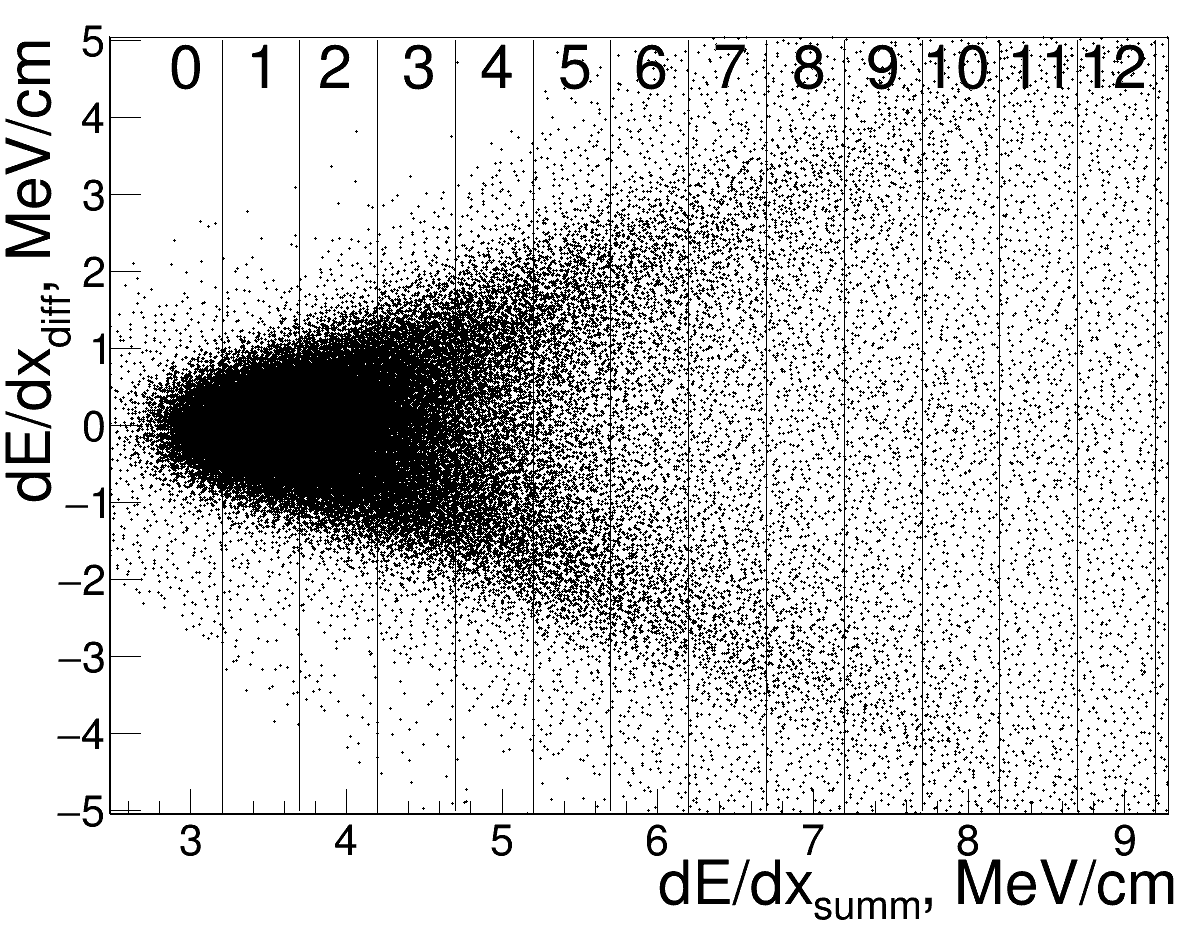}
    \caption{The $dE/dx_{\rm diff}$ vs. $dE/dx_{\rm summ}$ distribution 
    for the cosmic muons in the innermost double layer in the experiment. 
    The slicing on the $dE/dx_{\rm summ}$ is also shown.
    \label{fig:dedx_diff_vs_summ_cosmics}}
  \end{center}
\end{figure}

\begin{figure}[hbtp]
\begin{center}
\includegraphics[width=\textwidth]{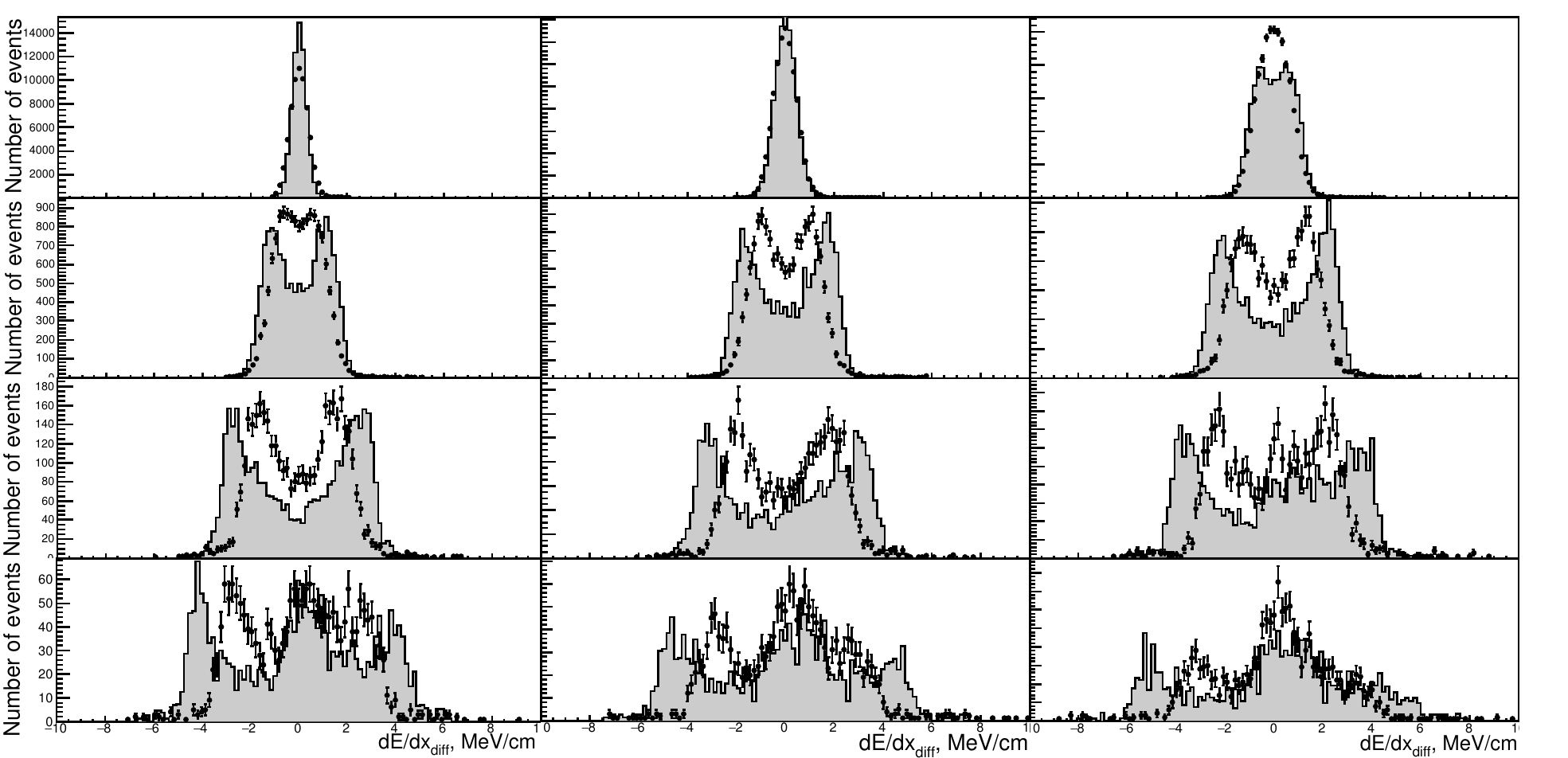}
\includegraphics[width=\textwidth]{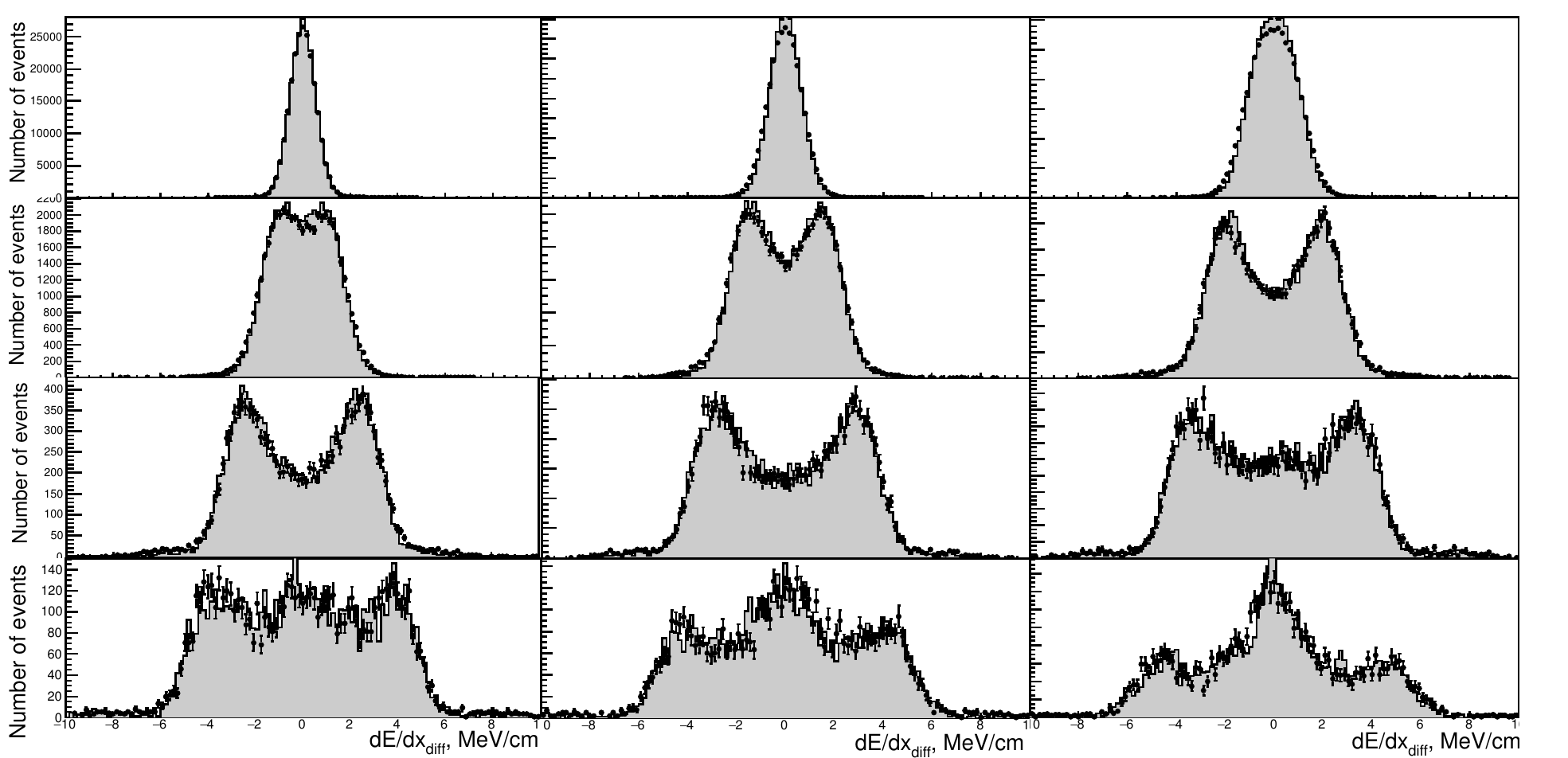}
\end{center}
\caption{The $dE/dx_{\rm diff}$ spectra for the cosmic muons in the third double layer 
in all slices in the experiment (markers) and MC (gray histogram) 
before (upper figure) and after (lower figure) $T_{l}$ tuning and addition of
the anticorrelated noise.
\label{fig:dedx_diff_layer_3_before_tuning}}
\end{figure}

Next, Fig.~\ref{fig:dedx_diff_vs_summ_cosmics} shows the distribution of 
the $dE/dx_{\rm diff}$ vs. $dE/dx_{\rm summ}$ for the 
cosmic muons in the innermost double layer in the experiment. 
The vertical lines show the slices of 
the distribution, inside which we perform the data/MC comparison of 
the $dE/dx_{\rm diff}$ spectra. For example,
such a comparison for the third double layer is shown 
in Fig.~\ref{fig:dedx_diff_layer_3_before_tuning}.
Since the position of the peaks in 
Fig.~\ref{fig:dedx_diff_layer_3_before_tuning}
 is mainly controlled by the $T_{l}$, the discrepancy between the data/MC 
peak positions  
 means that the {\it a priori} taken $T_{l}$ values are wrong. 
 We tune the $T_{l}$ values to achieve the coincidence of the peaks and 
thus obtain 
the true transparency coefficients $T_{1}=0.23$, $T_{2}=0.22$, $T_{3}=0.35$, 
$T_{4}=0.32$, $T_{5}=0.35$, 
$T_{6}=0.33$, $T_{7}=0.33$ with about 5\% uncertainty.

Apart from the shift of the peak positions we observe the relative broadening 
of the experimental $dE/dx_{\rm diff}$ spectra, presumably related 
to the variation of the transparency around its average value of $T_{l}$.
To account for this broadening, we add the anticorrelated Gaussian noise to the amplitudes, 
induced on the upper and lower strips in simulation. 
This means that the same random value is added to the amplitude of the upper strips and subtracted from 
the amplitude of the lower strips. This additional anticorrelated 
noise simulates the effect of the redistribution of the charge between 
upper and lower strips due to the $T_{l}$ variations. The variance of the 
additional noise is tuned individually in all double layers, 
the noise energy equivalents are ${\sim}0.6-0.12$~MeV depending on the layer.
After the applied corrections we observe a good data/MC agreement 
in the $dE/dx_{\rm diff}$ spectra, see Fig.~\ref{fig:dedx_diff_layer_3_before_tuning}.

\subsection{Tuning for electromagnetic shower}

\begin{figure}[hbtp]
  \begin{minipage}[t]{0.32\textwidth}
   \centerline{\includegraphics[width=0.98\textwidth]{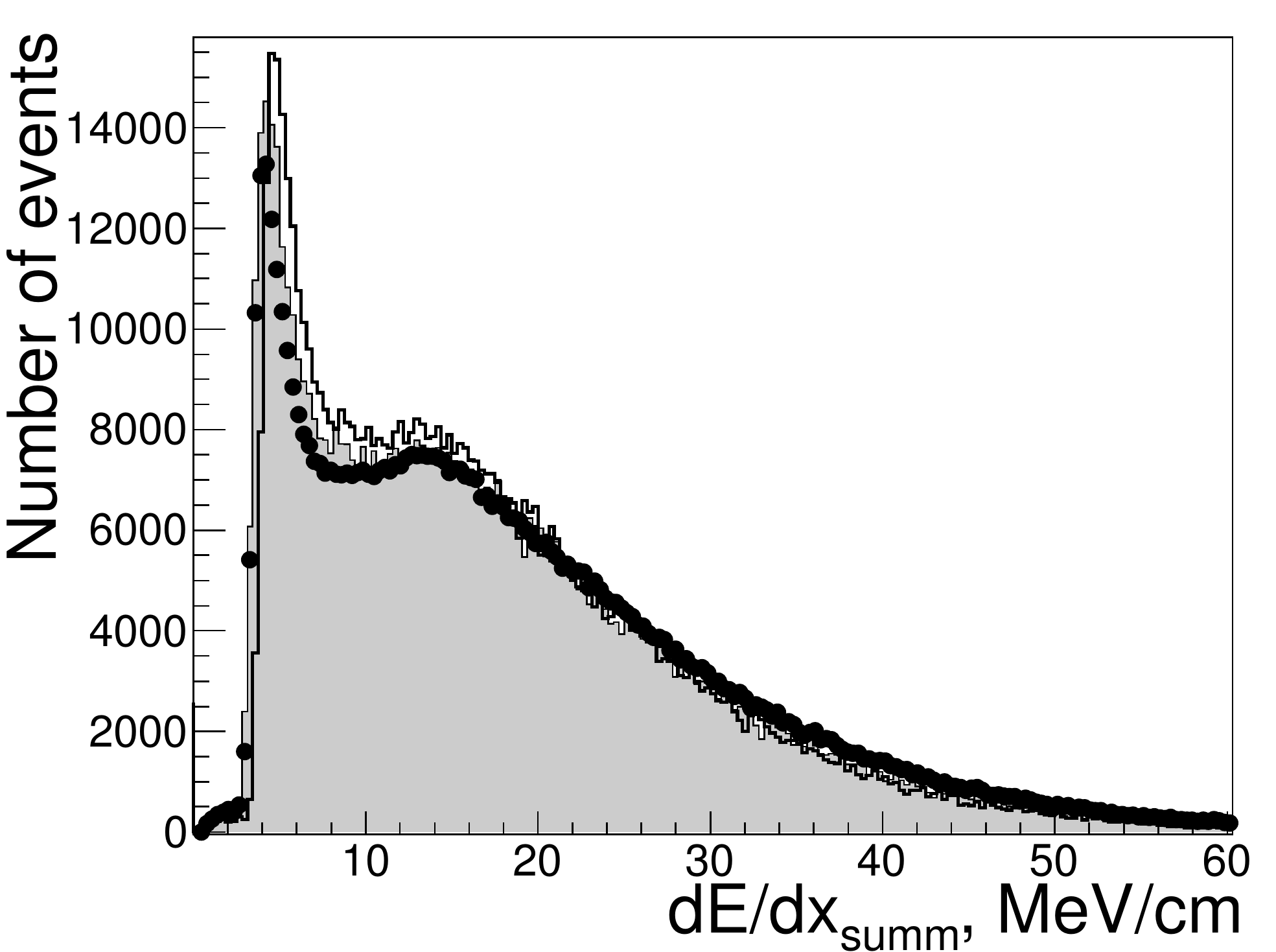}}    
  \end{minipage}\hfill\hfill
   \begin{minipage}[t]{0.32\textwidth}
   \centerline{\includegraphics[width=0.98\textwidth]{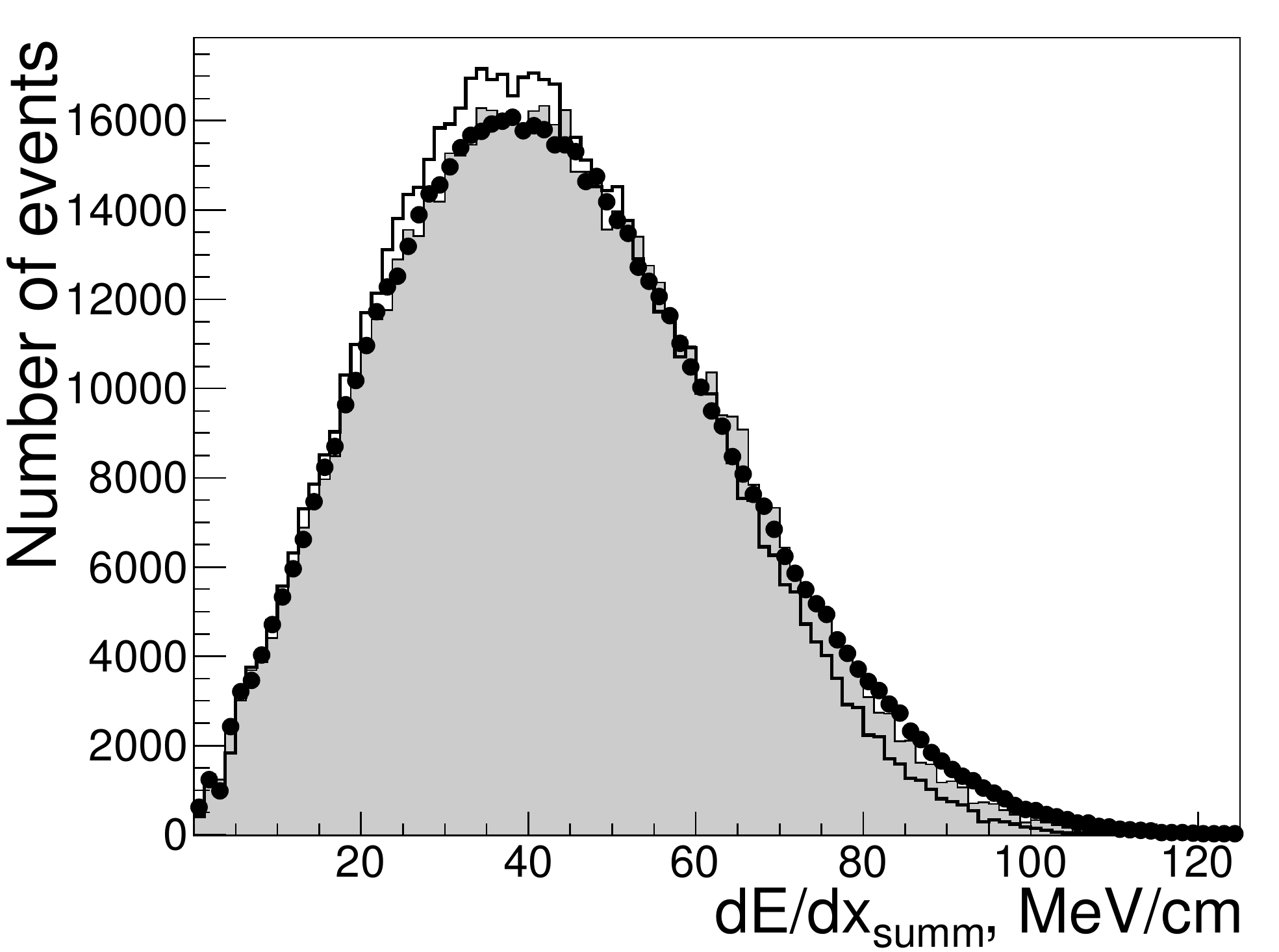}}
   \end{minipage}\hfill\hfill
  \begin{minipage}[t]{0.32\textwidth}
    \centerline{\includegraphics[width=0.98\textwidth]{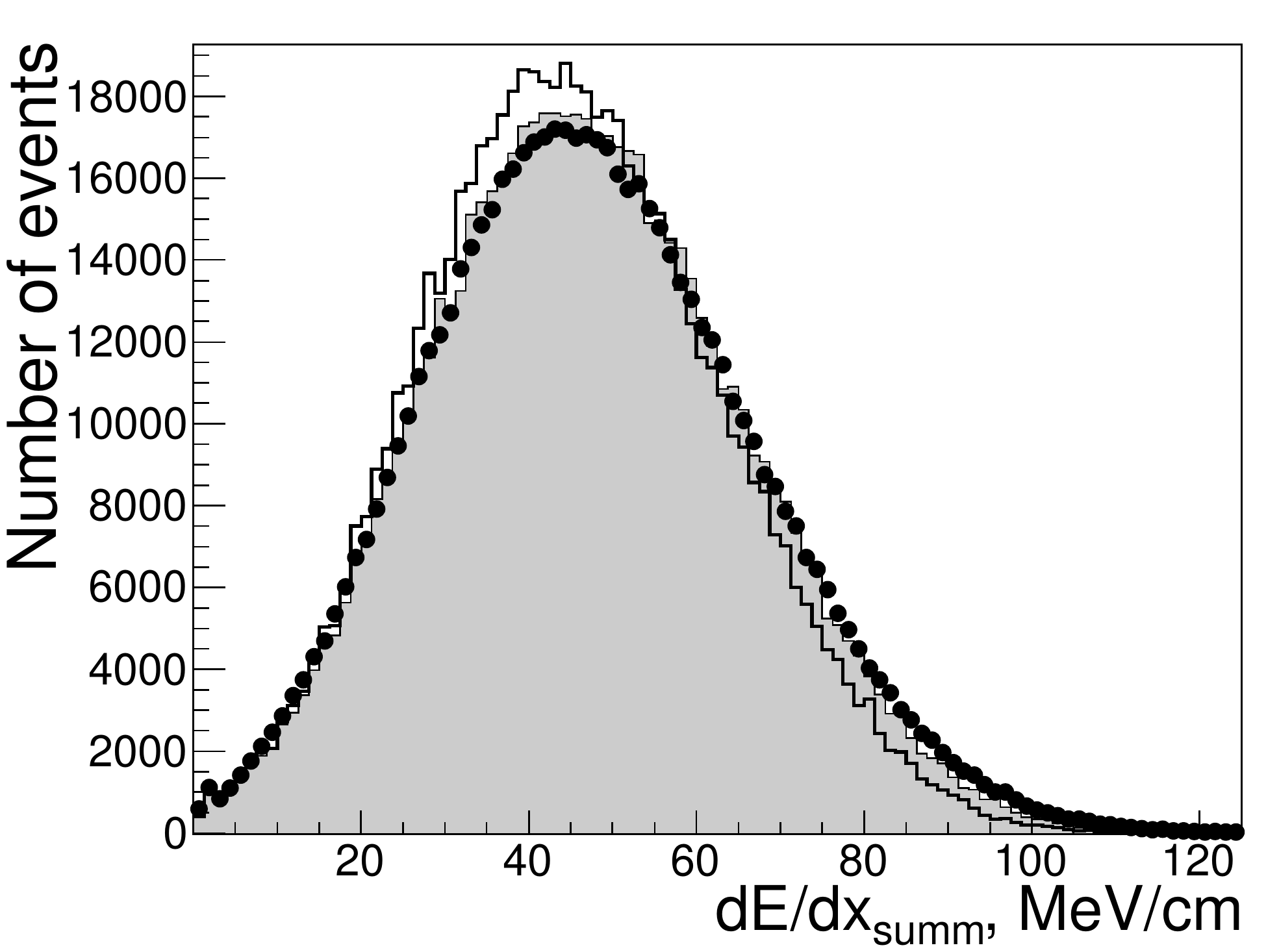}}    
  \end{minipage}\hfill\hfill  
  \caption{The $dE/dx_{\rm summ}$ spectra in the first (left), third (middle) and fifth (right) double layers
  for $e^{\pm}$ from  the process $e^{+}e^{-}{\to}e^{+}e^{-}$ 
  in the experiment (markers), MC before (open histogram) 
  and after (gray histogram) linear transformation. 
  The beam energy is 987.5~MeV.    
          \label{fig:dedx_summ_EpEm}}
\end{figure}

Another kind of the data/MC discrepancy is observed in the $dE/dx_{\rm summ}$ 
spectra  for the electromagnetic (e.m.) showers, produced in the calorimeter 
by electrons and positrons from the process $e^{+}e^{-}{\to}e^{+}e^{-}$, see 
Fig.~\ref{fig:dedx_summ_EpEm}. The additional noises used for tuning of the MIPs 
in simulation show no serious effect on the large amplitudes 
of the e.m. showers. The actual reasons of the observed discrepancy 
remain unclear, but many 
possible sources were studied, including the imprecise description 
of the dead material in front 
of the calorimeter, the influence of the electronegative admixtures in LXe, 
the inaccurate value 
of LXe density {\it etc.} Fortunatelly, the discrepancy can be mostly 
eliminated by the simple linear 
transformation of the simulated amplitudes 
$dE/dx^{\rm meas,\, corr} = a{\cdot}dE/dx^{\rm meas}-b$,
where $a=1.055$ is the ``additional calibration'' coefficient for 
the showers and $b=0.7$ is the shift introduced to reach the 
data/MC coincidence of the minimum ionizing peaks in the innermost 
double layer. The resulting data/MC good agreement (except the innermost 
double layer), 
shown in Fig.~\ref{fig:dedx_summ_EpEm}, is conserved for all $e^{\pm}$ 
momenta and angles.

\section{\boldmath Spectra of classifier response and signal/background separation power}
\hspace*{\parindent}

\begin{figure}[hbtp]
  \begin{minipage}[t]{0.49\textwidth}
   \centerline{\includegraphics[width=0.98\textwidth]{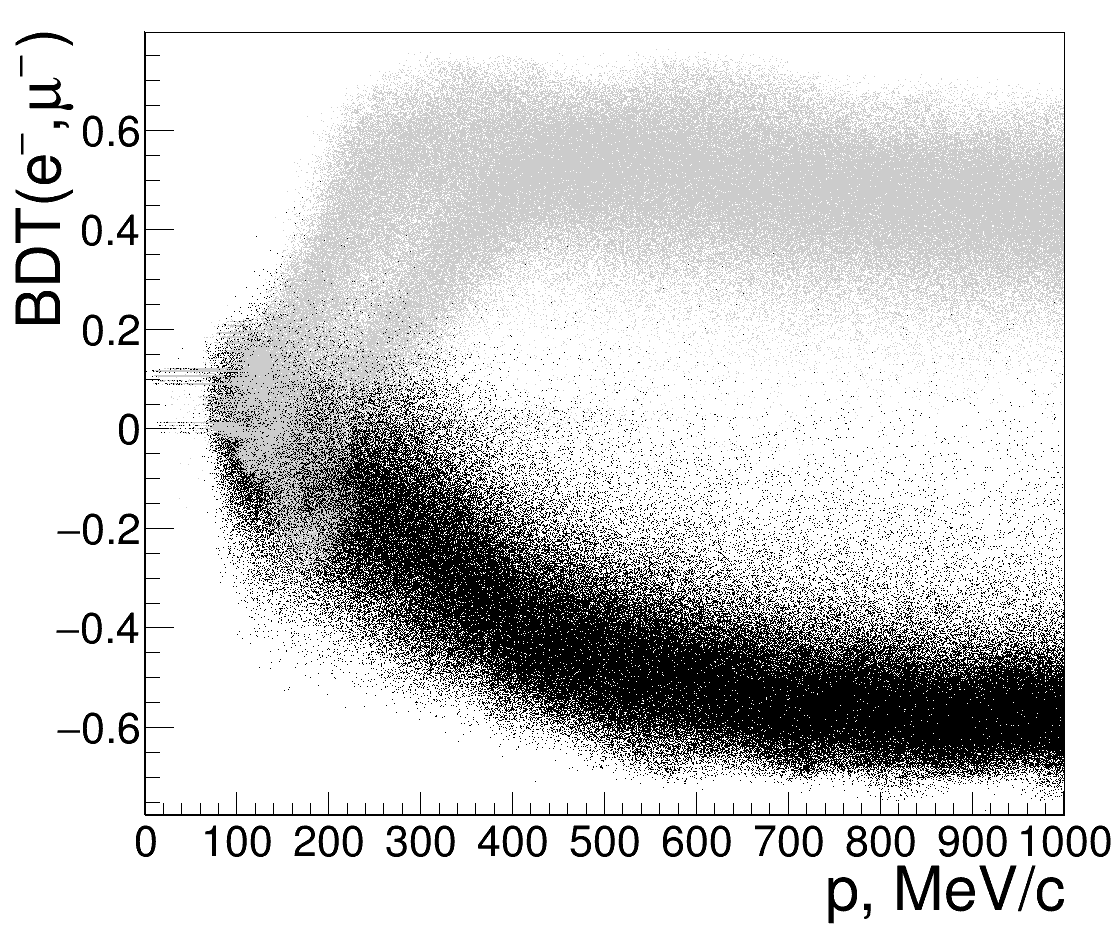}}
    \caption{The ${\rm BDT}(e^{-},\mu^{-})$ vs. particle momentum for 
    simulated $e^{-}$ (black) and $\mu^{-}$ (gray), uniformly distributed in $d_{\rm LXe}$. 
          \label{fig:bdt_e_mu_vs_p}}
  \end{minipage}\hfill\hfill
  \begin{minipage}[t]{0.49\textwidth}
    \centerline{\includegraphics[width=0.98\textwidth]{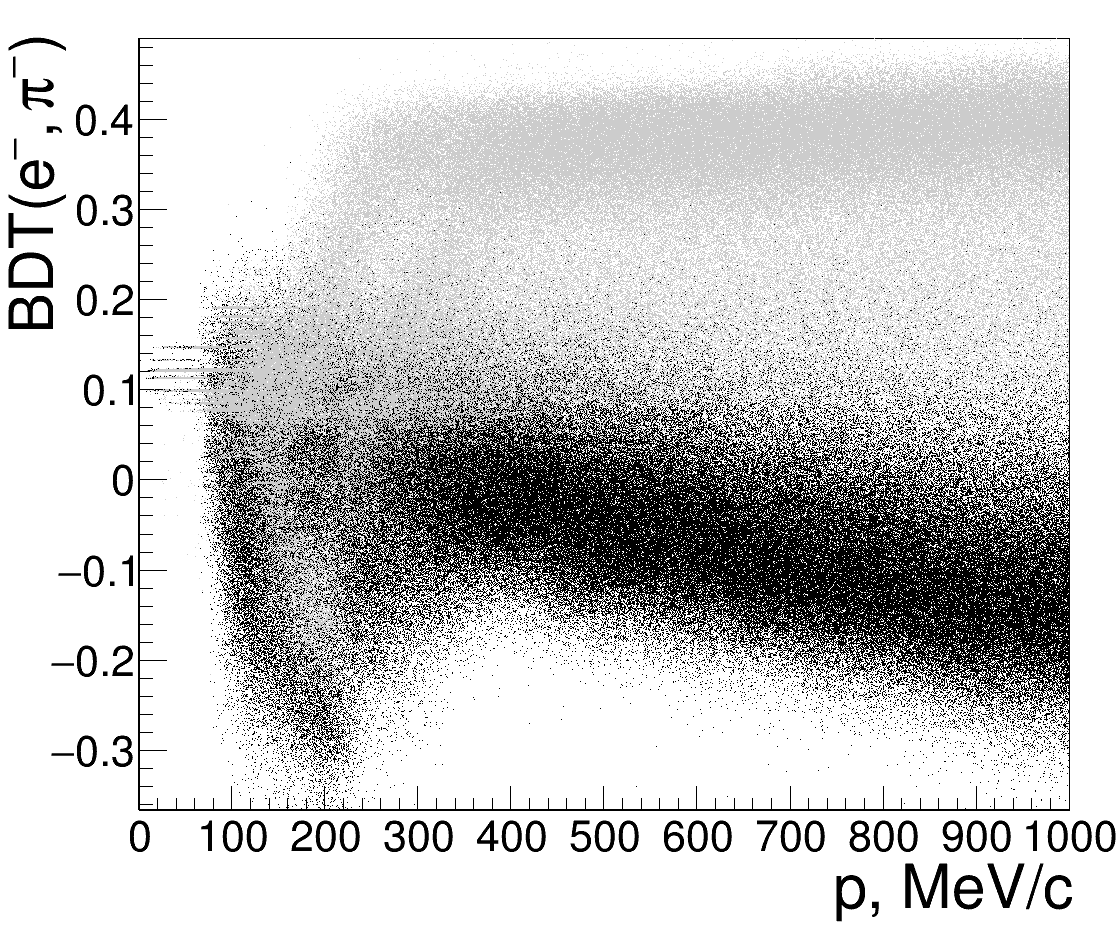}}
    \caption{The ${\rm BDT}(e^{-},\pi^{-})$ vs. particle momentum for 
    simulated $e^{-}$ (black) and $\pi^{-}$ (gray), uniformly distributed in $d_{\rm LXe}$.        
          \label{fig:bdt_e_pi_vs_p}}
  \end{minipage}\hfill\hfill
\end{figure}

\begin{figure}[hbtp]
  \begin{minipage}[t]{0.49\textwidth}
   \centerline{\includegraphics[width=0.98\textwidth]{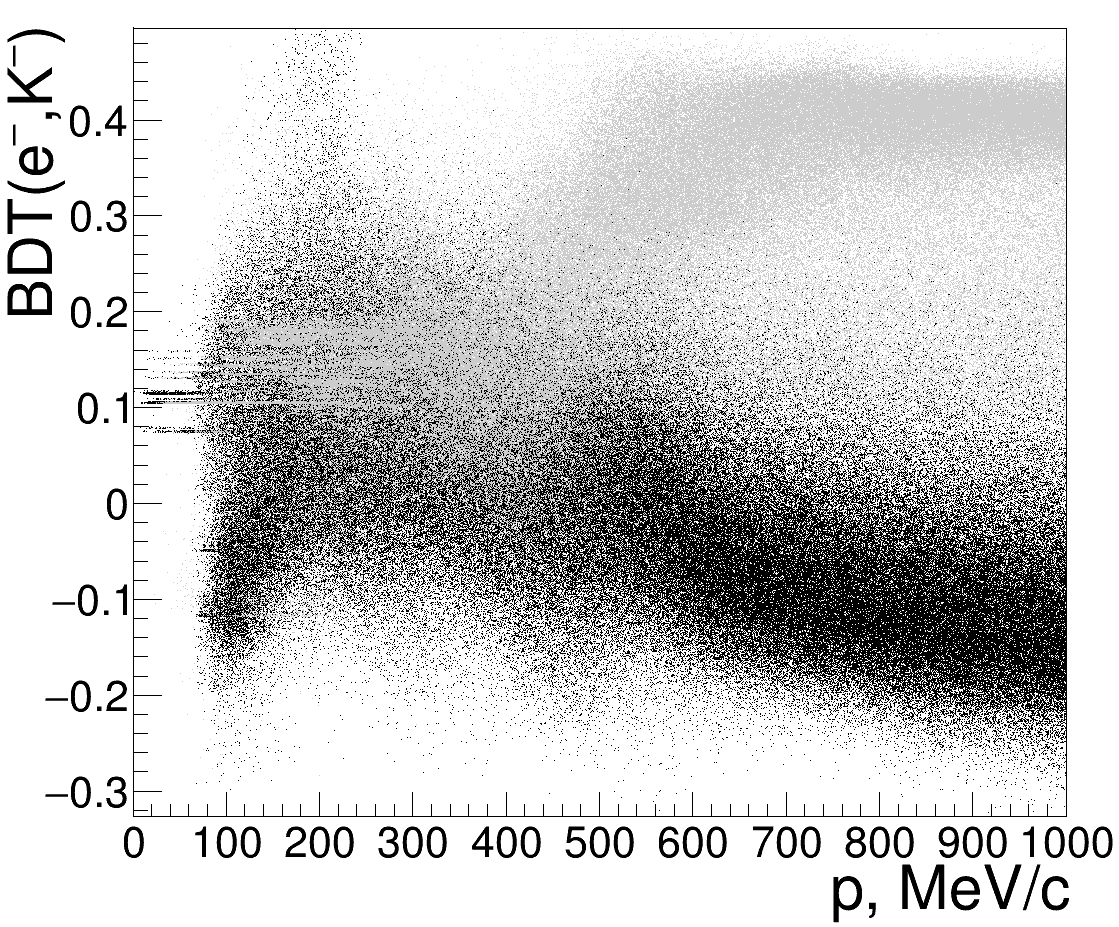}}
    \caption{The ${\rm BDT}(e^{-},K^{-})$ vs. particle momentum for the 
    simulated $e^{-}$ (black) and $K^{-}$ (gray), uniformly distributed in $d_{\rm LXe}$. 
          \label{fig:bdt_e_K_vs_p}}
  \end{minipage}\hfill\hfill
  \begin{minipage}[t]{0.49\textwidth}
    \centerline{\includegraphics[width=0.98\textwidth]{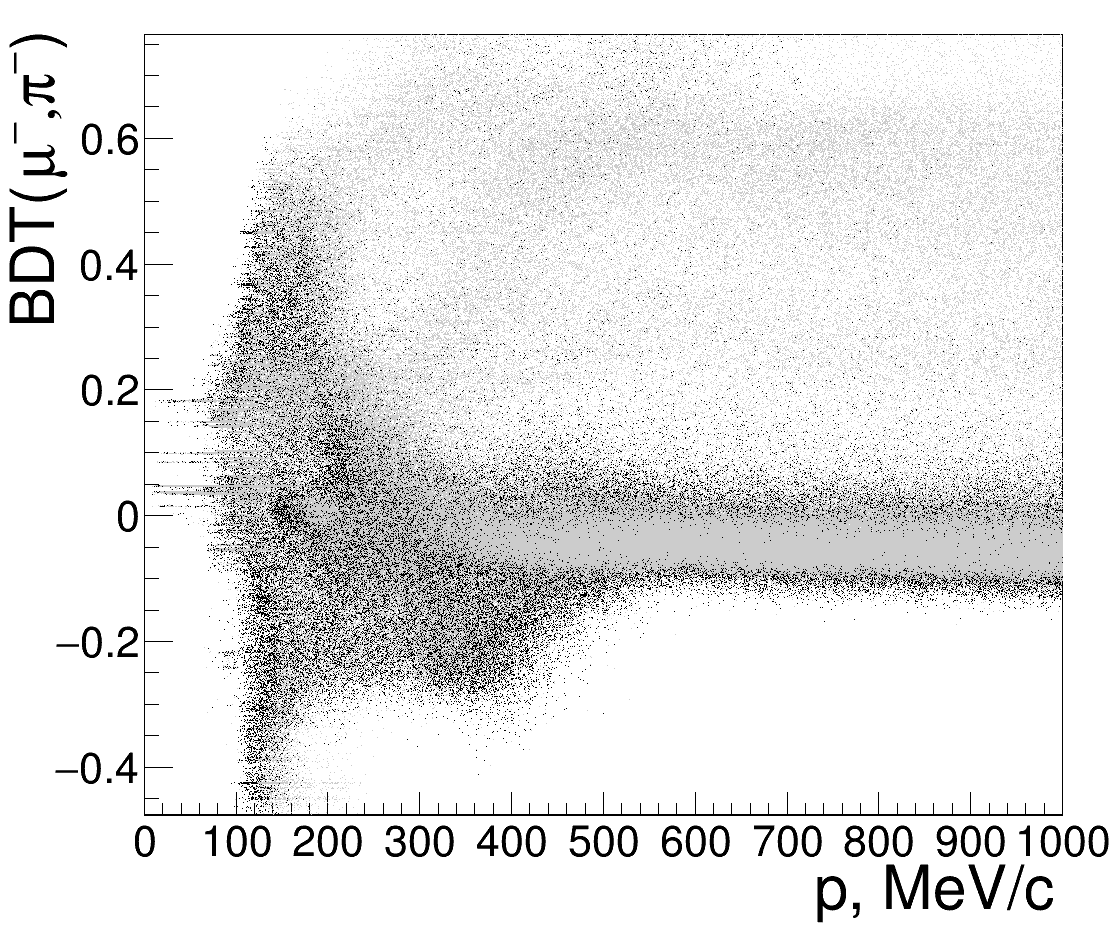}}
    \caption{The ${\rm BDT}(\mu^{-},\pi^{-})$ vs. particle momentum for 
    simulated $\mu^{-}$ (black) and $\pi^{-}$ (gray), uniformly distributed in $d_{\rm LXe}$.        
          \label{fig:bdt_mu_pi_vs_p}}
  \end{minipage}\hfill\hfill
\end{figure}

\begin{figure}[hbtp]
  \begin{minipage}[t]{0.49\textwidth}
   \centerline{\includegraphics[width=0.98\textwidth]{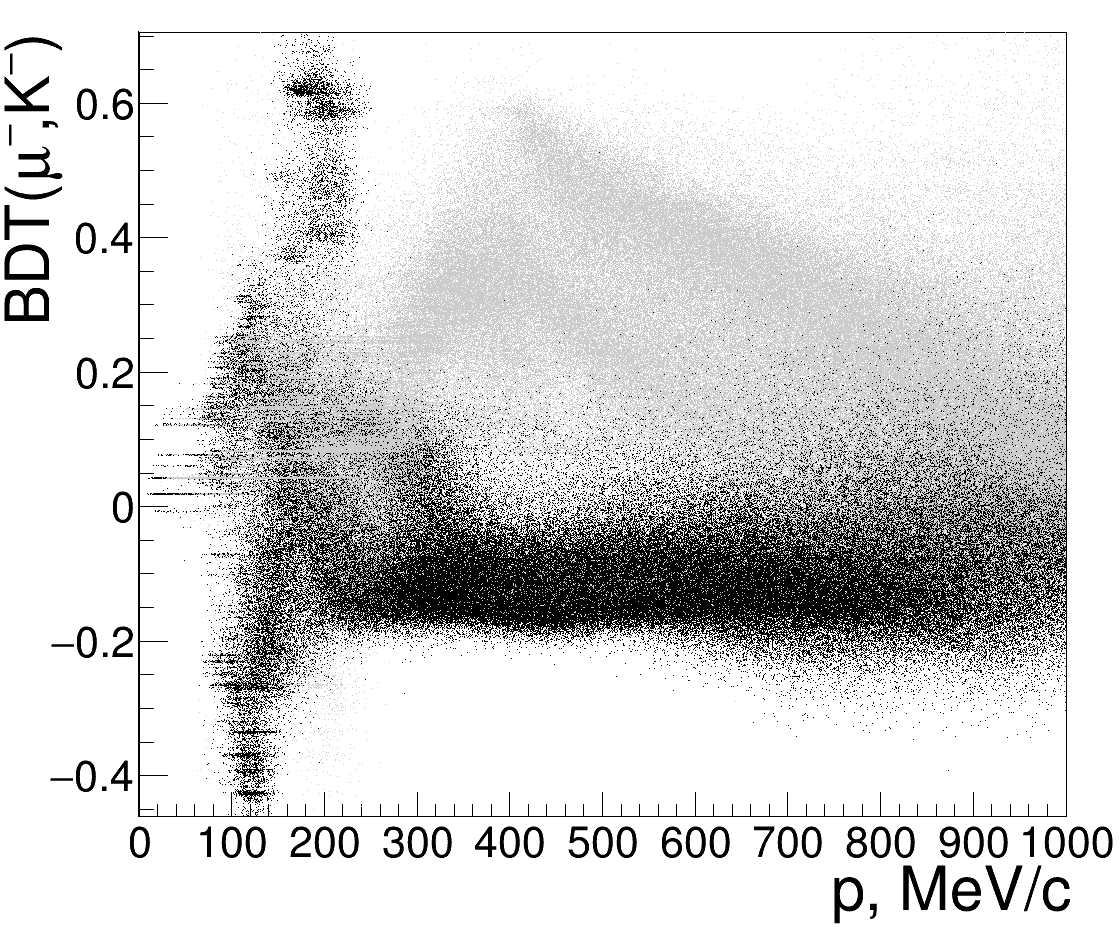}}
    \caption{The ${\rm BDT}(\mu^{-},K^{-})$ vs. particle momentum for 
    simulated $\mu^{-}$ (black) and $K^{-}$ (gray), uniformly distributed in $d_{\rm LXe}$. 
          \label{fig:bdt_mu_K_vs_p}}
  \end{minipage}\hfill\hfill
  \begin{minipage}[t]{0.49\textwidth}
    \centerline{\includegraphics[width=0.98\textwidth]{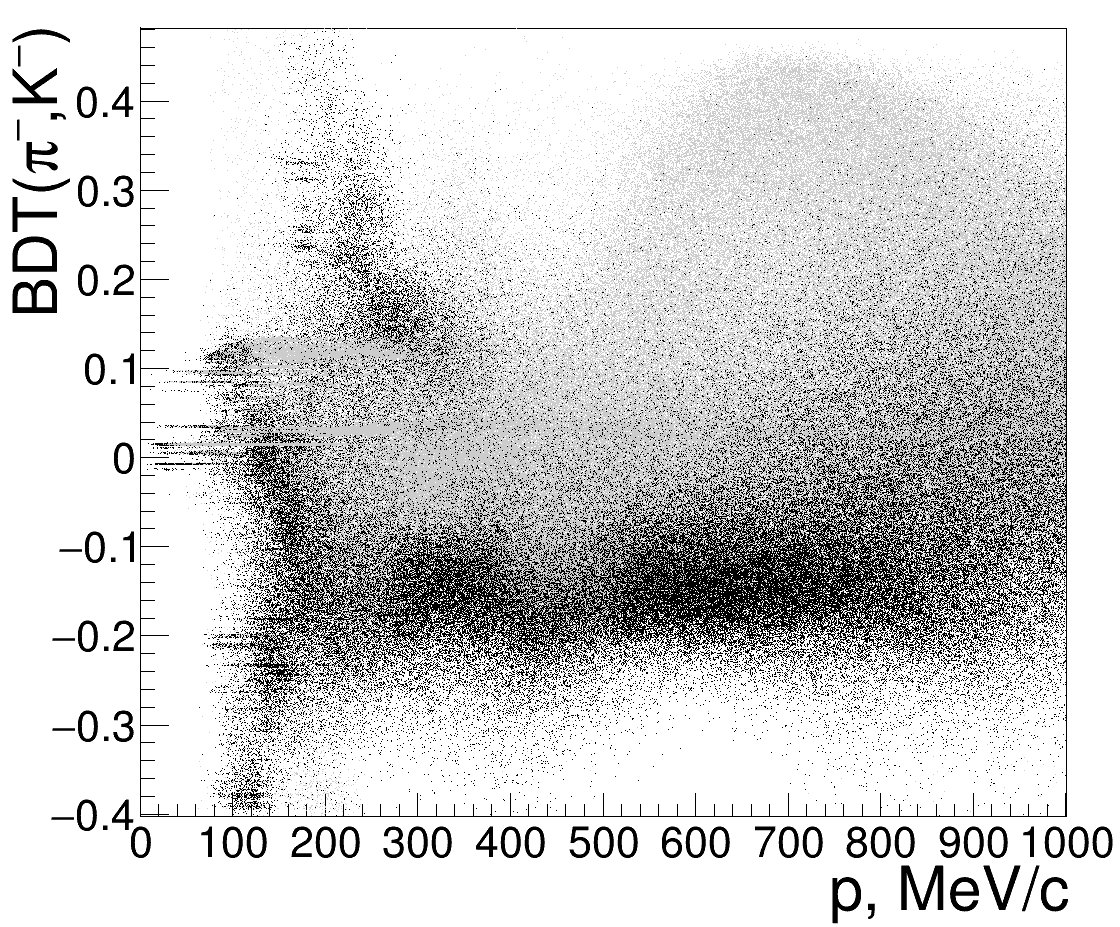}}
    \caption{The ${\rm BDT}(\pi^{-},K^{-})$ vs. particle momentum for 
    simulated $\pi^{-}$ (black) and $K^{-}$ (gray), uniformly distributed in $d_{\rm LXe}$.        
          \label{fig:bdt_pi_K_vs_p}}
  \end{minipage}\hfill\hfill
\end{figure}

In this section we perform the data/MC comparison
of the resulting BDT response spectra for different types of particles.
Figures~\ref{fig:bdt_e_mu_vs_p}--\ref{fig:bdt_pi_K_vs_p} provide a 
general view of the potential effectiveness of all six types of classifiers
as a function of particle momentum according to simulation. 
The ``comb'' in the BDT spectra at low momentum corresponds to the 
cases when all input variables of the classifiers are zero.
It is seen, that $\mu / \pi$ separation (Fig.~\ref{fig:bdt_mu_pi_vs_p}) is not effective at all, 
whereas separation of $e^{\pm}$ from $\mu^{\pm}$, $\pi^{\pm}$ and $K^{\pm}$ 
(Figs.~\ref{fig:bdt_e_mu_vs_p}--\ref{fig:bdt_e_K_vs_p}) is effective starting from some threshold momentum.

\subsection{Electrons/positrons}

\begin{figure} [hbtp]
\centering
\begin{tabular}{cccc}
\includegraphics[width=0.31\textwidth]{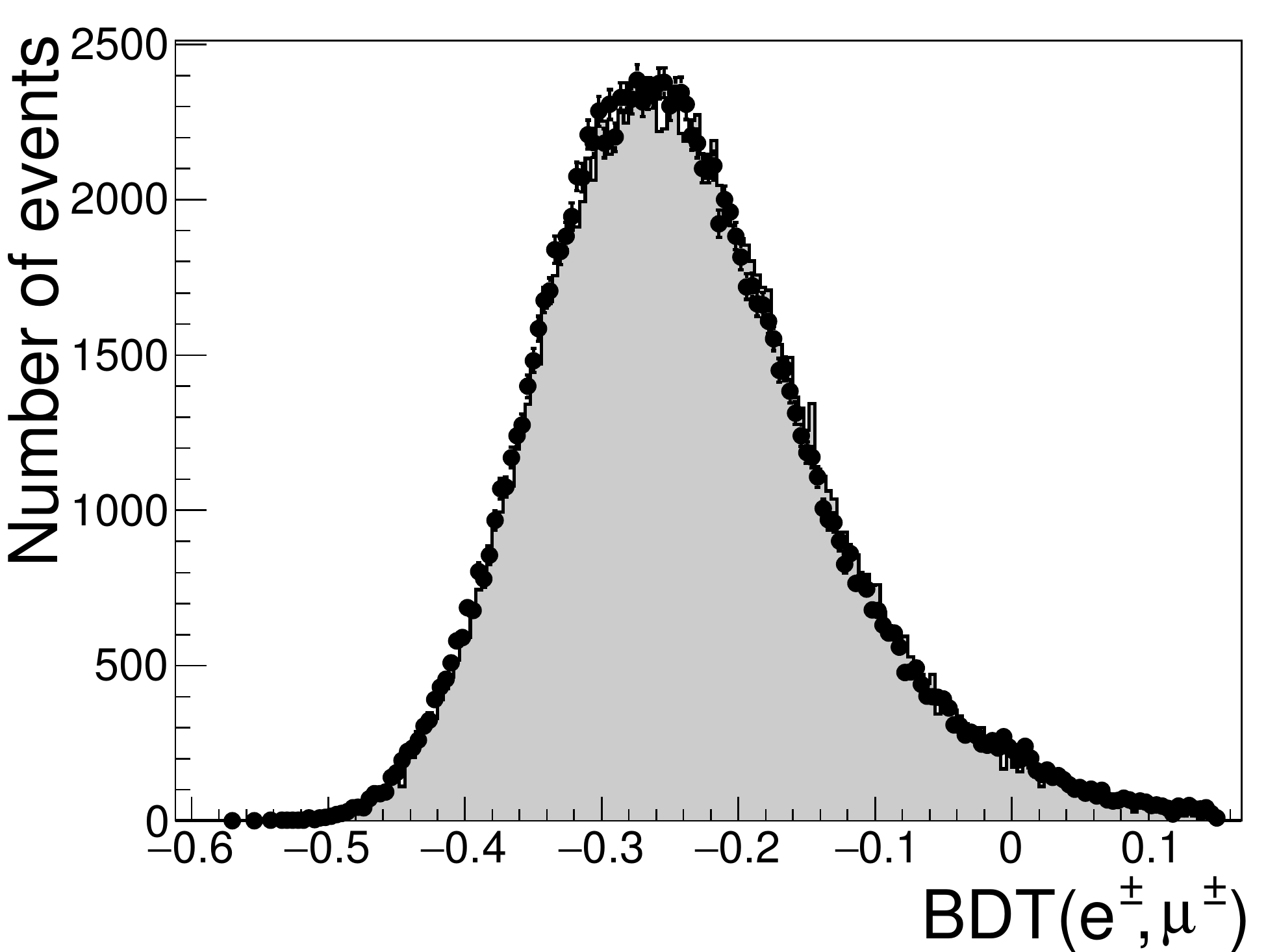} &
\includegraphics[width=0.31\textwidth]{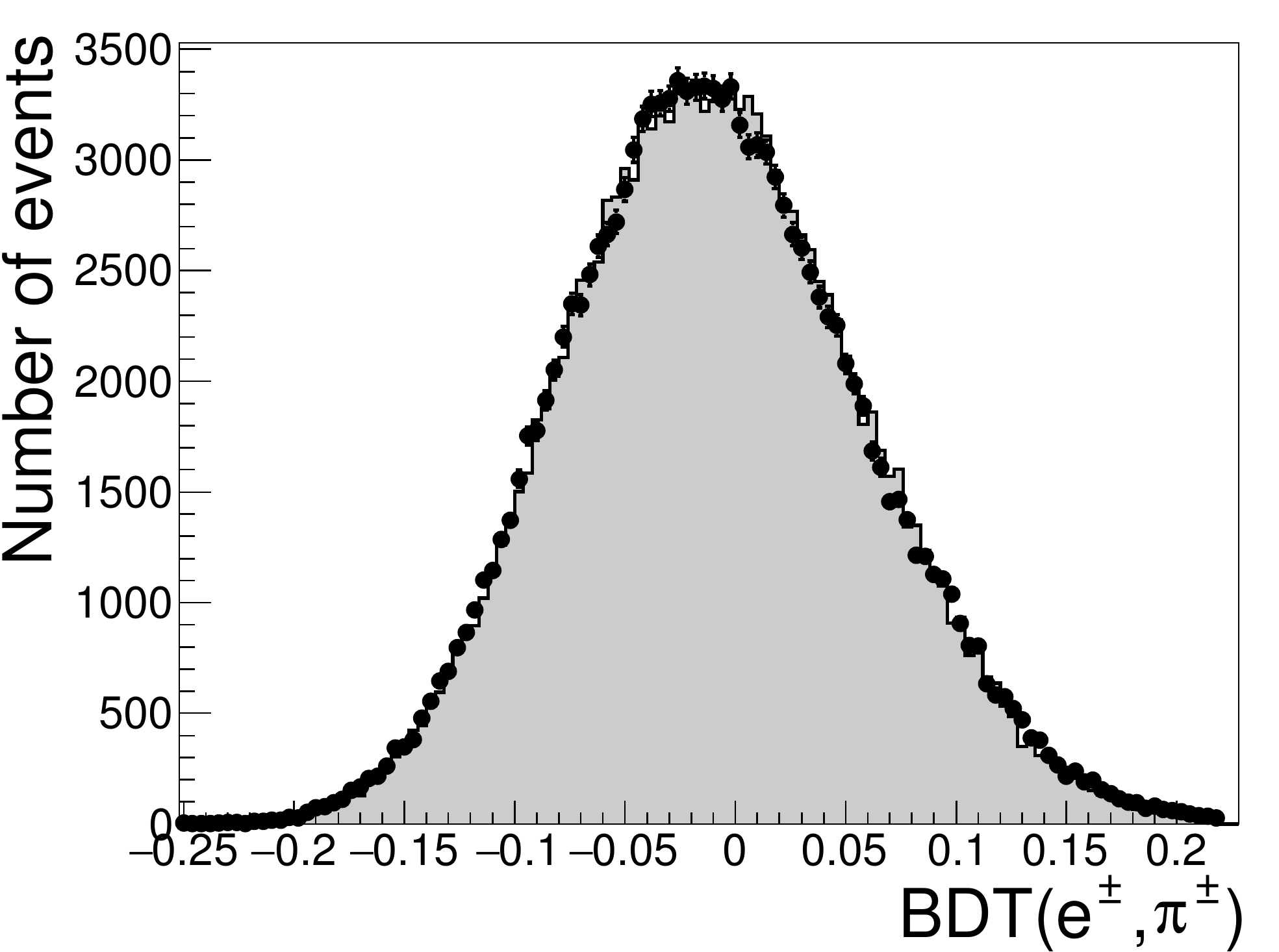} &
\includegraphics[width=0.31\textwidth]{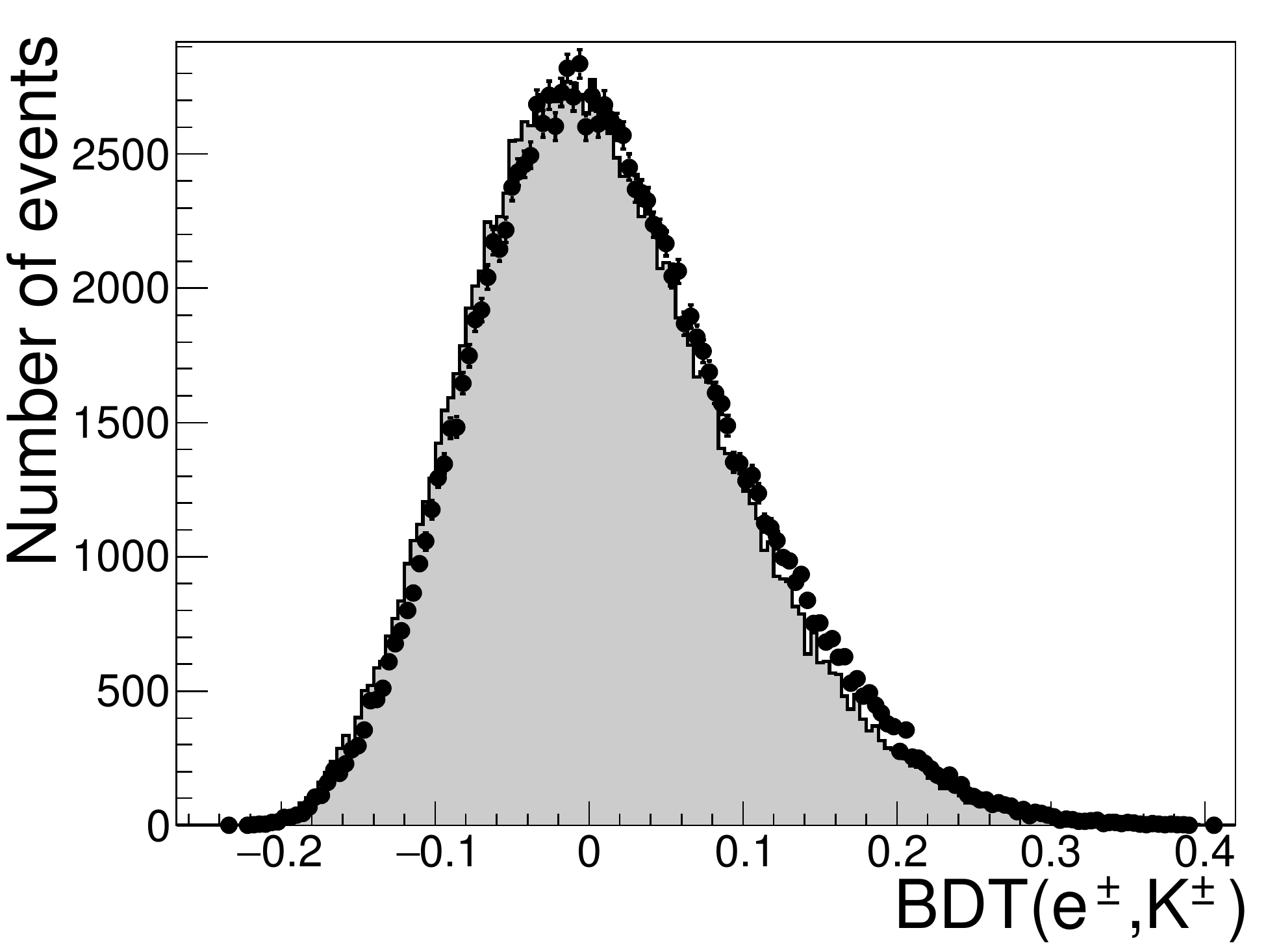} \\
\end{tabular}
\begin{tabular}{cccc}
\includegraphics[width=0.31\textwidth]{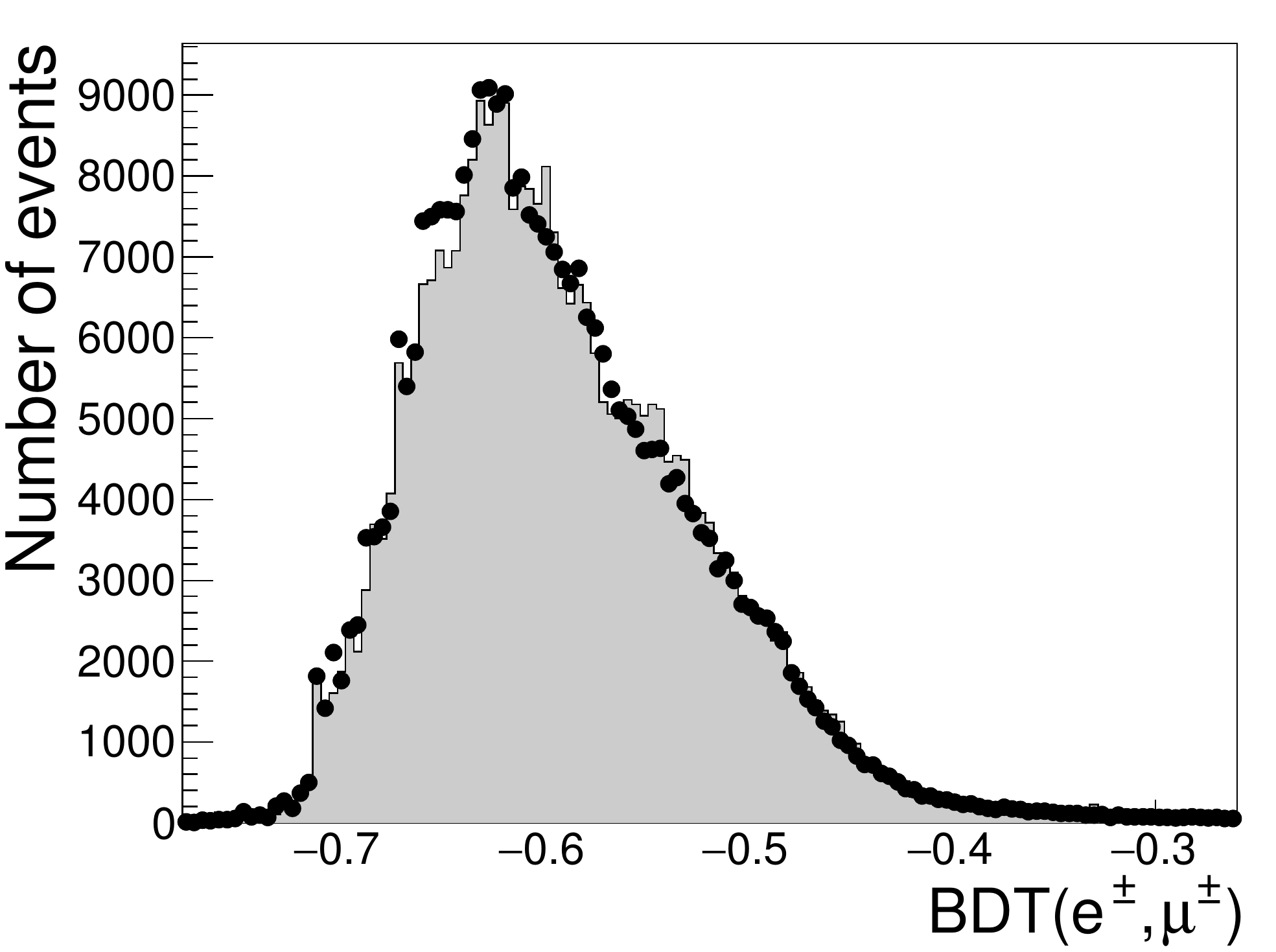} &
\includegraphics[width=0.31\textwidth]{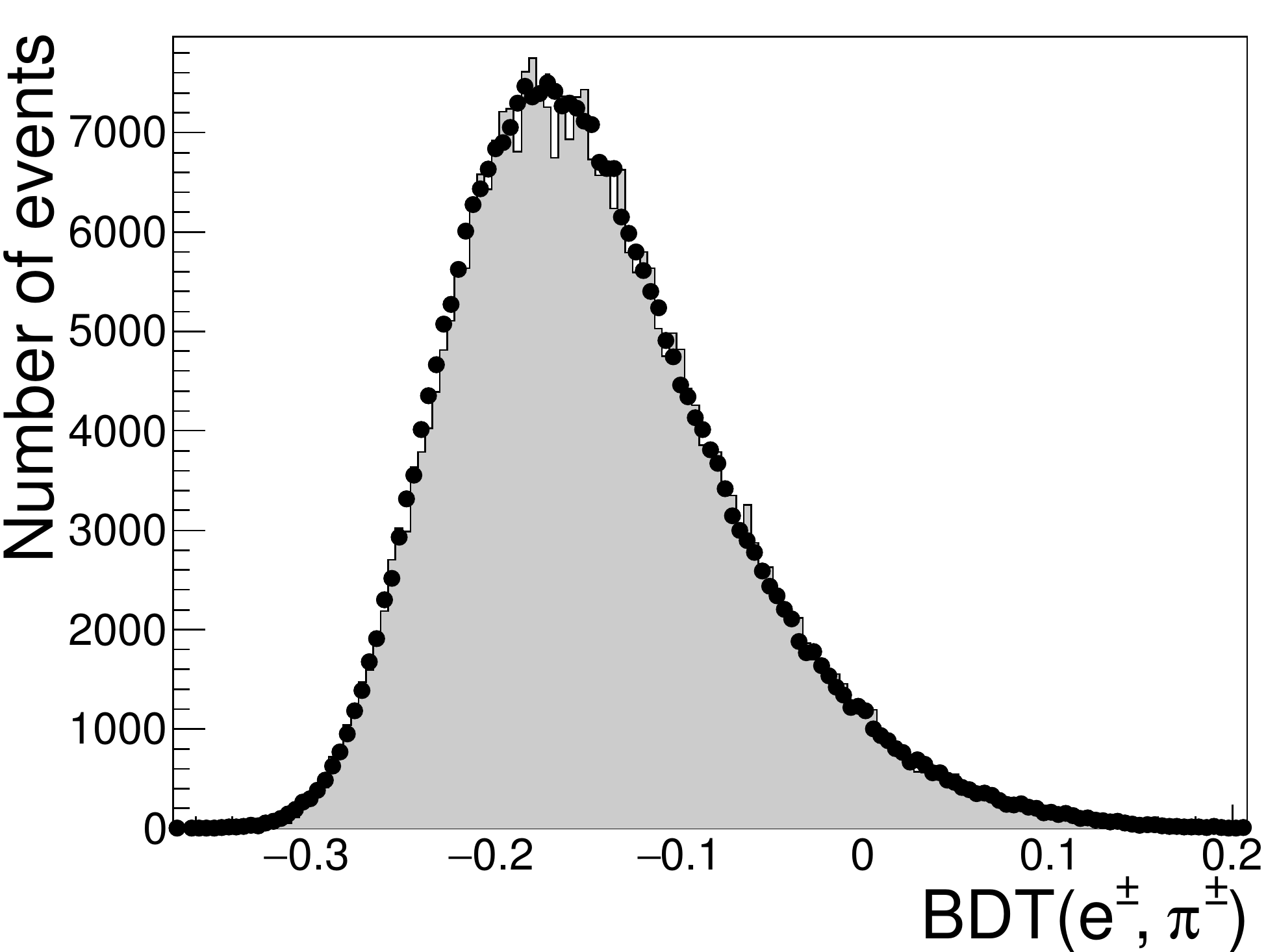} &
\includegraphics[width=0.31\textwidth]{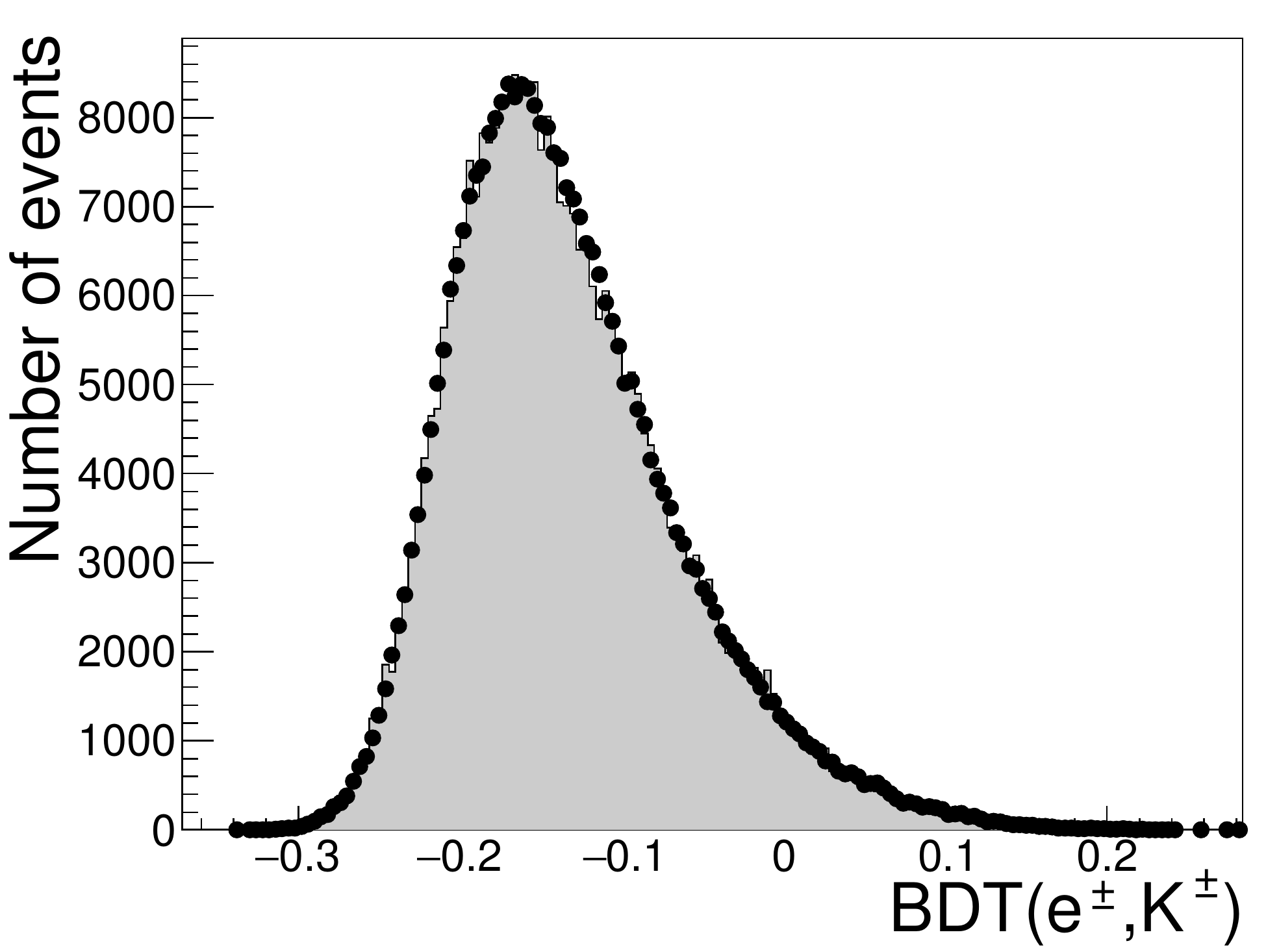} \\
\end{tabular}
\caption{The ${\rm BDT}(e^{\pm},\mu^{\pm})$ (left), ${\rm BDT}(e^{\pm},\pi^{\pm})$ 
(middle) and ${\rm BDT}(e^{\pm},K^{\pm})$ 
(right) spectra for the $e^{\pm}$ selected from 
$e^{+}e^{-}{\to}e^{+}e^{-}$ events in the experiment (markers) 
and MC (gray histrogram) at 
$E_{\rm beam}=280$~MeV (top figures) and $E_{\rm beam}=987.5$~MeV (bottom figures).
\label{fig:bdt_EpEm}}
\end{figure}

We select $e^{\pm}$ from $e^{+}e^{-}{\to}e^{+}e^{-}$ events 
using the following criteria: 
1) there are exactly two DC-tracks with the opposite charges; 
2) the $|\rho|$ and $|z|$ of the track point of the 
closest approach to the beam axis should be less than
0.6 and 12~cm, respectively;
3) the polar angles of tracks should be in the range from 0.9 to $\pi-0.9$~rad; 
4) the tracks are collinear: $|\theta_{1}+\theta_{2}-\pi|<0.15$~rad
and $||\varphi_{1}-\varphi_{2}|-\pi|<0.15$~rad; 
5) the energy deposition of each particle in the barrel calorimeter (LXe and CsI) is larger 
than the half beam energy ($E_{\rm beam}$).
Data/MC comparison for the ${\rm BDT}(e^{\pm},\mu^{\pm})$, ${\rm 
BDT}(e^{\pm},\pi^{\pm})$ and ${\rm BDT}(e^{\pm},K^{\pm})$ 
spectra for the selected $e^{\pm}$ at the low (280~MeV) and high (987.5~MeV) 
beam energies is shown in 
Fig.~\ref{fig:bdt_EpEm}. Agreement is good in all cases.

\subsection{Muons}

\begin{figure}[hbtp]
  \begin{minipage}[t]{0.33\textwidth}
   \centerline{\includegraphics[width=0.98\textwidth]{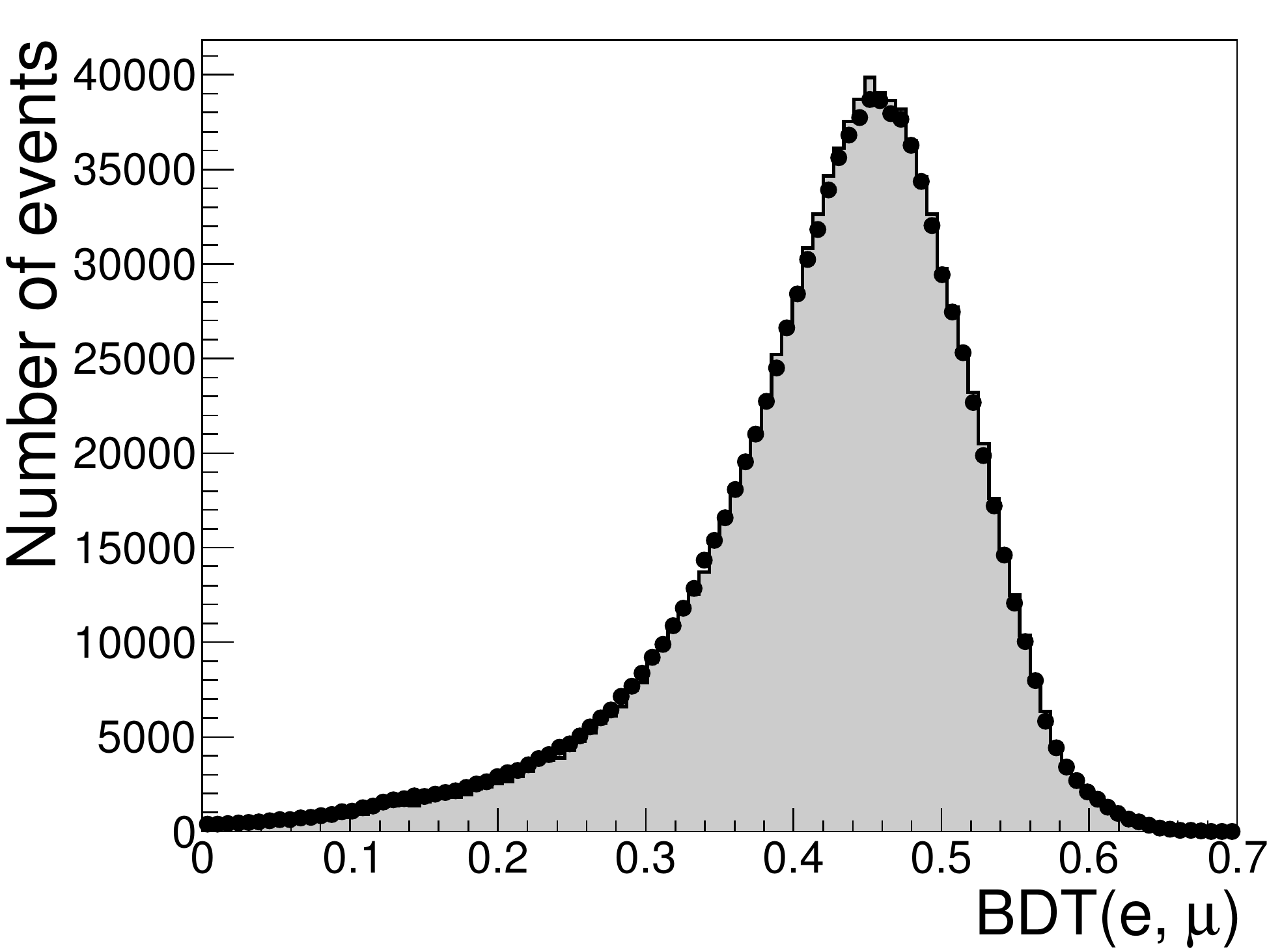}}
  \end{minipage}\hfill\hfill
   \begin{minipage}[t]{0.33\textwidth}
   \centerline{\includegraphics[width=0.98\textwidth]{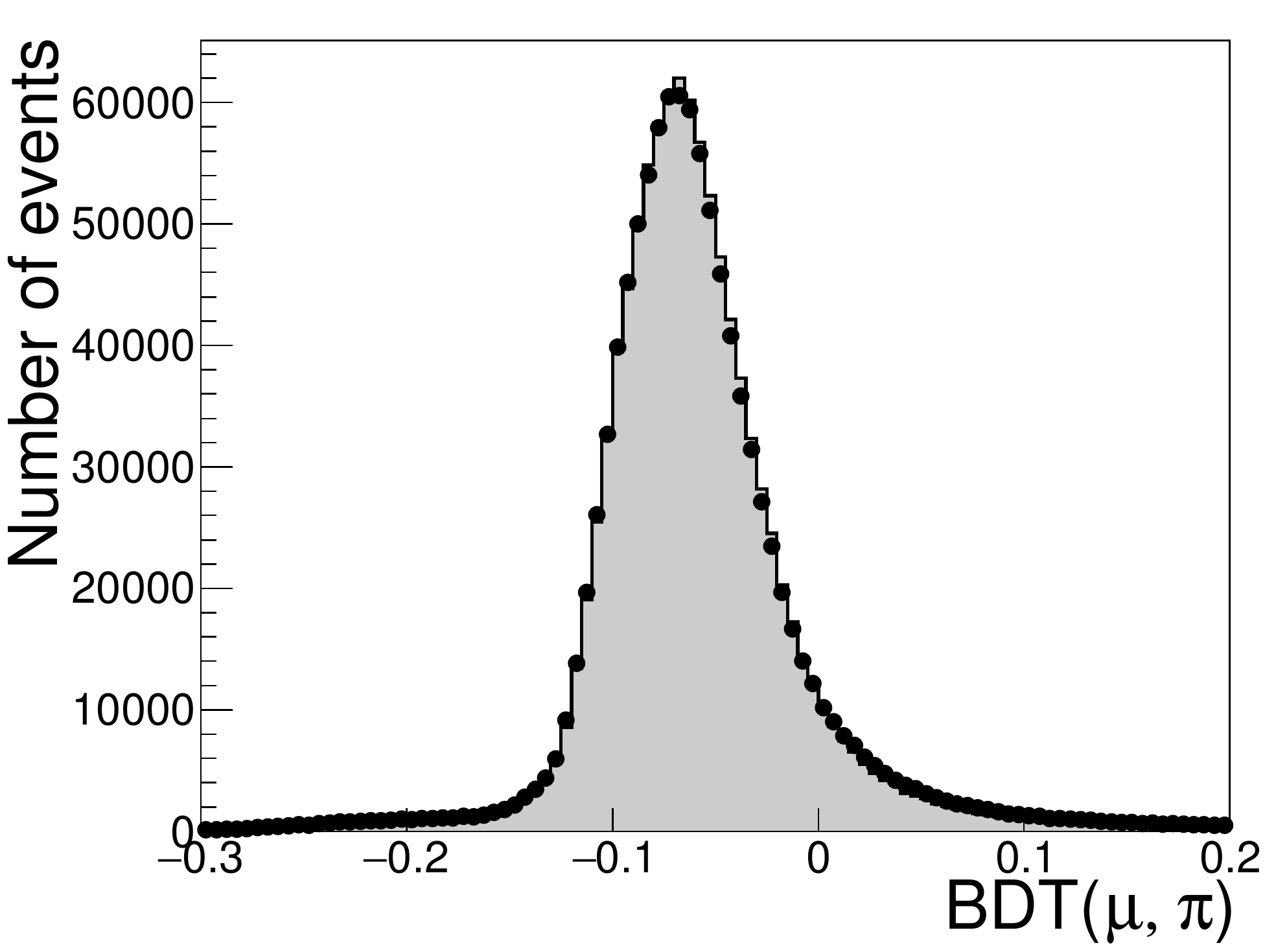}}
  \end{minipage}\hfill\hfill
  \begin{minipage}[t]{0.33\textwidth}
    \centerline{\includegraphics[width=0.98\textwidth]{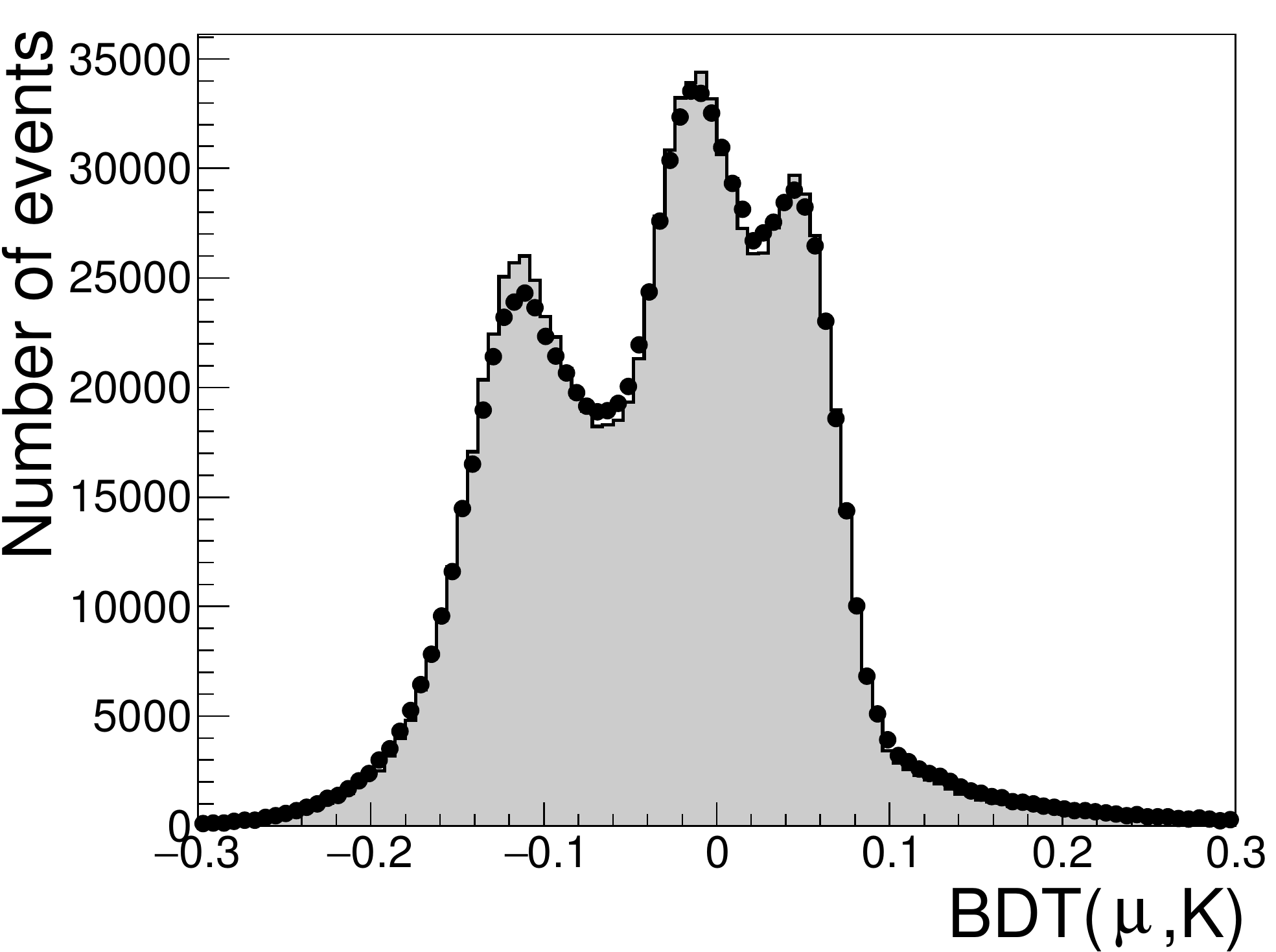}}
  \end{minipage}\hfill\hfill
  \caption{The ${\rm BDT}(e^{\pm},\mu^{\pm})$ (left), ${\rm 
BDT}(\mu^{\pm},\pi^{\pm})$ (middle) and ${\rm BDT}(\mu^{\pm},K^{\pm})$ 
(right) spectra for the cosmic $\mu^{\pm}$ in the experiment (markers) 
and MC (gray histrogram). The muon momentum is in the range from 100 
to 1200~MeV/{\rm c}. 
          \label{fig:bdt_cosmics}}
\end{figure}

We select a sample of $\mu^{\pm}$ from events with cosmic muons 
using the following criteria:
1) there is only one DC-track; 2) the track momentum is in the 
range from 100 to 1200~MeV/$c$; 3) track is not central: the minimal distance 
from the track to the beam axis is in the range from 3 to 15~cm; 
4) the energy deposition of the particle in 
the calorimeter is less than 400~MeV. 
Reasonable data/MC agreement for the ${\rm BDT}(e^{\pm},\mu^{\pm})$, 
${\rm BDT}(\mu^{\pm},\pi^{\pm})$ and ${\rm BDT}(\mu^{\pm},K^{\pm})$ 
spectra can be seen in Fig.~\ref{fig:bdt_cosmics}. 

\subsection{Pions}

The clean $\pi^{\pm}$ sample with well-predicted angular-momentum 
distributions can be 
obtained by selection of $e^{+}e^{-}{\to}\phi(1020){\to}\pi^{+}\pi^{-}\pi^{0}$ 
events. To do this, we search for events with exactly two DC-tracks with 
opposite charges and momenta larger than 100~MeV/{\rm c}. 
Then, there should be not less than two photons with energies larger 
than 40~MeV. Sorting over all the pairs of such photons, we perform the 
4C-kinematic fit for two tracks and the photon pair 
assuming energy-momentum conservation and choose 
the pair giving the smallest $\chi^{2}_{\rm 4C}$. If the invariant mass of the photon 
pair $m_{\rm 2\gamma}$ satisfies the $|m_{\rm 2\gamma}-m_{\pi^{0}}|<40\,{\rm MeV/{\rm c}^{2}}$ condition, 
we consider the $\pi^{+}\pi^{-}\pi^{0}$ event as reconstructed. 

\begin{figure} [hbtp]
\centering
\begin{tabular}{cccc}
\includegraphics[width=0.31\textwidth]{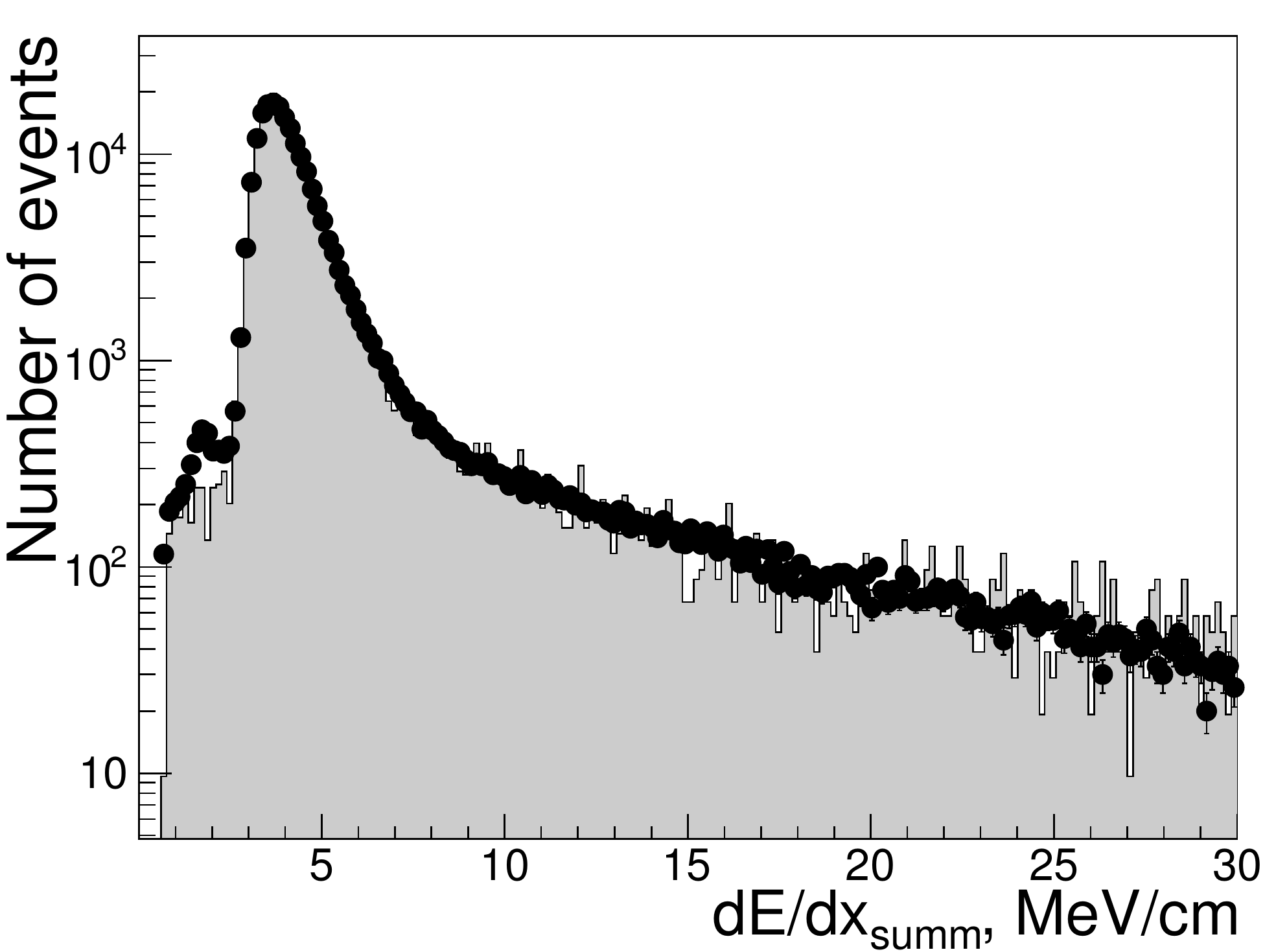} &
\includegraphics[width=0.31\textwidth]{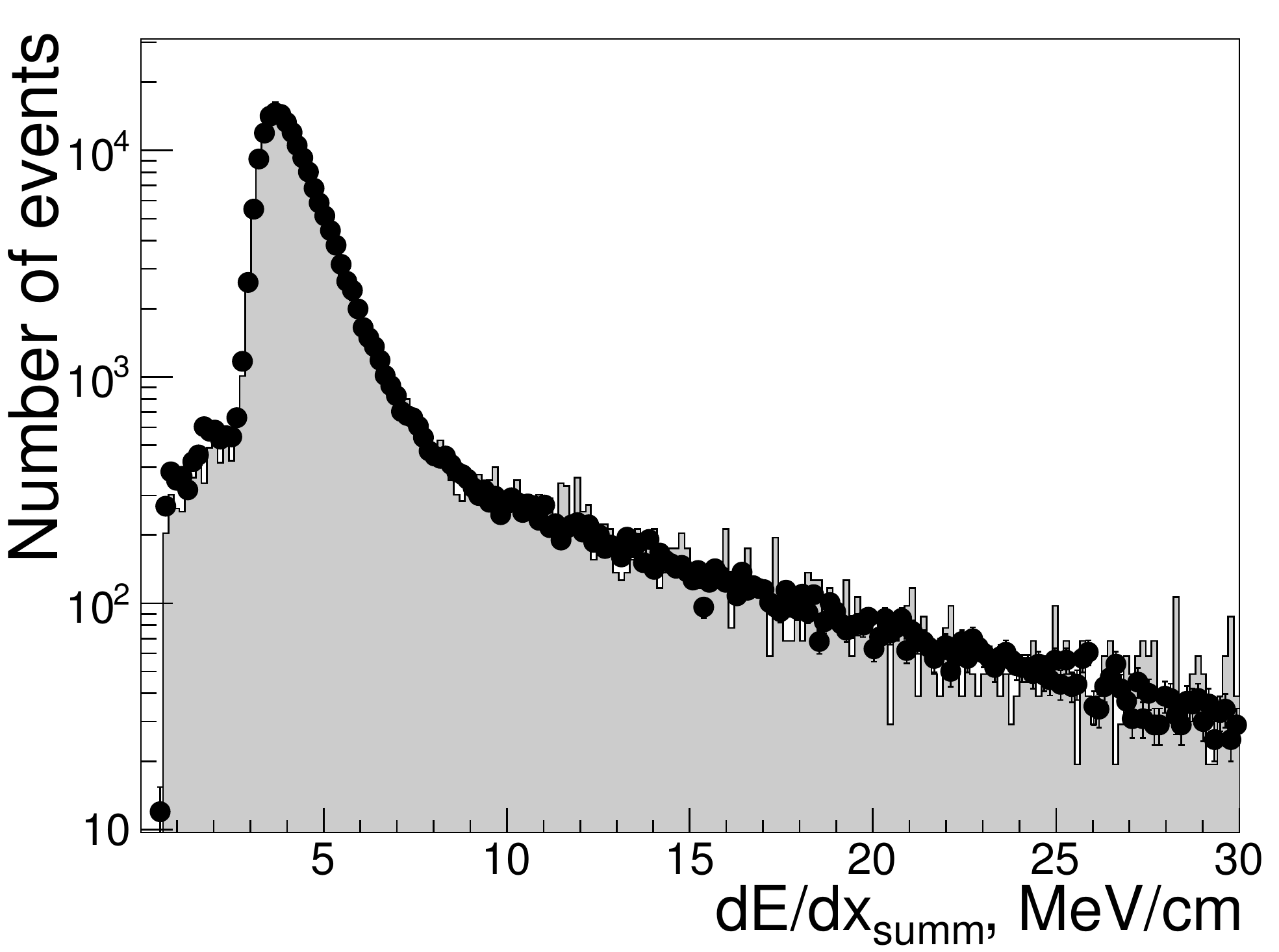} &
\includegraphics[width=0.31\textwidth]{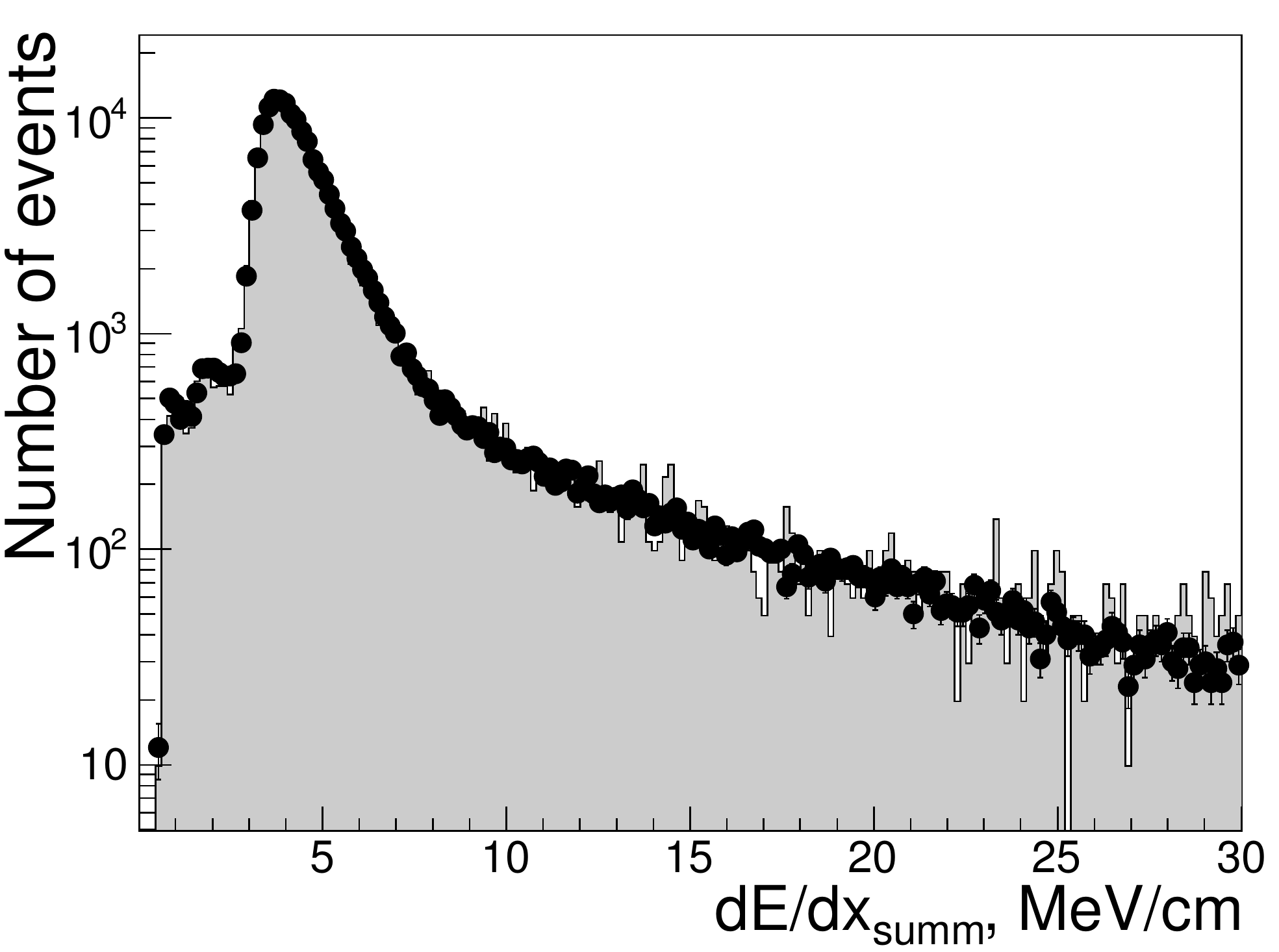} \\
\end{tabular}
\begin{tabular}{cccc}
\includegraphics[width=0.31\textwidth]{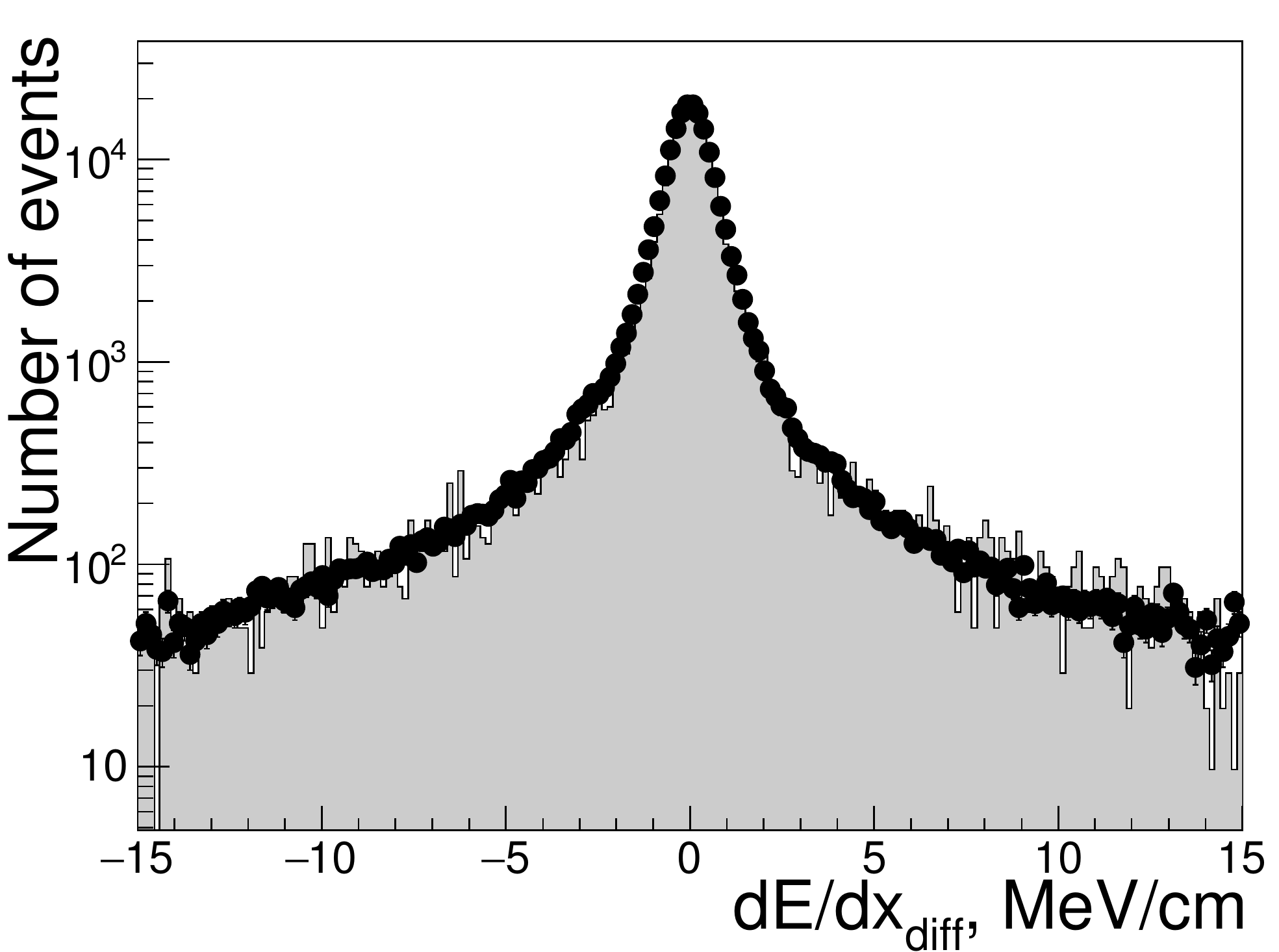} &
\includegraphics[width=0.31\textwidth]{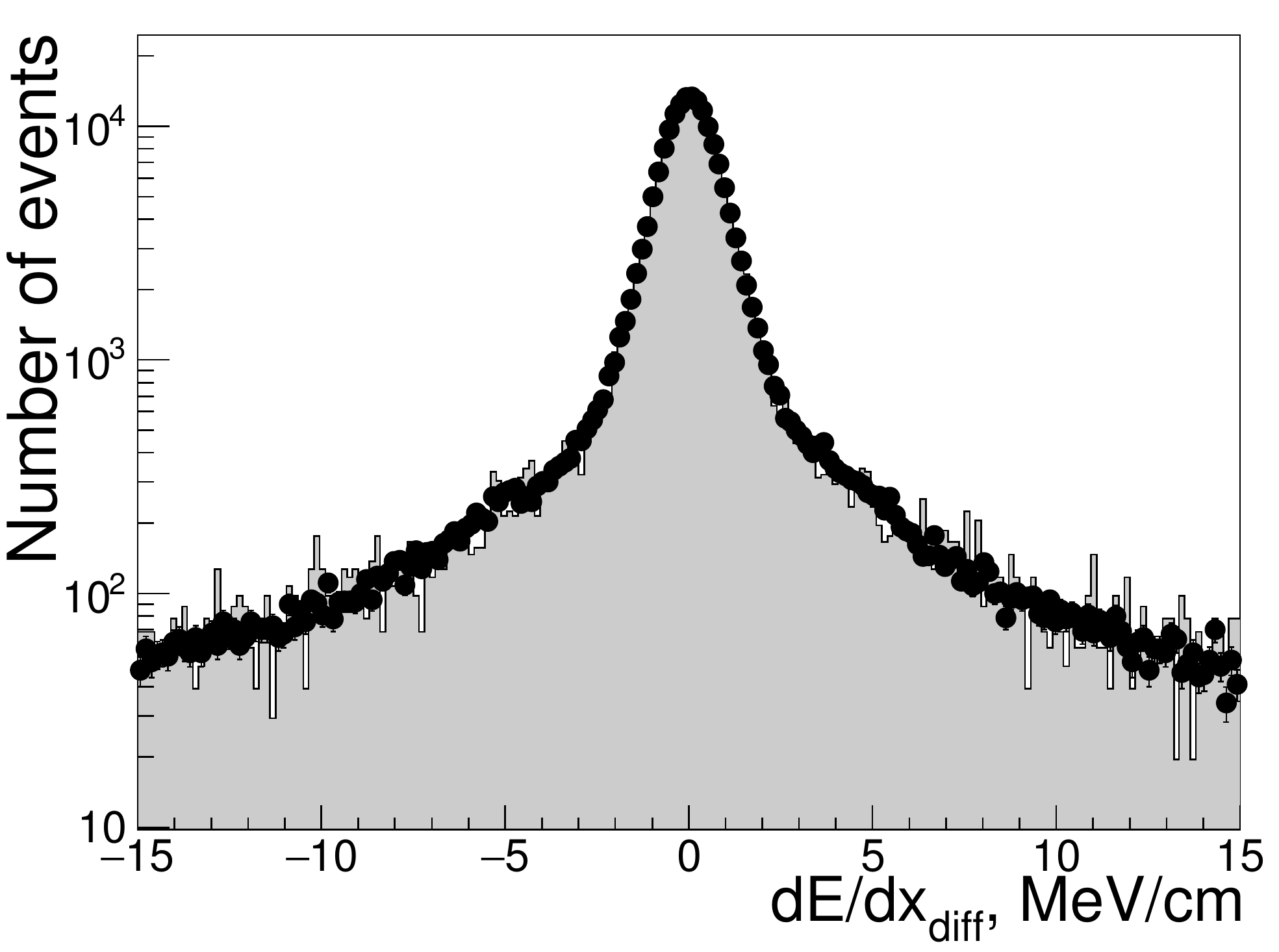} &
\includegraphics[width=0.31\textwidth]{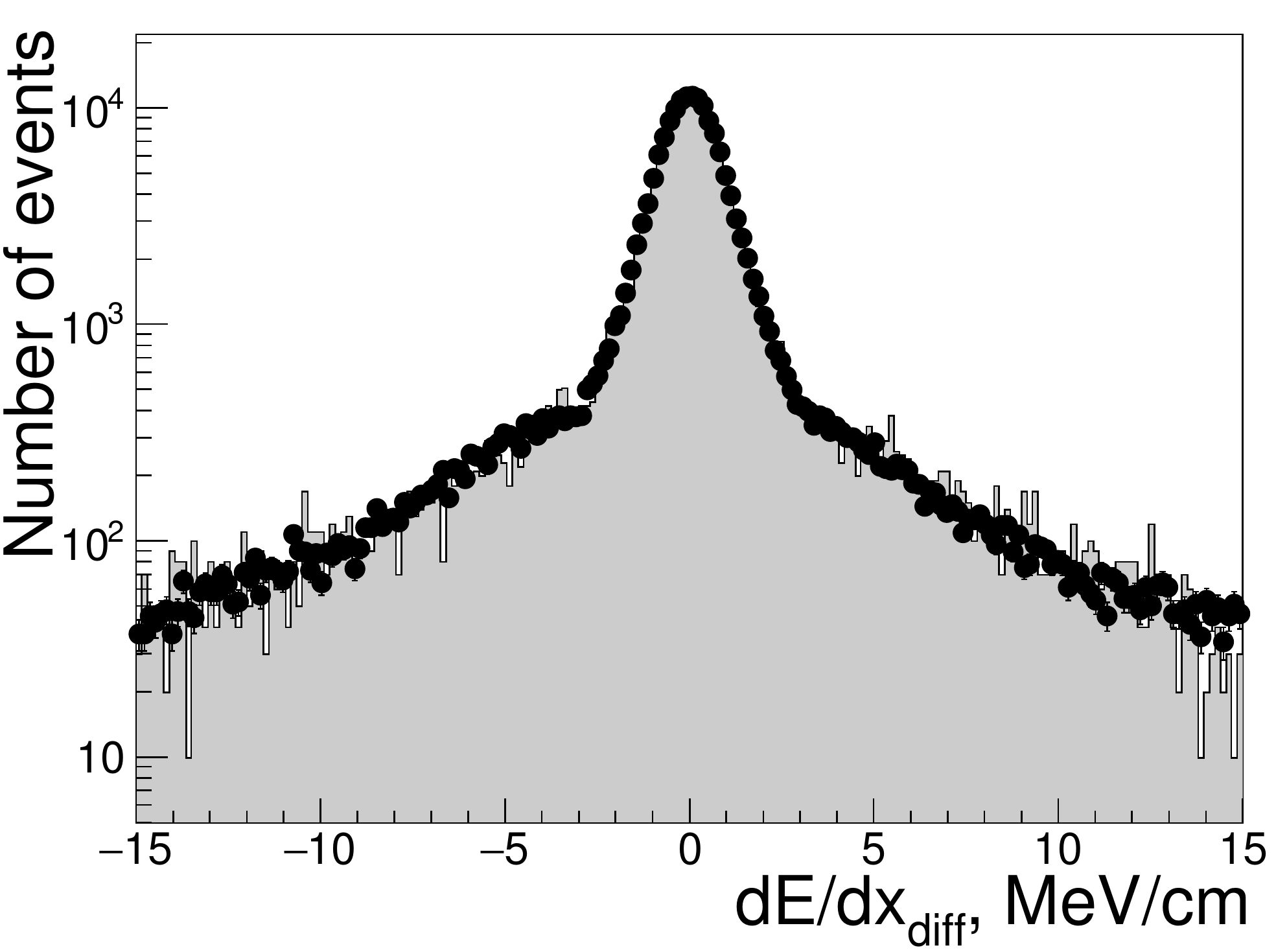} \\
\end{tabular}
\caption{The $dE/dx_{\rm summ}$ (top figures) and $dE/dx_{\rm diff}$ (bottom 
figures) in the 1st (left), 3rd (middle) and 5th (right) double layers for the $\pi^{\pm}$ selected from  
$e^{+}e^{-}{\to}\pi^{+}\pi^{-}\pi^{0}$ events in the experiment (markers) 
and simulation (gray histogram). The c.m. energy is 1019~MeV ($\phi(1020)$ 
meson peak). \label{fig:pions_from_3pi_summ}}
\end{figure}

\begin{figure}[hbtp]
  \begin{minipage}[t]{0.32\textwidth}
   \centerline{\includegraphics[width=0.98\textwidth]{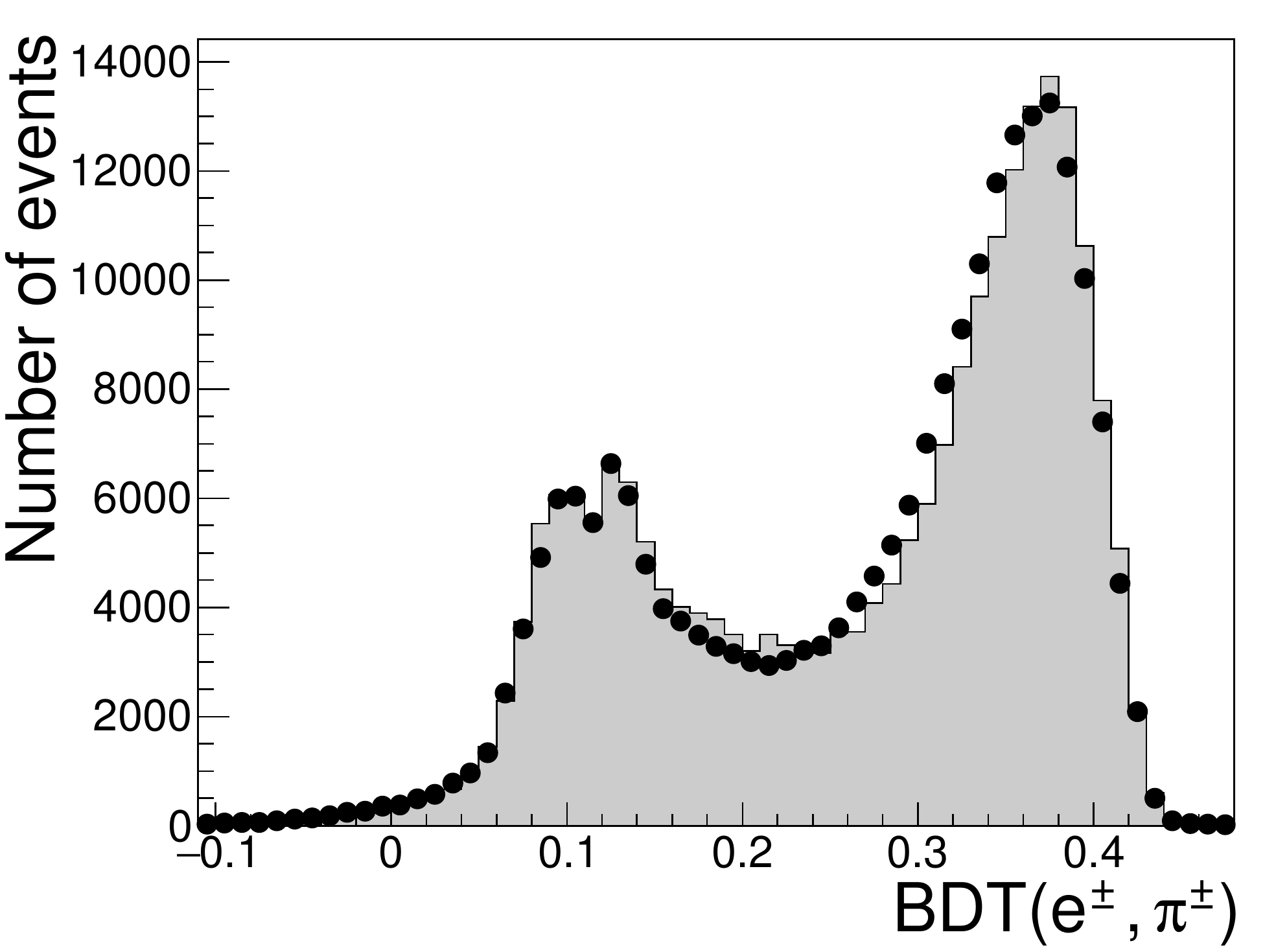}}    
  \end{minipage}\hfill\hfill
   \begin{minipage}[t]{0.32\textwidth}
   \centerline{\includegraphics[width=0.98\textwidth]{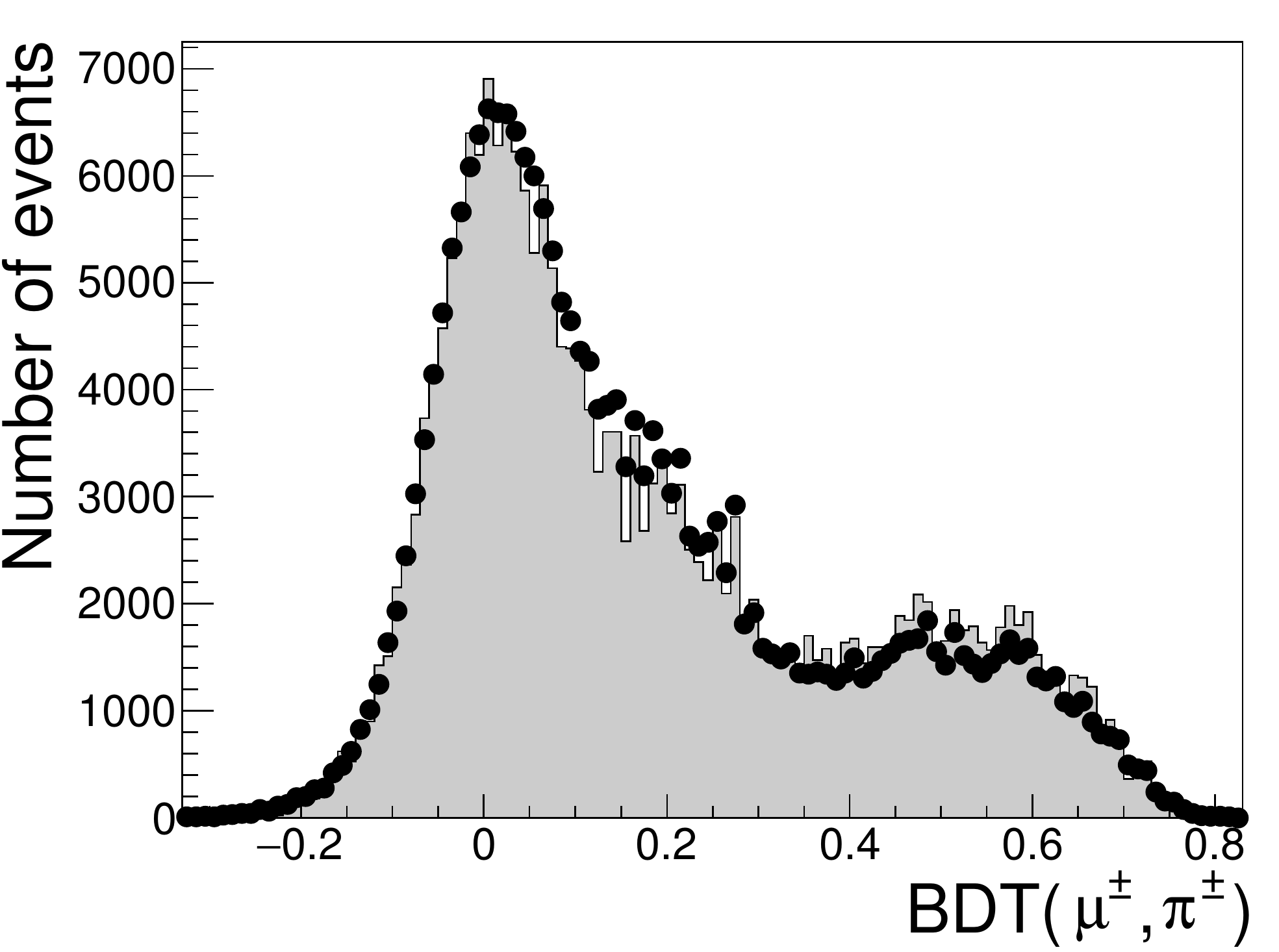}}    
  \end{minipage}\hfill\hfill
  \begin{minipage}[t]{0.32\textwidth}
    \centerline{\includegraphics[width=0.98\textwidth]{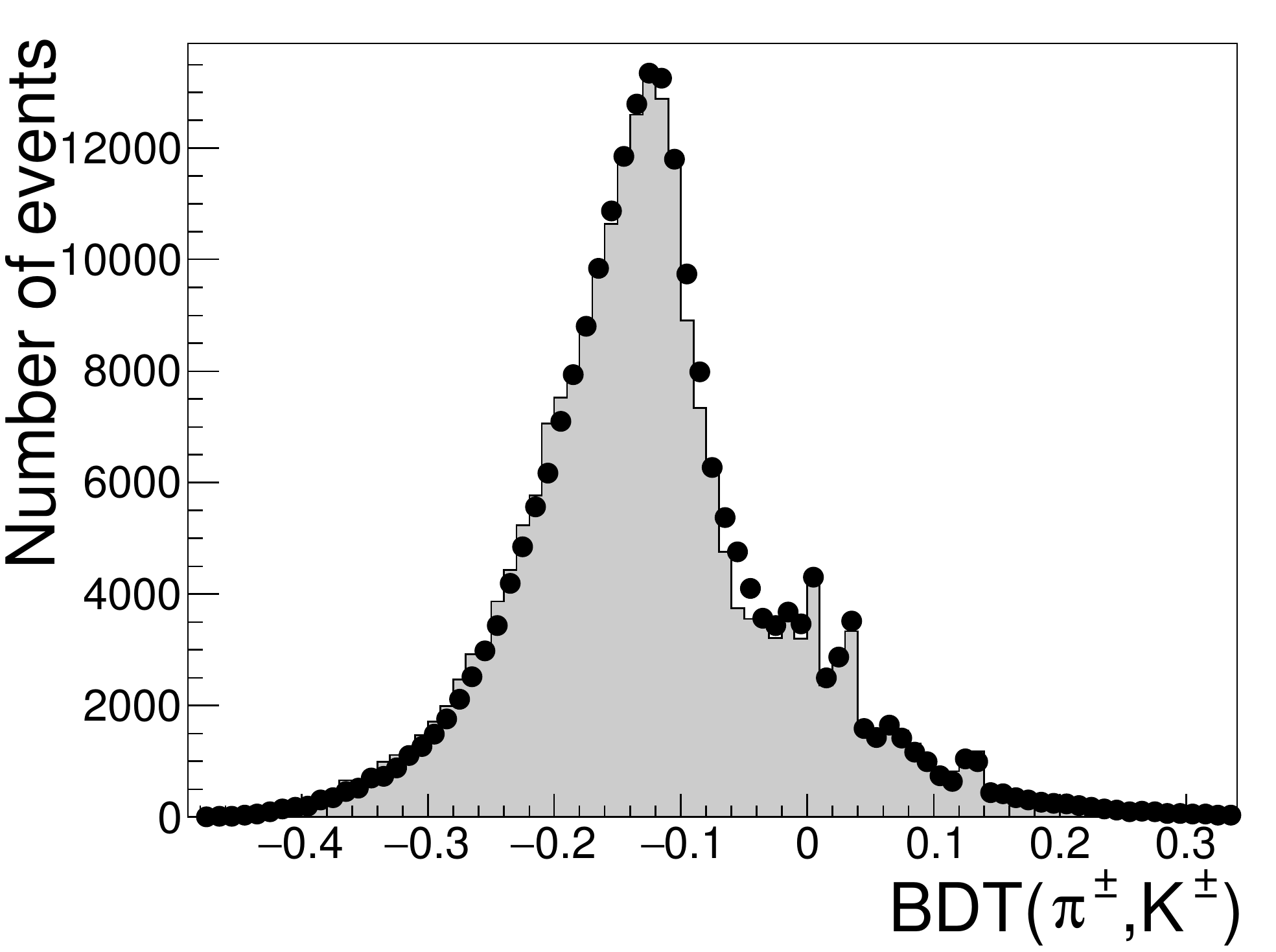}}    
  \end{minipage}\hfill\hfill
  \caption{The ${\rm BDT}(e^{\pm},\pi^{\pm})$ (left), ${\rm 
BDT}(\mu^{\pm},\pi^{\pm})$ (middle) and ${\rm BDT}(\pi^{\pm},K^{\pm})$ 
(right) spectra for the $\pi^{\pm}$ selected from  
$e^{+}e^{-}{\to}\pi^{+}\pi^{-}\pi^{0}$ events in the experiment (markers) 
and MC (gray histrogram). The c.m. energy is 1019~MeV ($\phi(1020)$ 
meson peak).
          \label{fig:bdt_pions_from_3pi_509_5}}
\end{figure}

\begin{figure}
  \begin{center}
    \includegraphics[width=0.5\textwidth]{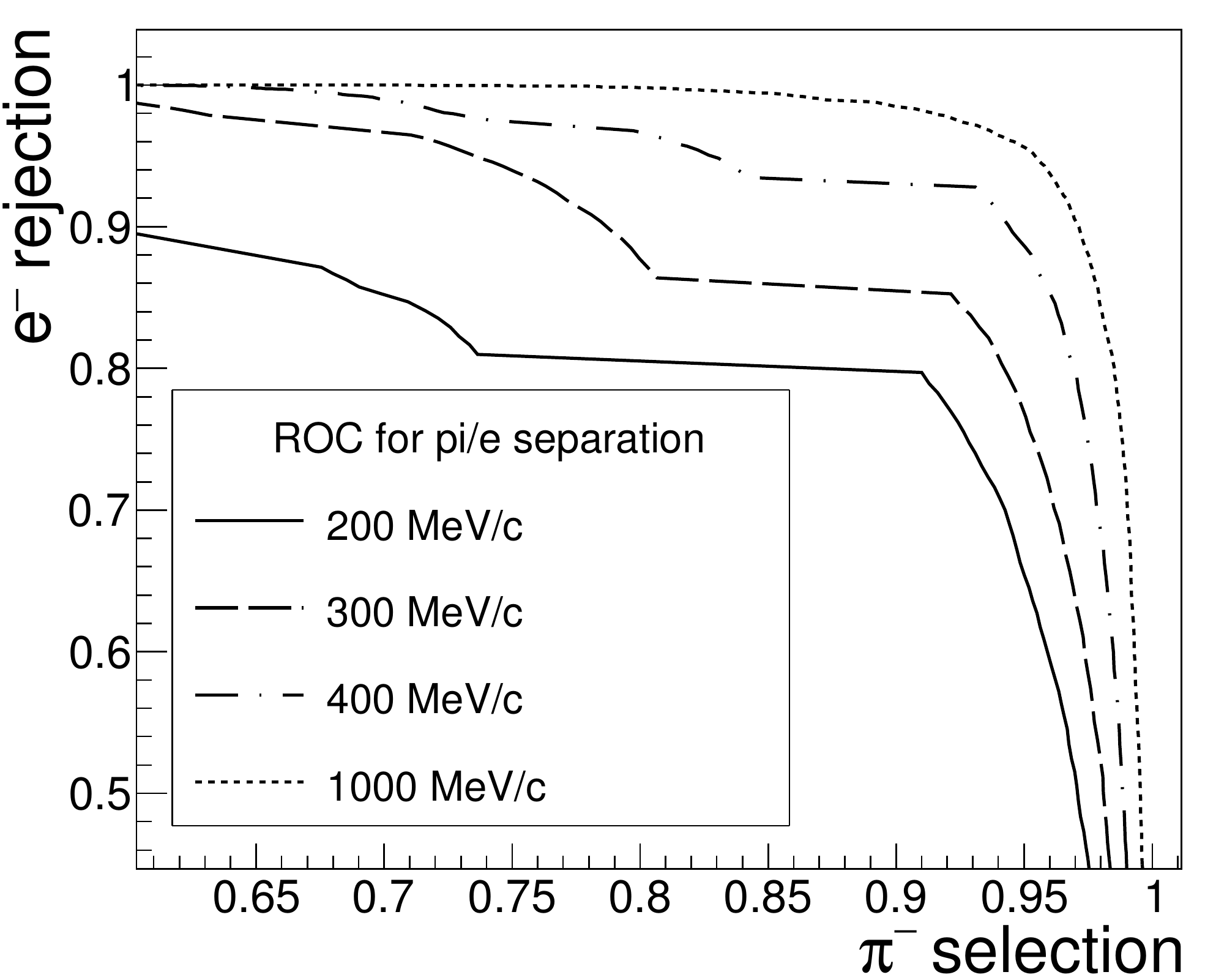}
    \caption{The ROC-curves for the ${\rm BDT}(e^{-},\pi^{-})$ 
    classifier for different particle momenta (see legend) 
    according to simulation. \label{fig:ROC_e_pi}}
  \end{center}
\end{figure}

First of all, since simulation of nuclear interactions of pions is not perfect,
we check the data/MC agreement in the $dE/dx_{\rm summ}$ and $dE/dx_{\rm diff}$
spectra for selected $\pi^{\pm}$, see Fig.~\ref{fig:pions_from_3pi_summ}. 
The agreement is good for all pion momenta.
Then, Fig.~\ref{fig:bdt_pions_from_3pi_509_5} shows  
good data/MC agreement for the ${\rm BDT}(e^{\pm},\pi^{\pm})$, 
${\rm BDT}(\mu^{\pm},\pi^{\pm})$ and ${\rm BDT}(\pi^{\pm},K^{\pm})$ 
spectra. The agreement is good for both pion charges.
The efficiency of $e^{-}$ rejection vs. the 
efficiency of $\pi^{-}$ selection (ROC-curve) for the ${\rm BDT}(e^{-},\pi^{-})$ 
at different pion momenta is shown in Fig.~\ref{fig:ROC_e_pi}.

\subsection{Kaons \label{kaons}}

The clean $K^{\pm}$ sample can be selected from the four-track 
$e^{+}e^{-}{\to}K^{+}K^{-}\pi^{+}\pi^{-}$ events. 
We select these events on the base of ${\sim}60\, {\rm pb^{-1}}$ of integrated 
luminosity collected in the 2019 runs and use data from all energy points above the reaction threshold. 
The event selection procedure involves the kinematic energy-momentum selections along 
with the cuts on the value of the likelihood function, 
based on the $dE/dx_{\rm DC}$ of tracks, as described in~\cite{shemyakin_kkpipi,kkpipi_at_nbarn}. 
However, a large part of selected kaons has the momenta 
lower than the $p^{\rm K}_{\rm trh} \sim 300$~MeV/{\rm c}, and 
for such momenta only the products of kaon decay or nuclear 
interaction can reach the LXe. 

\begin{figure} [hbtp]
\centering
\begin{tabular}{cccc}
\includegraphics[width=0.31\textwidth]{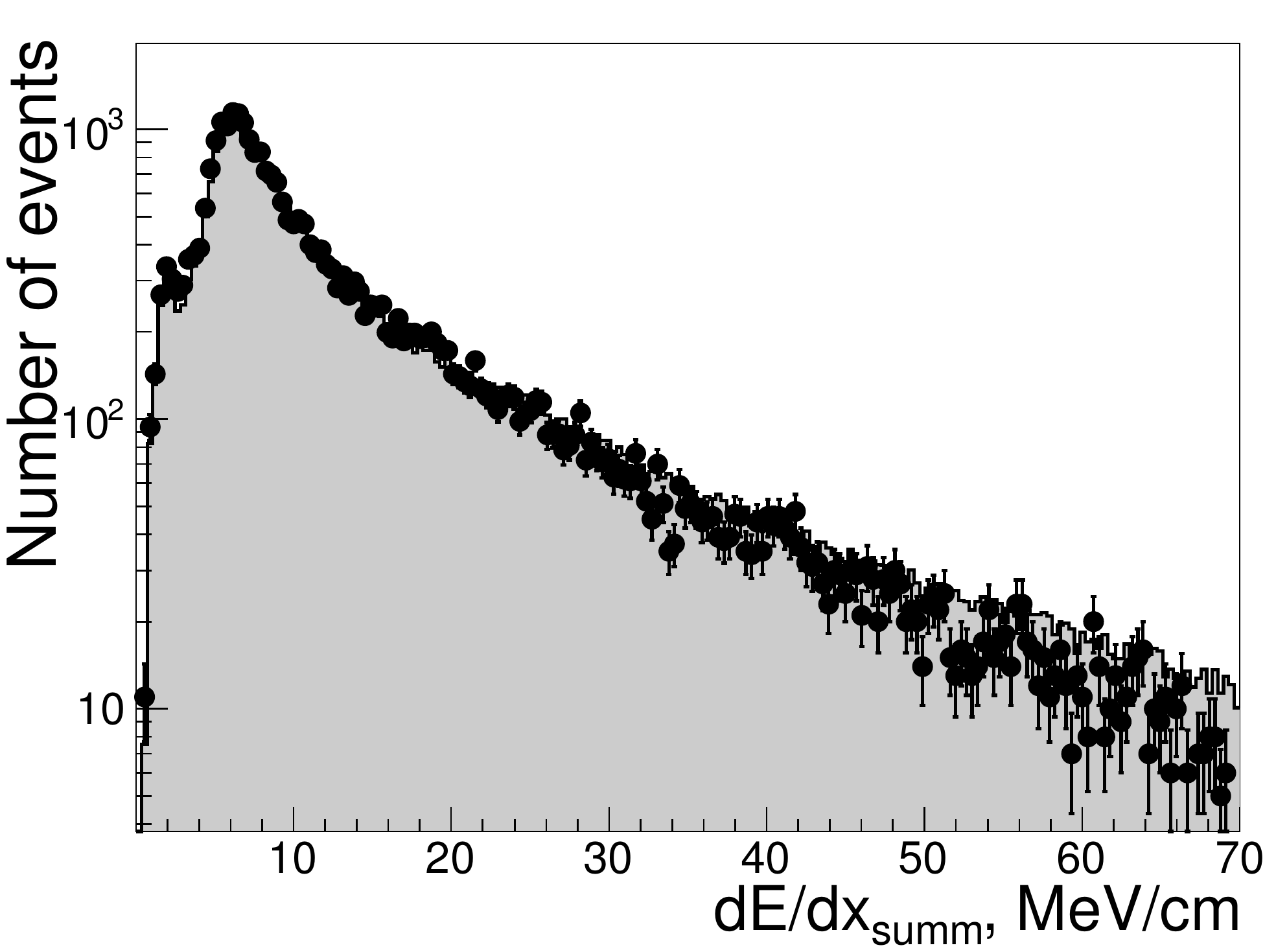} &
\includegraphics[width=0.31\textwidth]{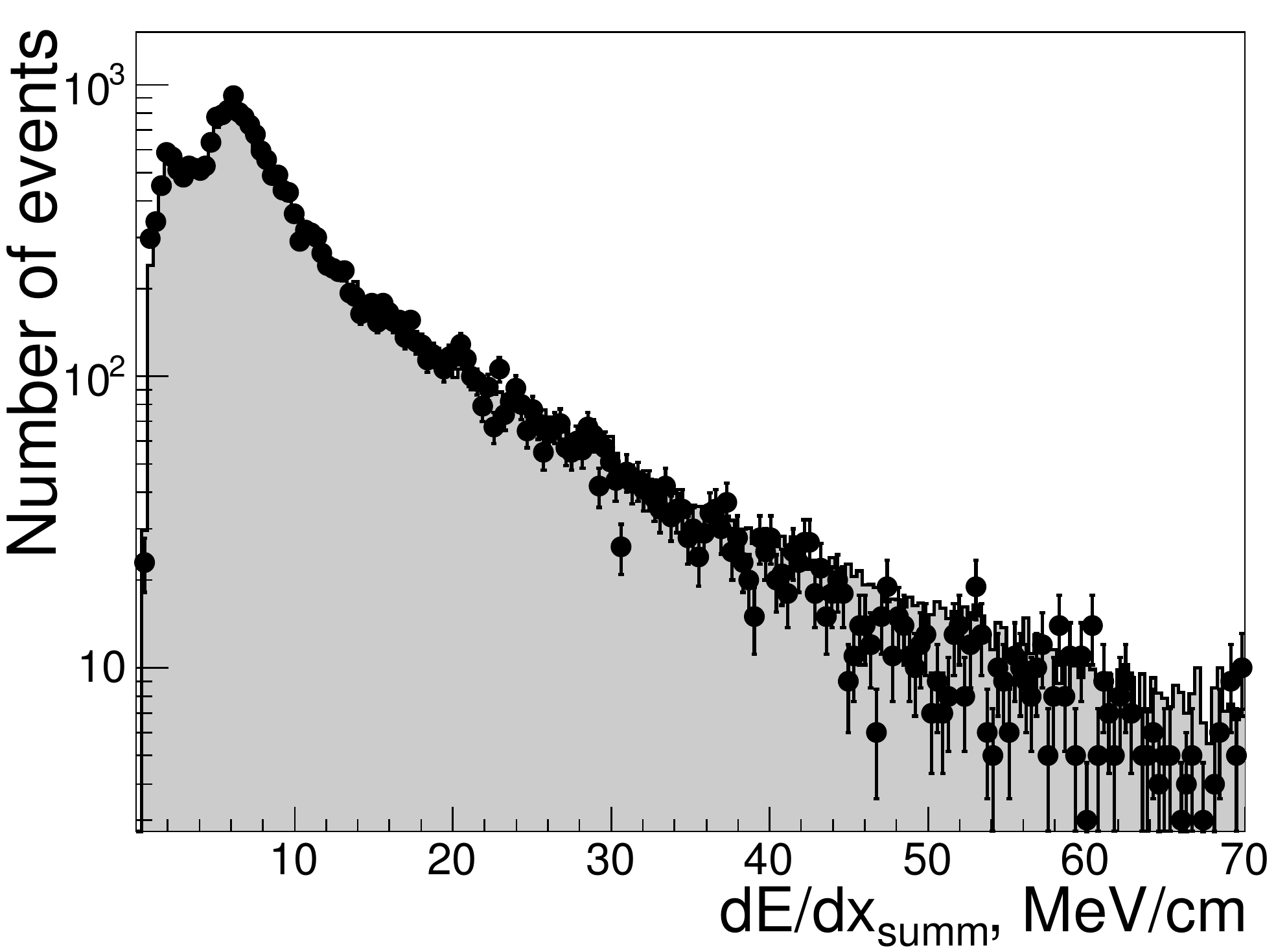} &
\includegraphics[width=0.31\textwidth]{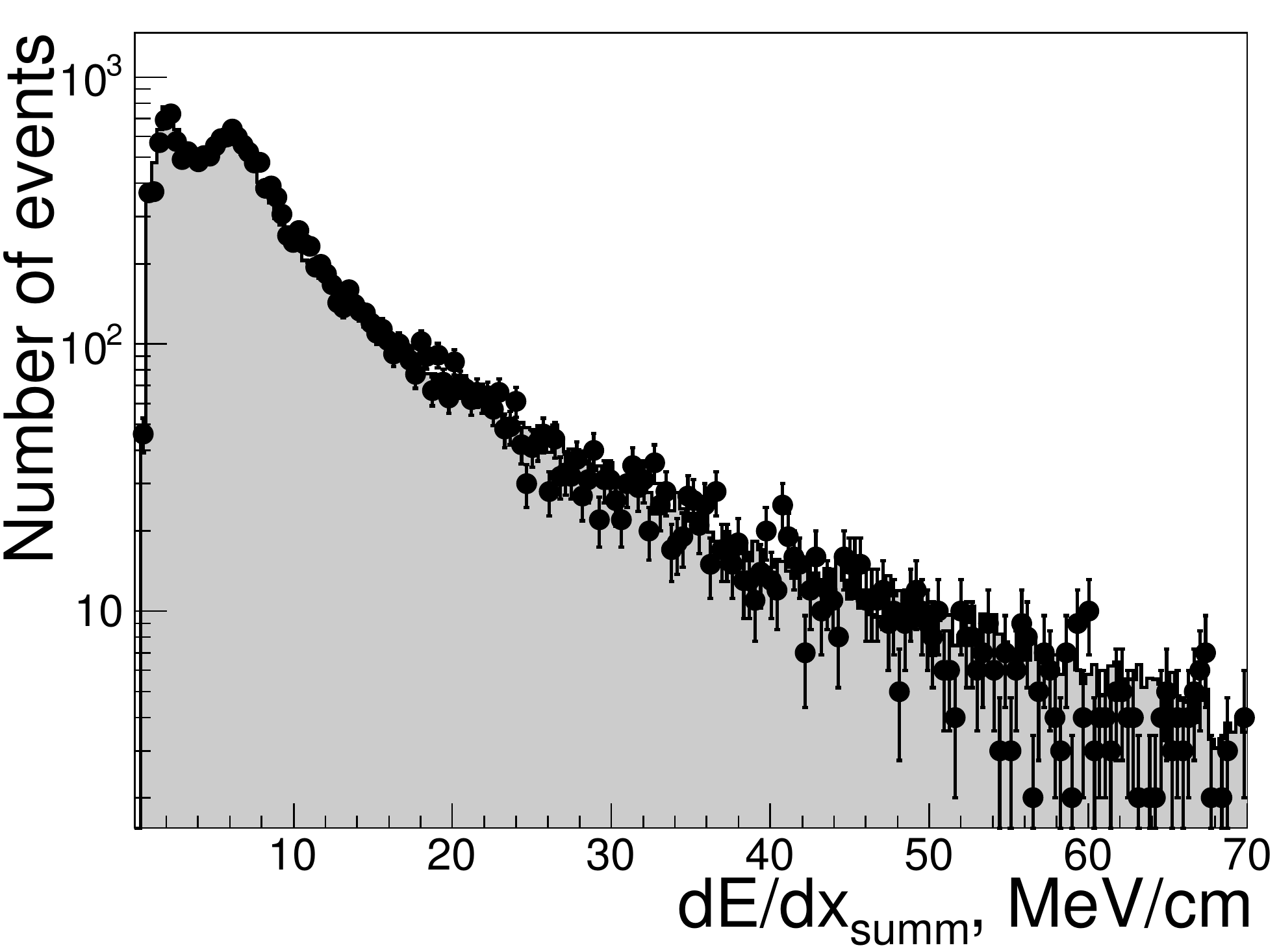} \\
\end{tabular}
\begin{tabular}{cccc}
\includegraphics[width=0.31\textwidth]{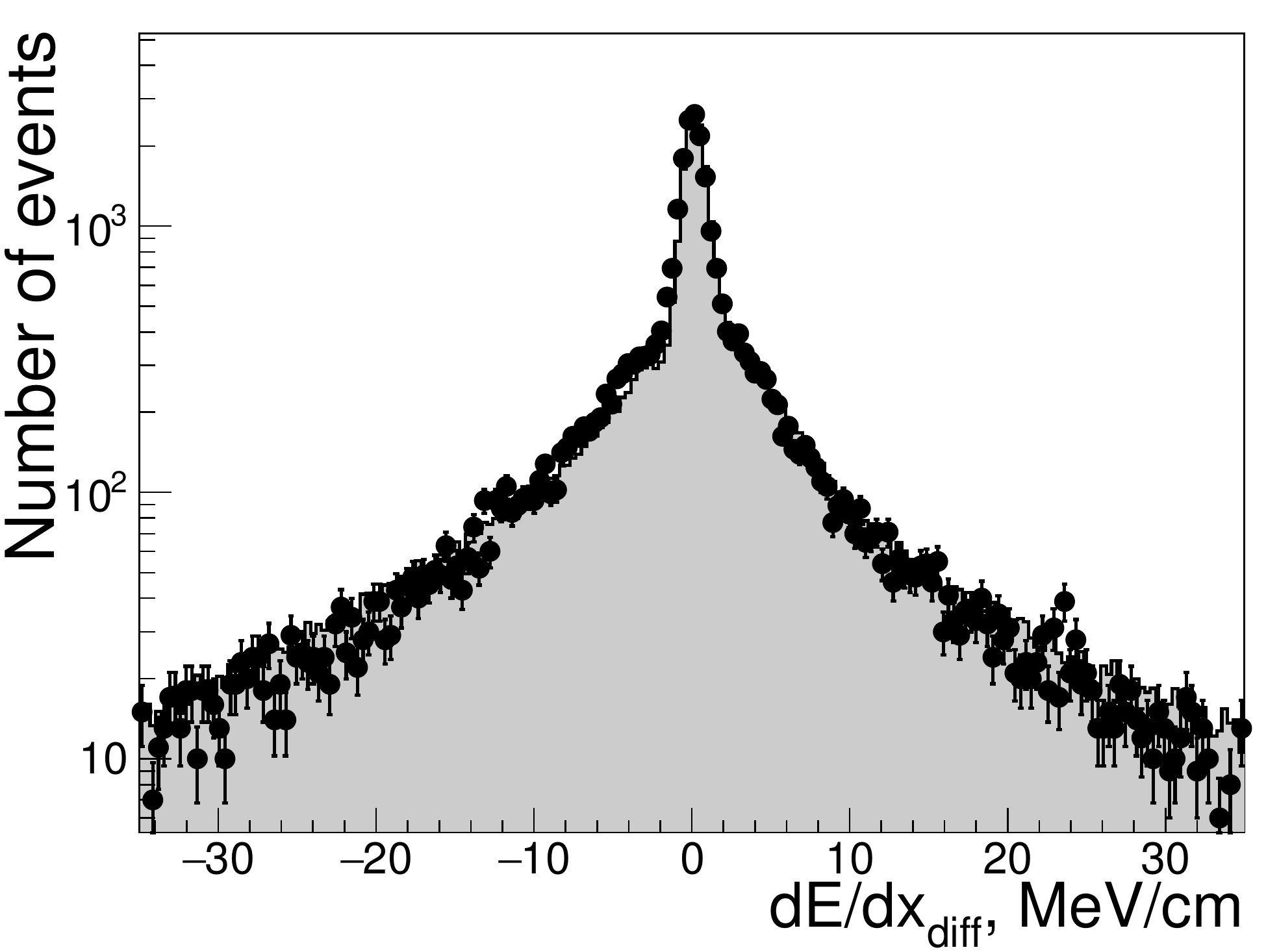} &
\includegraphics[width=0.31\textwidth]{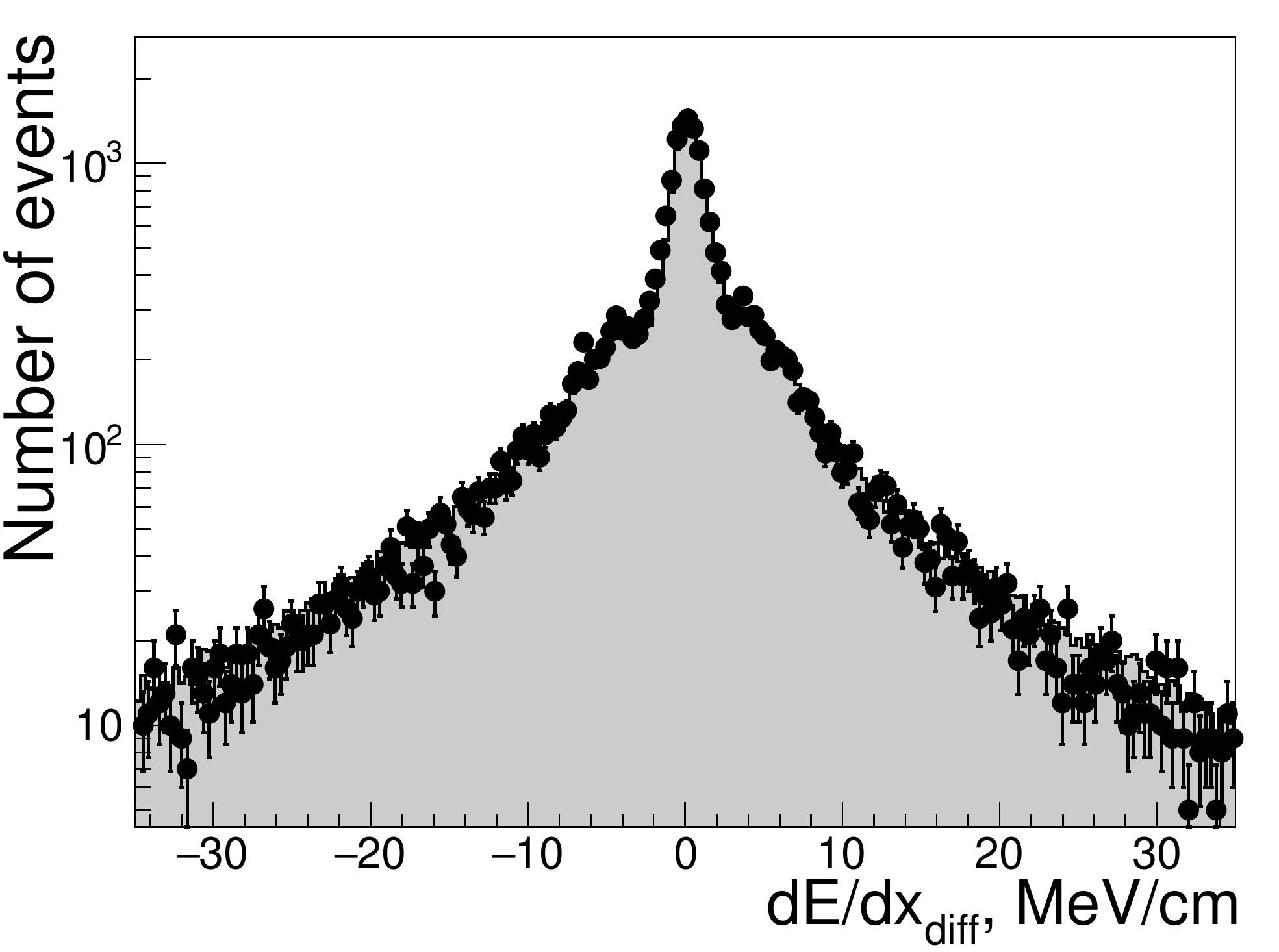} &
\includegraphics[width=0.31\textwidth]{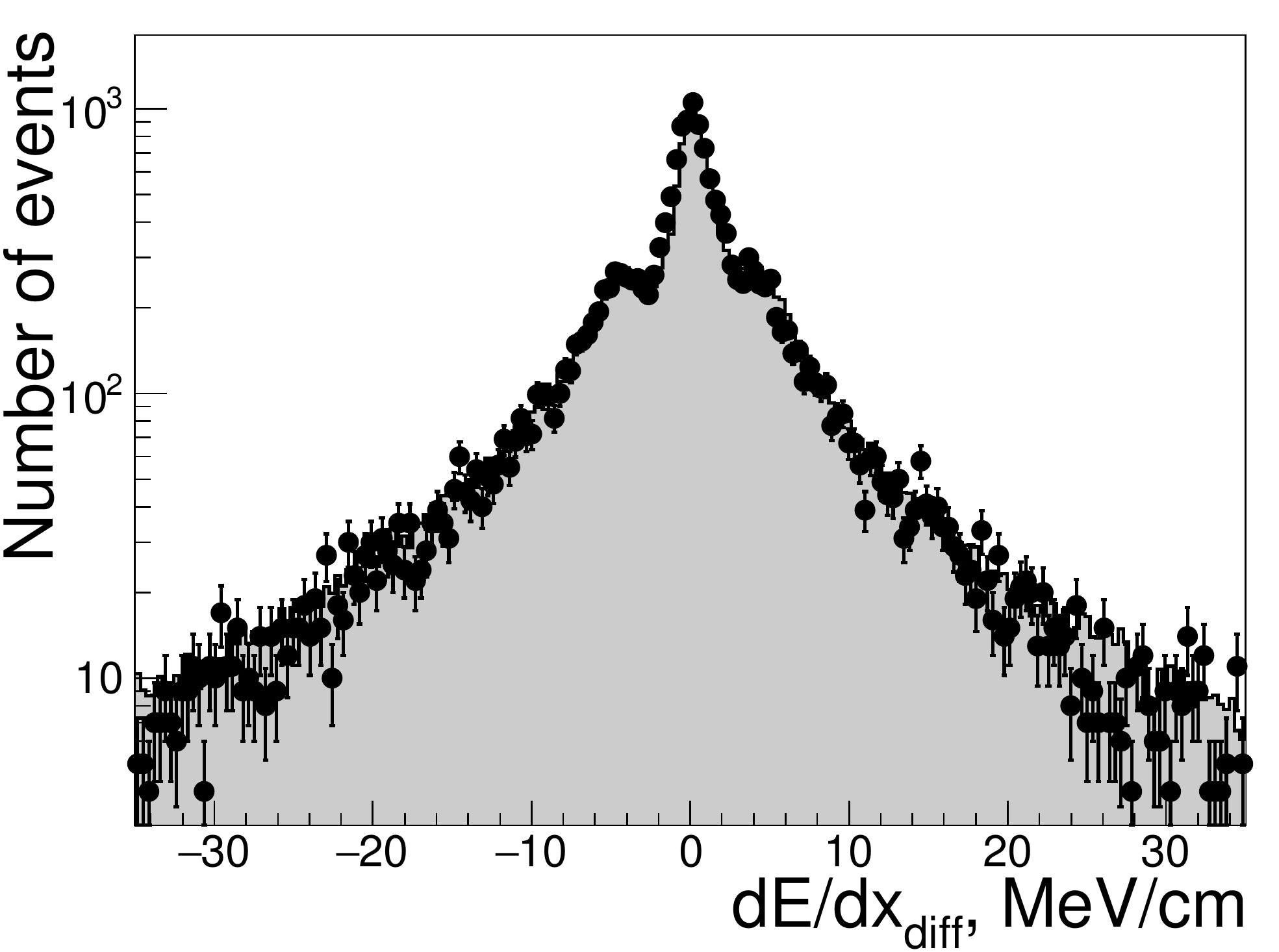} \\
\end{tabular}
\caption{The $dE/dx_{\rm summ}$ (top figures) and $dE/dx_{\rm diff}$ 
(bottom figures) 
in the 1st (left), 3rd (middle) and 5th (right) double layers 
for the $K^{\pm}$ selected from  
$e^{+}e^{-}{\to}K^{+}K^{-}\pi^{+}\pi^{-}$ events in the experiment 
(markers) and simulation (gray histogram). 
The data from all experimental runs of 2019 are used.
\label{fig:kaons_from_kkpipi_summ}}
\end{figure}

\begin{figure} [hbtp]
\centering
\begin{tabular}{ccc}
\includegraphics[width=0.48\textwidth]{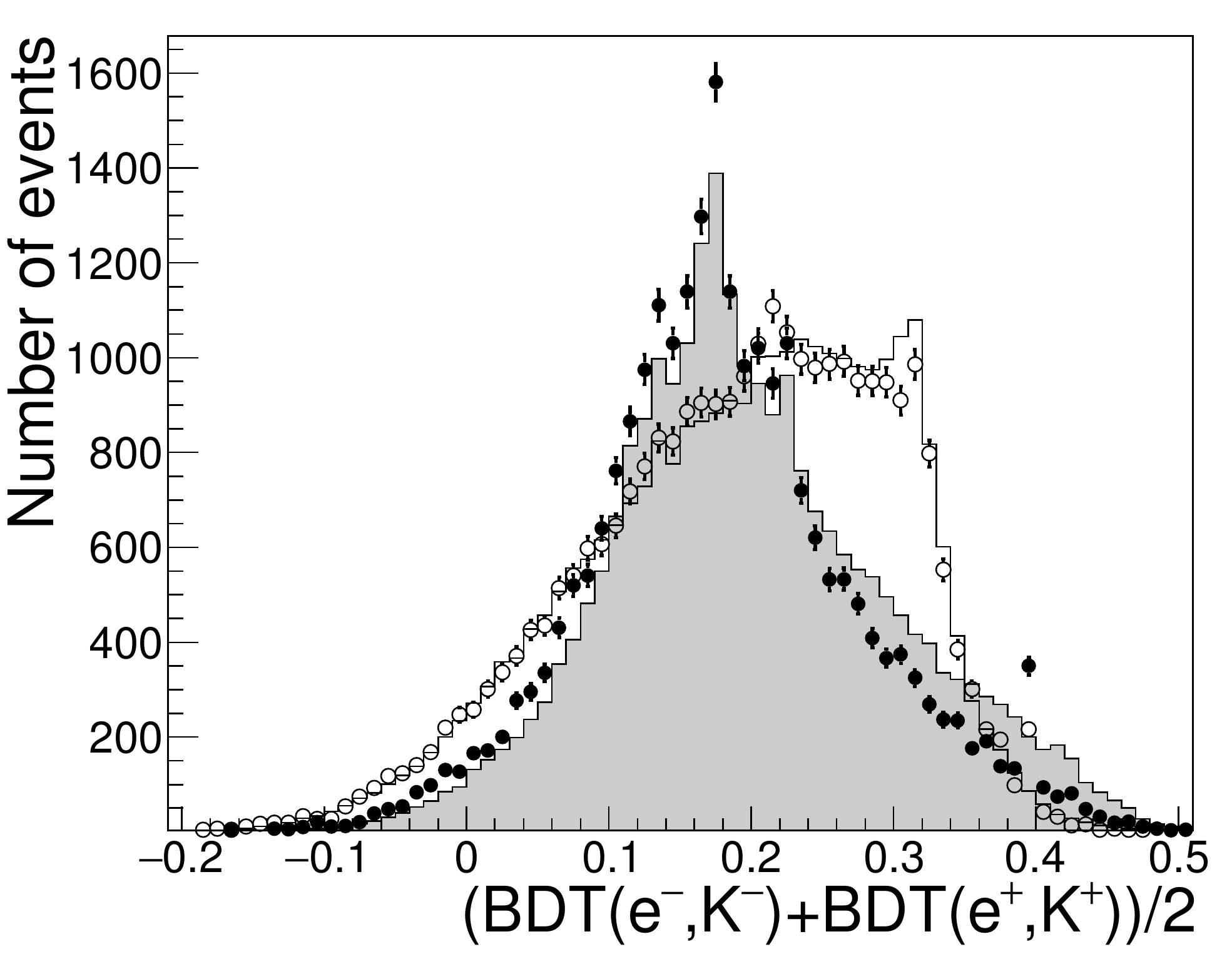} &
\includegraphics[width=0.48\textwidth]{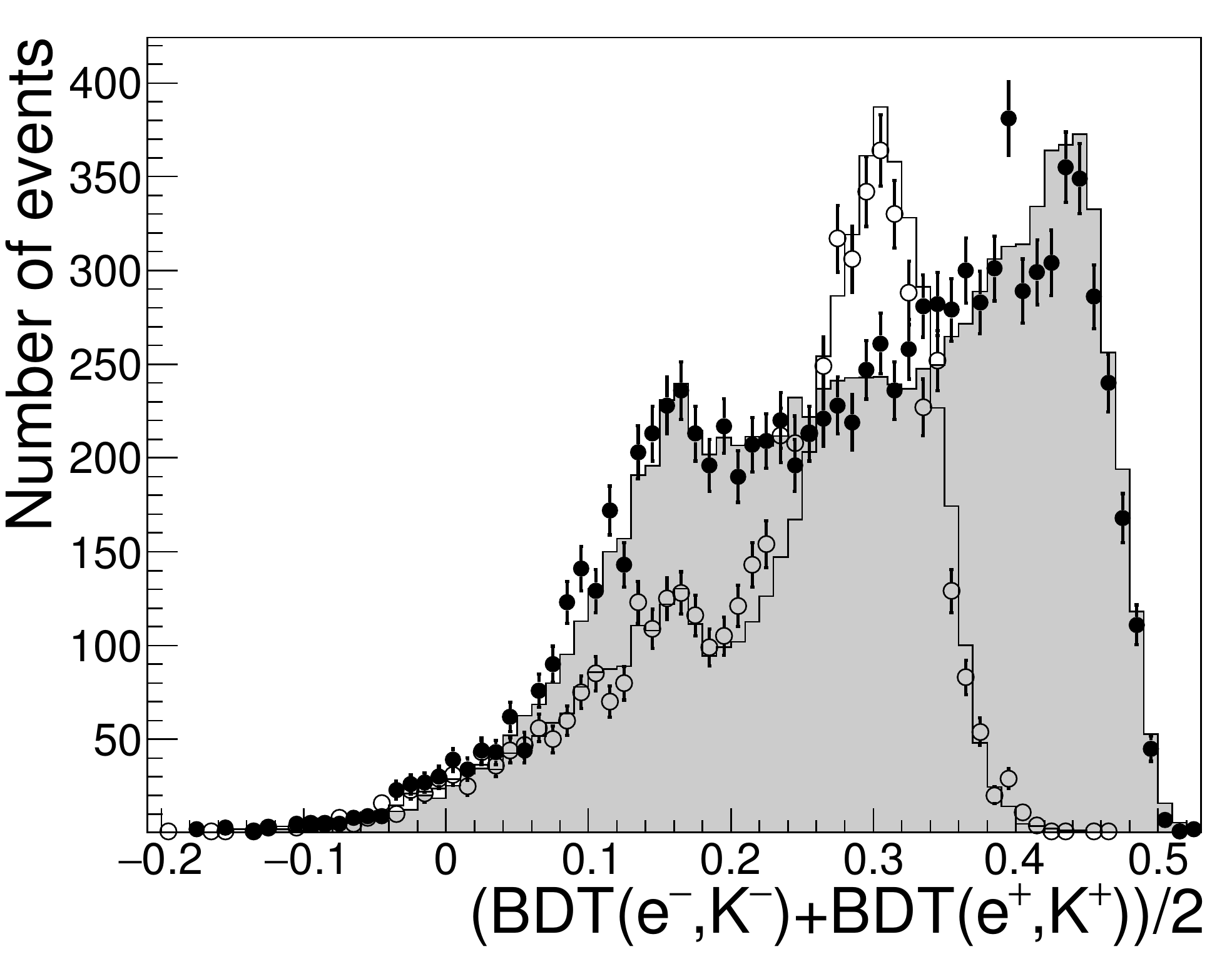} \\
\includegraphics[width=0.48\textwidth]{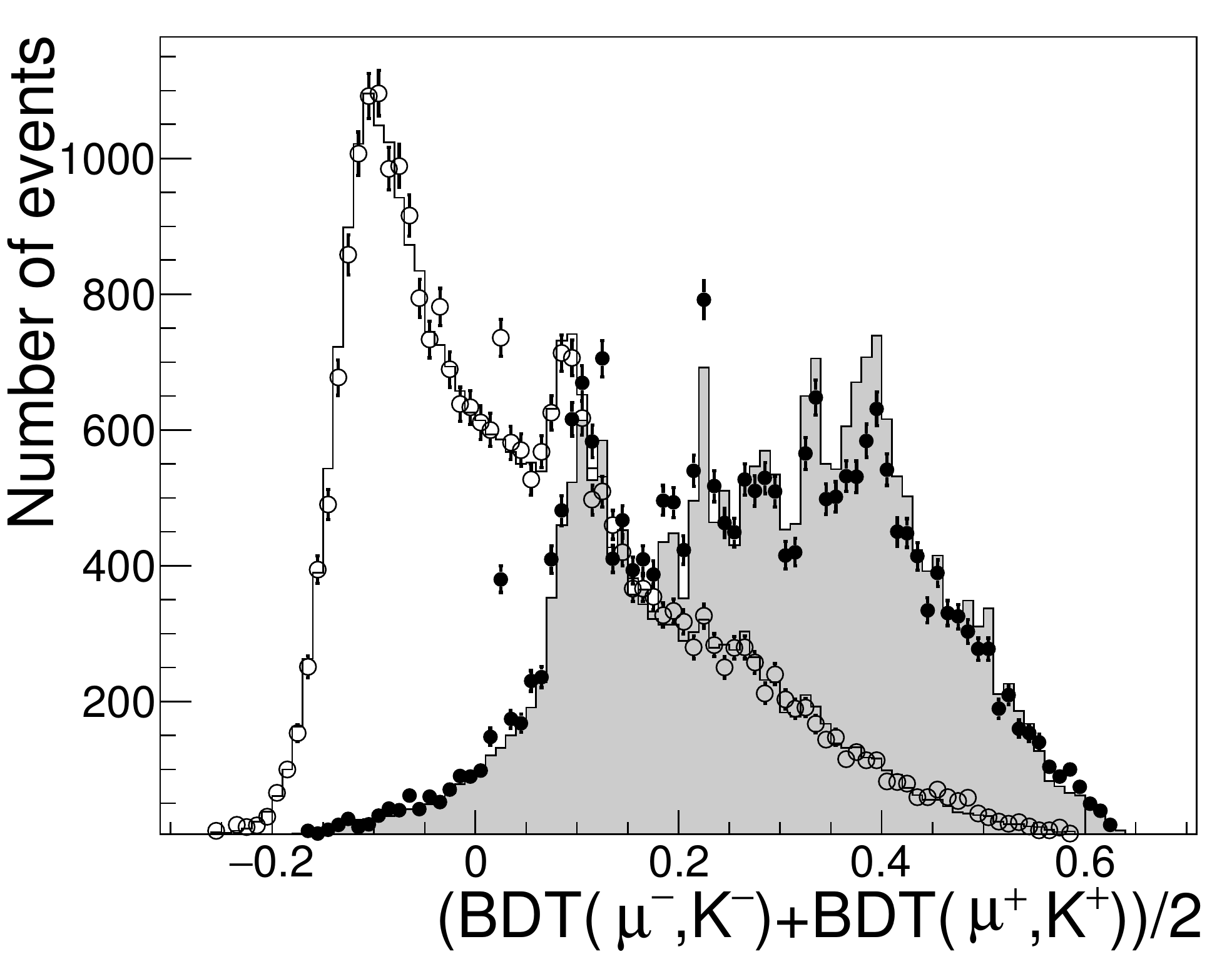} &
\includegraphics[width=0.48\textwidth]{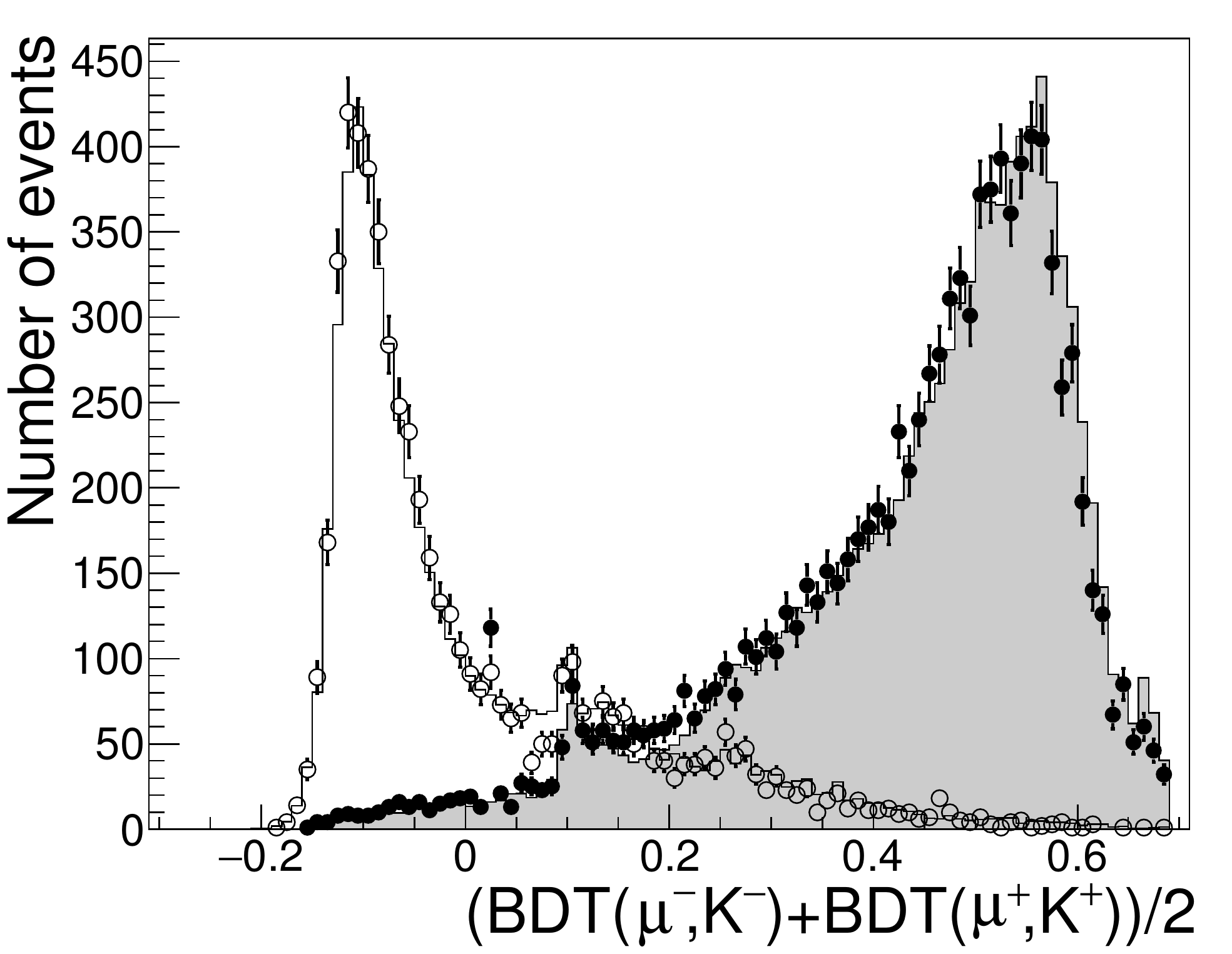} \\
\includegraphics[width=0.48\textwidth]{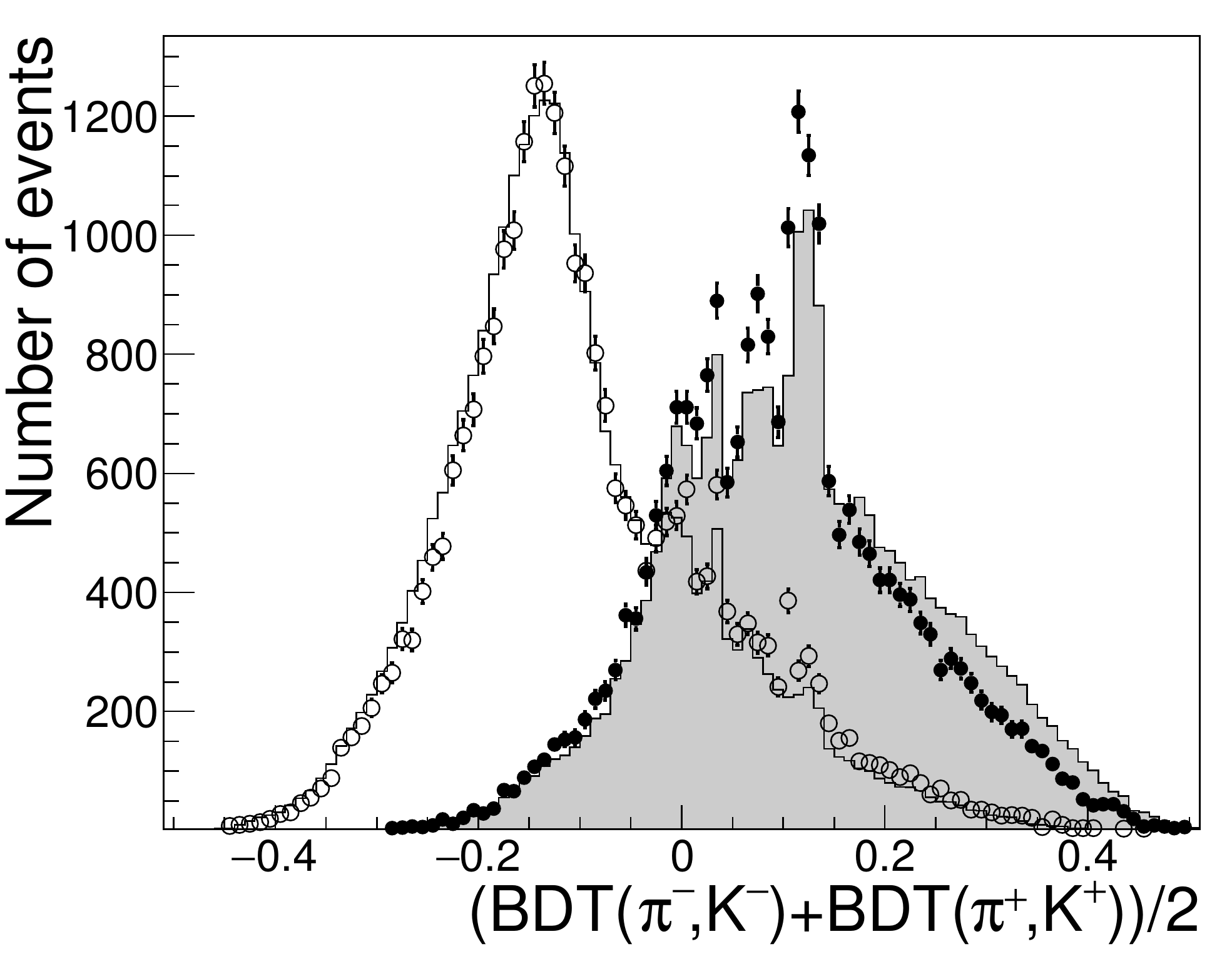} &
\includegraphics[width=0.48\textwidth]{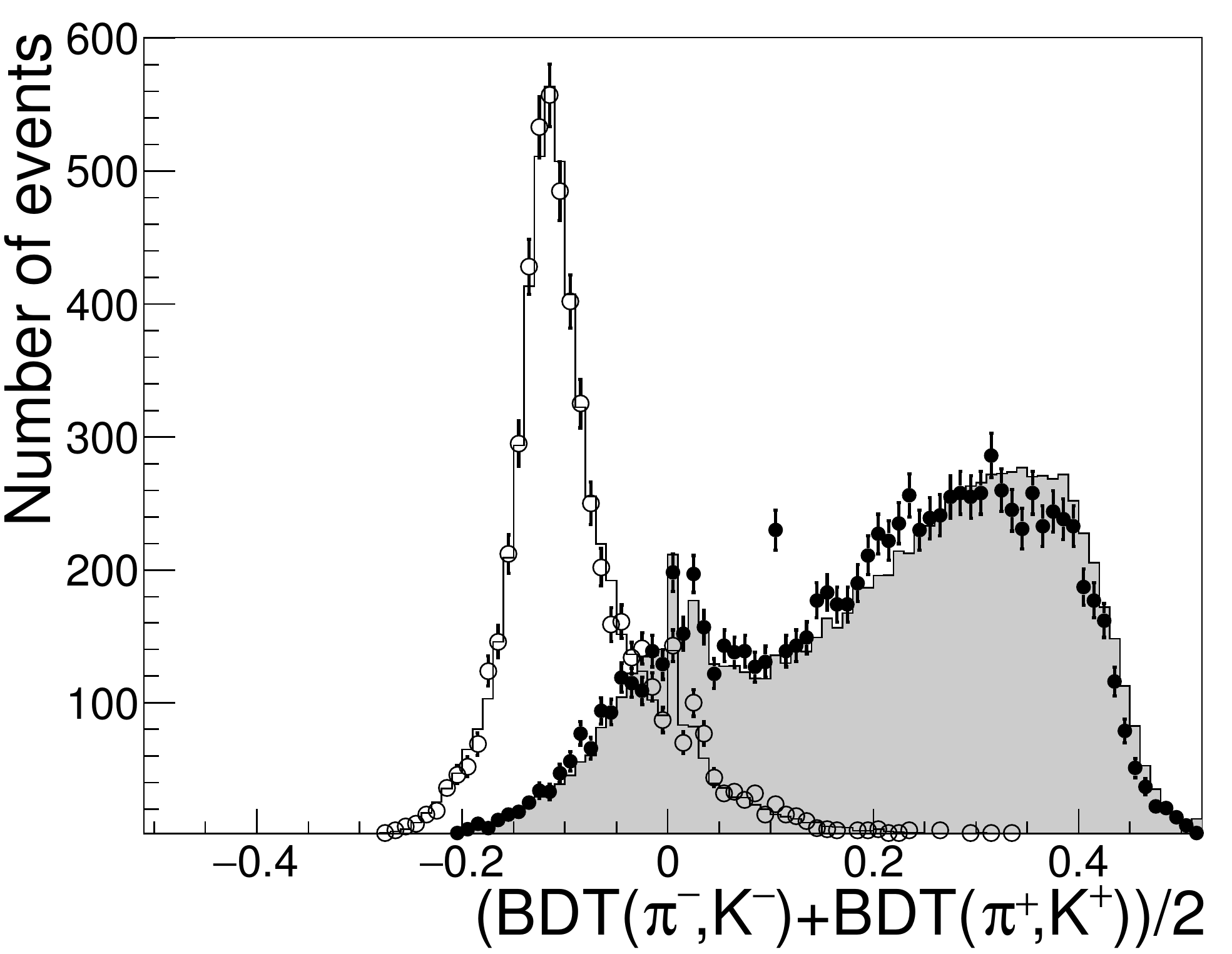} \\
\end{tabular}
\begin{tabular}{ccc}
\end{tabular}

\caption{The ${\rm BDT}(e^{\pm},K^{\pm})$ (top), 
${\rm BDT}(\mu^{\pm},K^{\pm})$ (middle) and ${\rm BDT}(\pi^{\pm},K^{\pm})$ 
(bottom) spectra for the $K^{\pm}$ and $\pi^{\pm}$ selected from 
$e^{+}e^{-}{\to}K^{+}K^{-}\pi^{+}\pi^{-}$ events 
in the experiment (filled circles for $K^{\pm}$ and empty circles for $\pi^{\pm}$) 
and simulation (gray histogram for $K^{\pm}$ and open histogram for $\pi^{\pm}$). 
The left figures are drawn for particles with momenta lower than 400~MeV,
the right -- larger than 400~MeV. The data from all experimental runs of 2019 are used.
\label{fig:bdt_kaons}}
\end{figure}

Similarly to the case of pions, we check the accuracy of simulation of the 
nuclear interactions of kaons by the data/MC comparison for the $dE/dx_{\rm summ}$
and $dE/dx_{\rm diff}$ spectra for selected $K^{\pm}$, see Fig.~\ref{fig:kaons_from_kkpipi_summ}.
The agreement is reasonable for all kaon momenta.
The data/MC comparison for the ${\rm BDT}(e^{\pm},K^{\pm})$, 
${\rm BDT}(\mu^{\pm},K^{\pm})$ and ${\rm BDT}(\pi^{\pm},K^{\pm})$ 
spectra is shown in Fig.~\ref{fig:bdt_kaons}.
The simulated ${\rm BDT}(\pi^{\pm},K^{\pm})$ spectrum seems somewhat distorted 
for $K^{\pm}$ with low momenta (lower left picture in Fig.~\ref{fig:bdt_kaons}), 
presumably due to the inacurracy in the simulation of nuclear interactions. 
However, the distortion mostly disappears at large kaon momenta, see
lower right picture in Fig.~\ref{fig:bdt_kaons}.

\begin{figure}
  \begin{center}
    \includegraphics[width=0.5\textwidth]{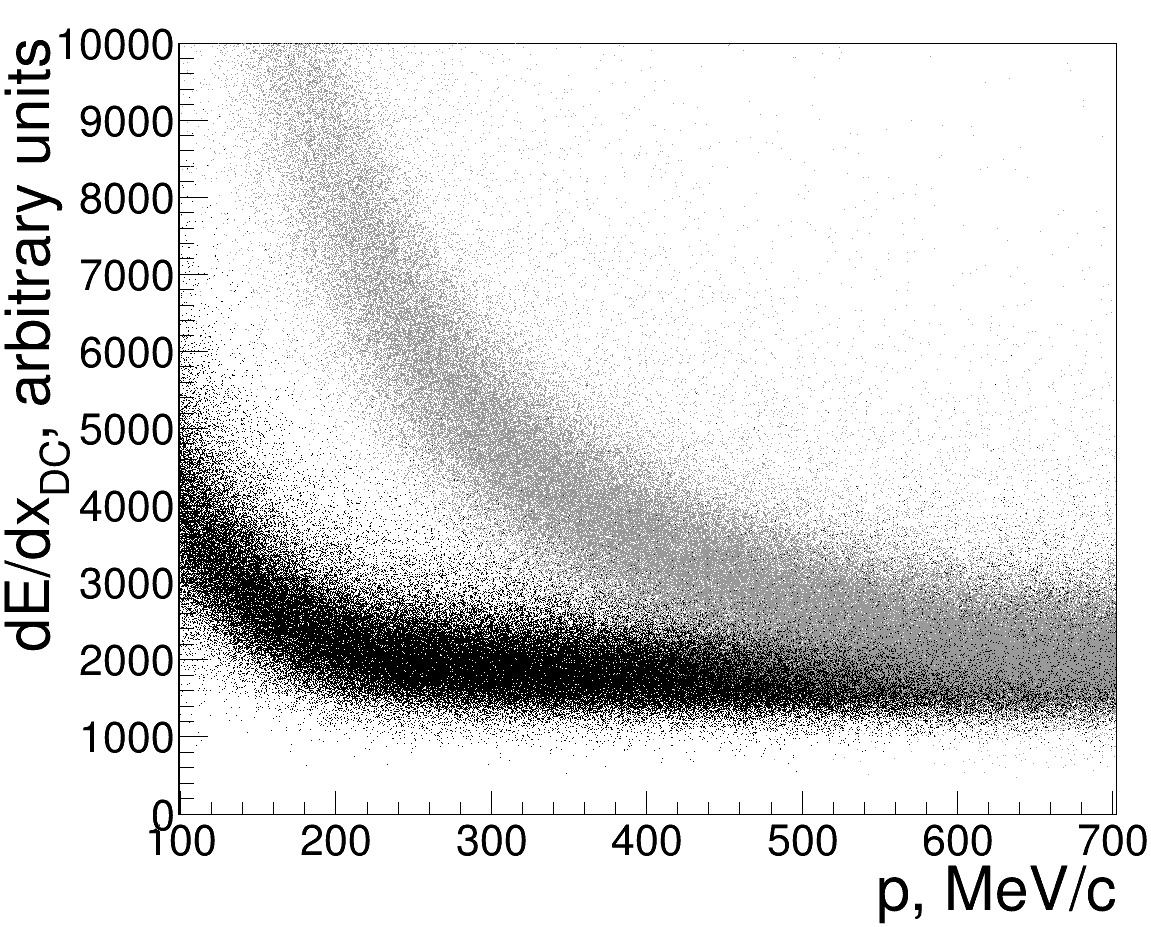}
    \caption{$dE/dx_{\rm DC}$ as a function of particle momentum 
    for simulated $K^{\pm}$ (gray) and $\pi^{\pm}$ (black).
    \label{fig:dedx_dc_pi_k}}
  \end{center}
\end{figure}

\begin{figure}[hbtp]
\centering
\begin{tabular}{cccc}
\includegraphics[width=0.48\textwidth]{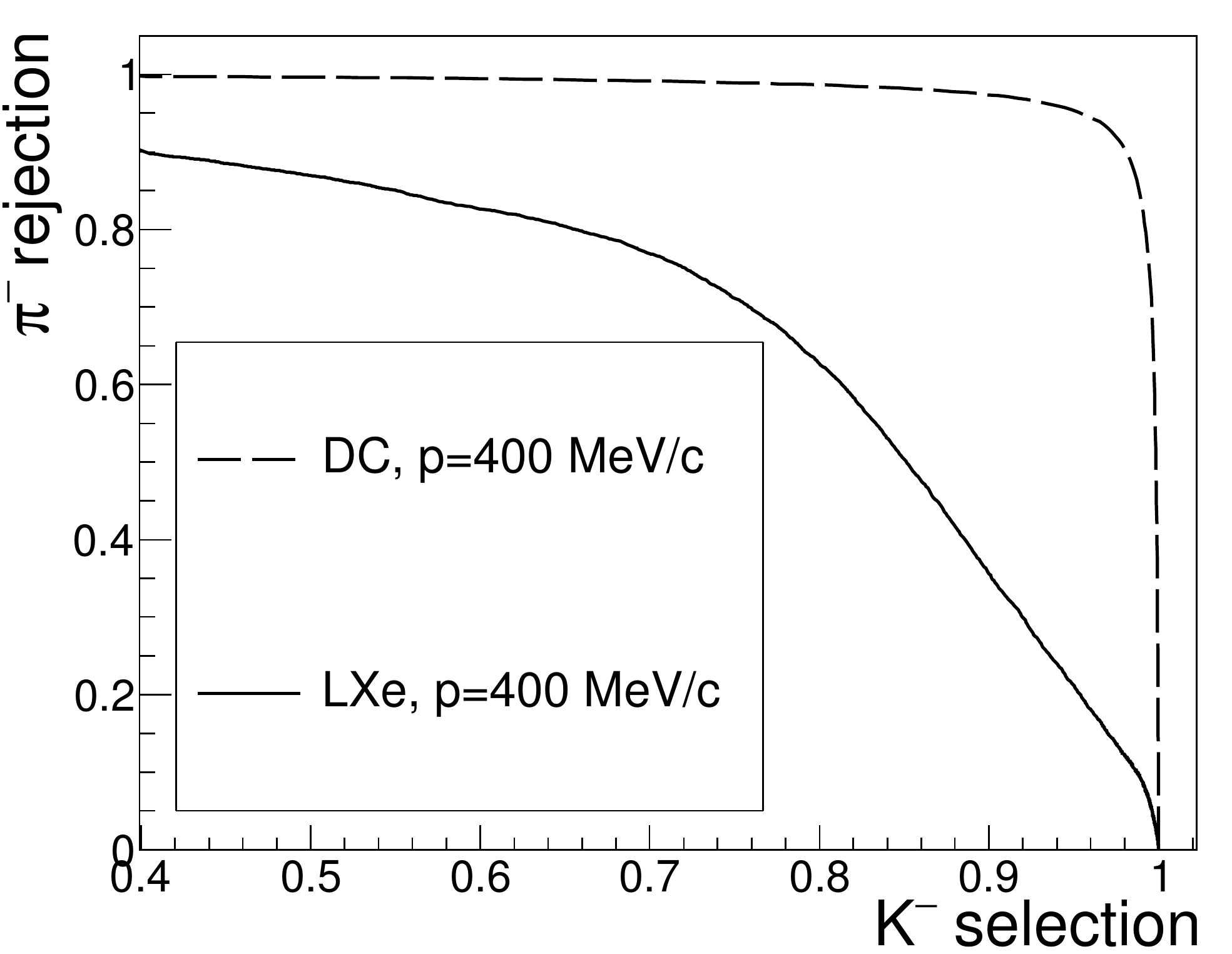} &
\includegraphics[width=0.48\textwidth]{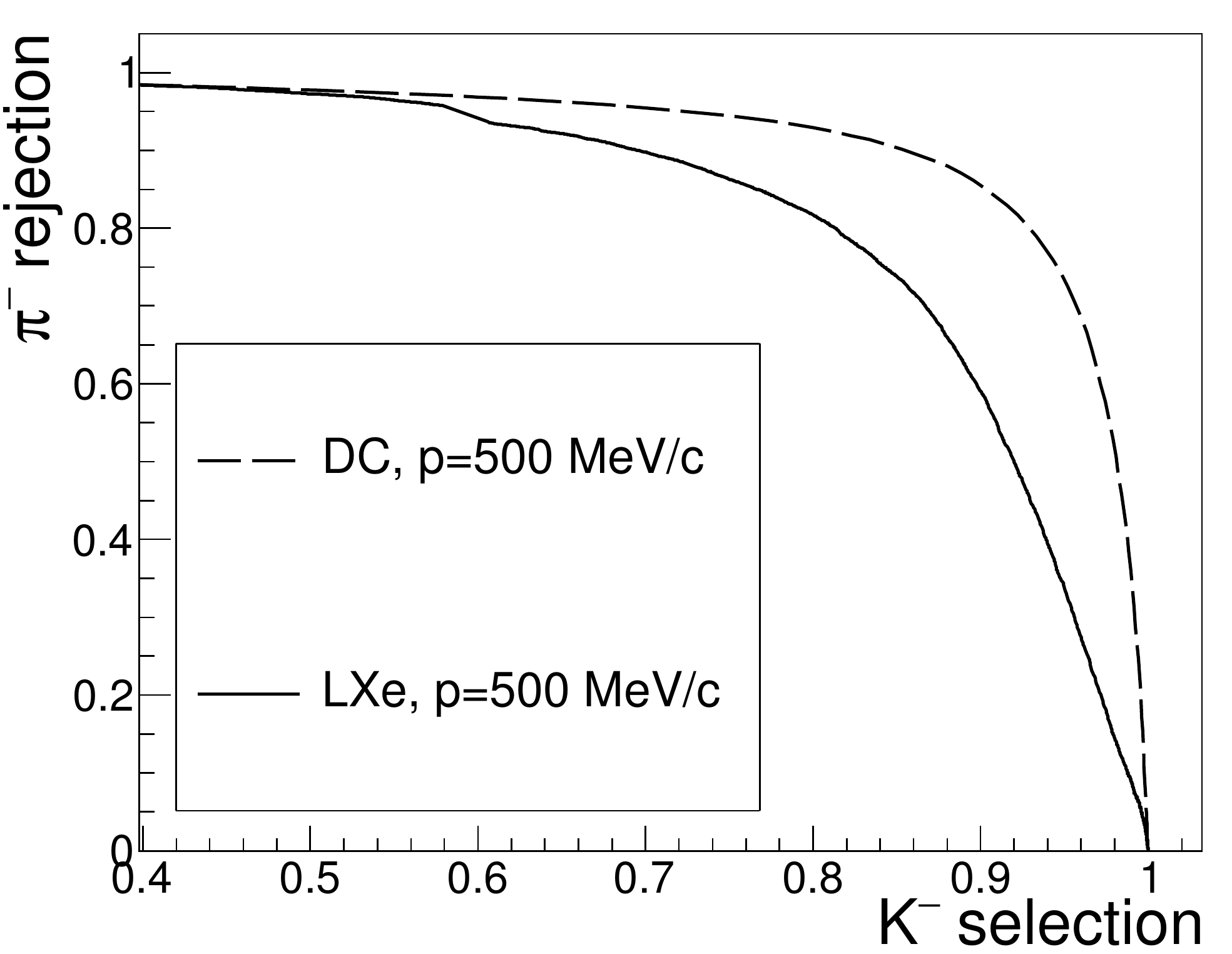} \\
\end{tabular}
\begin{tabular}{cccc}
\includegraphics[width=0.48\textwidth]{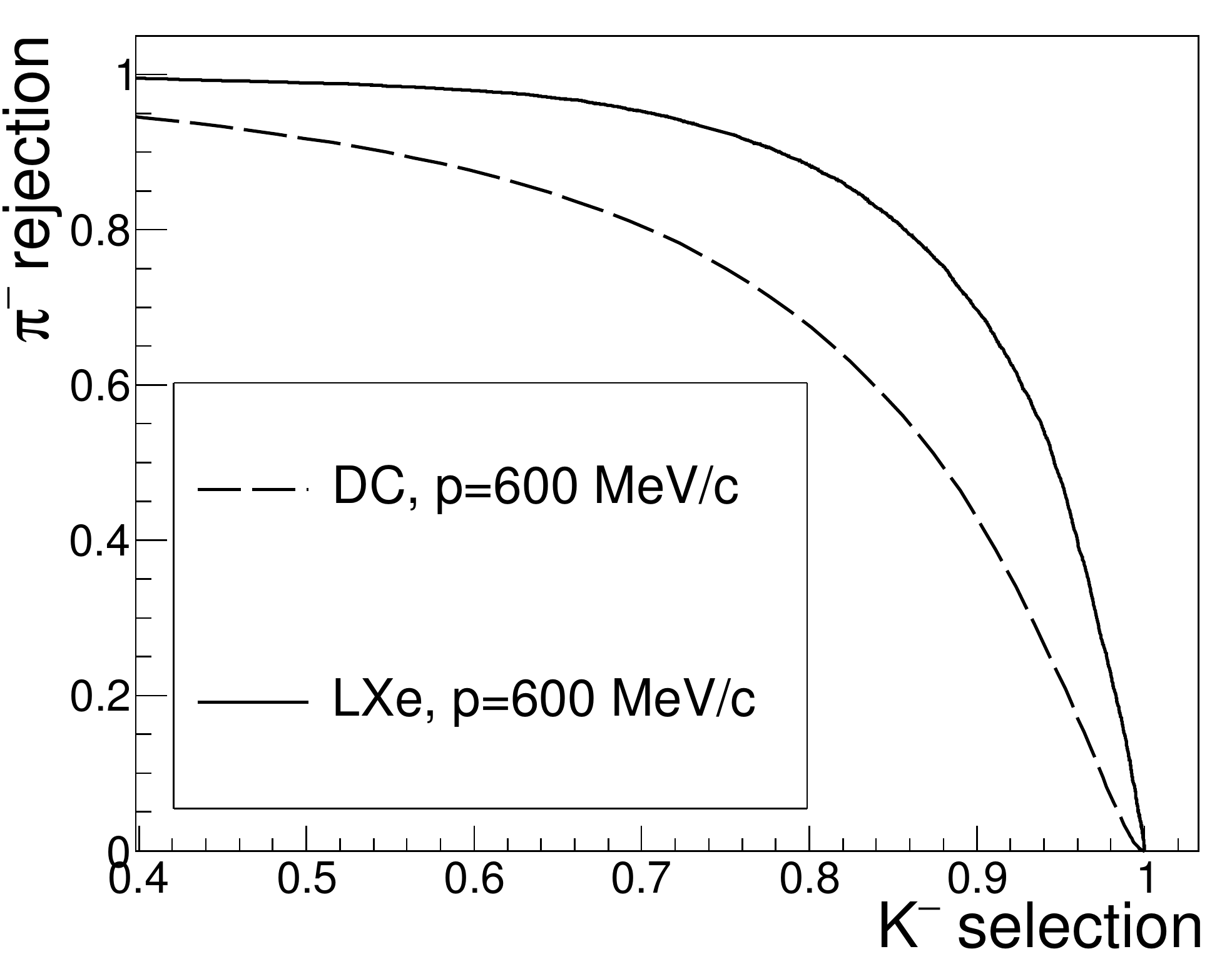} &
\includegraphics[width=0.48\textwidth]{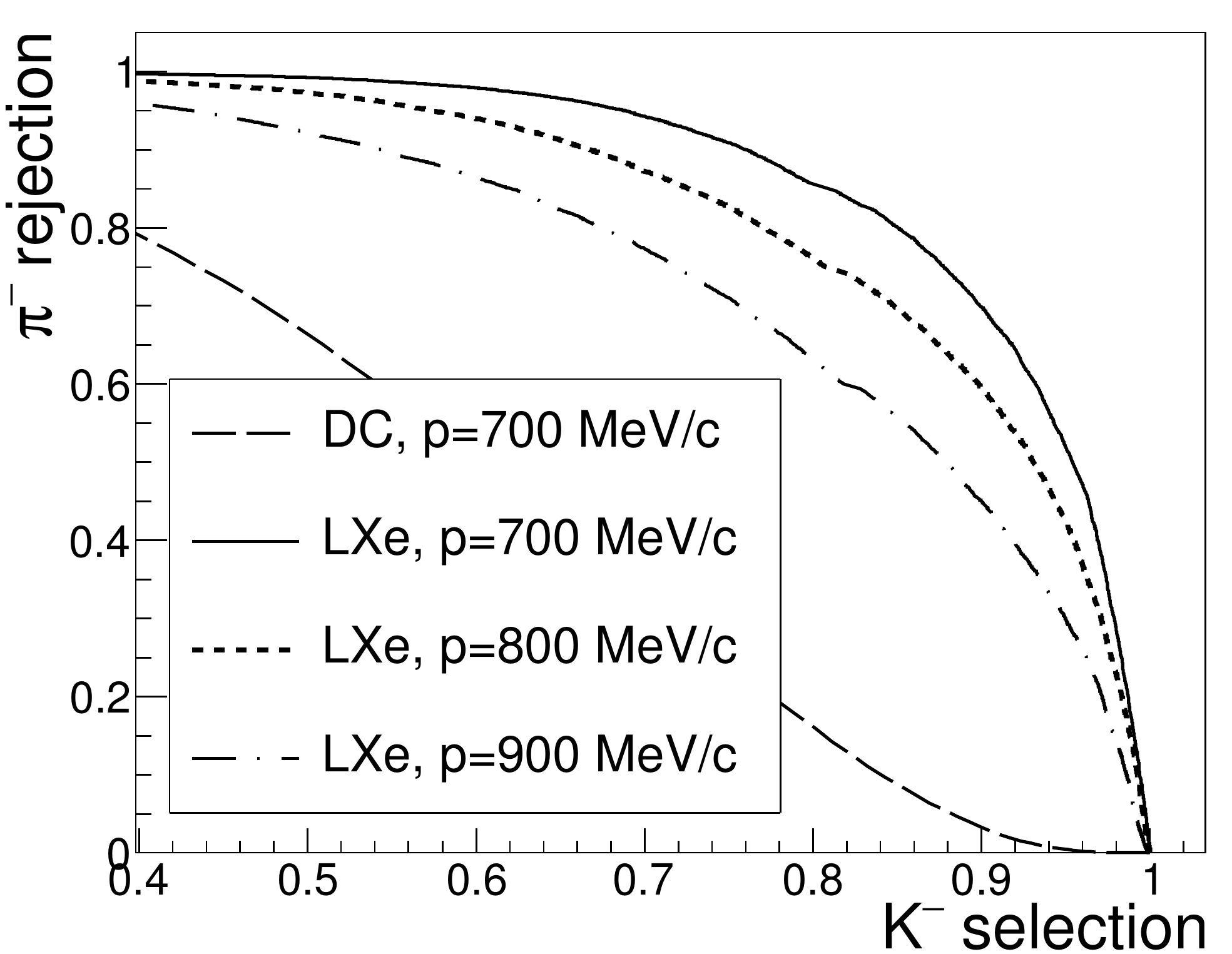} \\
\end{tabular}
\caption{The ROC-curves for the ${\rm BDT}(\pi^{-},K^{-})$ 
    classifier and $dE/dx_{\rm DC}$-based $\pi^{-}/K^{-}$ separation for different particle momenta according to simulation. 
    The classifier types and particle momenta are shown in the legends. 
    \label{fig:ROC_pi_k_dc_and_lxe}}
\end{figure}

The LXe-based $\pi/K$ separation is of special importance 
in the studies of the hadronic processes with $K^{\pm}$,
and it should be compared with the separation based on $dE/dx_{\rm DC}$. 
Figure~\ref{fig:dedx_dc_pi_k} shows the distributions of 
the $dE/dx_{\rm DC}$ vs. momentum for the simulated $K^{\pm}$ and $\pi^{\pm}$.
The ROC-curves for both types of classification at 
different particle momenta are shown in Fig.~\ref{fig:ROC_pi_k_dc_and_lxe}. 
At the momenta below 400~MeV/{\rm c} the LXe-based classifier has poor efficiency. 
At the largest momenta its efficiency gradually reduces due to the 
decrease of the difference between kaon and pion ionization losses, see Fig.~\ref{fig:dedx_k_pi}.
Hovewer, the LXe-based $\pi/K$ separation remains effective at the momenta 650--900~MeV/{\rm c}, 
where the $dE/dx_{\rm DC}$-based separation does not work. 

\section{Examples of the application of the LXe-based PID}
\hspace*{\parindent}

\subsection{Separation of the $e^+e^-(\gamma)$ and $\pi^+\pi^-(\gamma)$ final states at $E_{\rm beam}<500$~MeV \label{sec:PipPim}}

The developed PID procedure can be used for the important
task of the pion form factor $|F_{\pi}|^{2}$ measurement~\cite{ignatov}.
To calculate the $|F_{\pi}|^{2}$ at the given $E_{\rm c.m.}$ point
one needs to determine the number of events of the $\pi^{+}\pi^{-}(\gamma)$ final state, $N_{\pi^{+}\pi^{-}}$.
The major background sources for
$\pi^{+}\pi^{-}(\gamma)$ are the $e^+e^-(\gamma)$, $\mu^+\mu^-(\gamma)$ final states and cosmic muons.
The effective separation of the $\pi^+\pi^-(\gamma)$ and $\mu^+\mu^-(\gamma)$ final states at the CMD-3 is a 
difficult task at the energies $E_{\rm beam}>350$~MeV.
However, since the cross sections of the 
$e^{+}e^{-} \to e^{+}e^{-}(\gamma)$ and $e^{+}e^{-} \to \mu^{+}\mu^{-}(\gamma)$
processes are precisely calculated in the frame of QED, 
the number of events $N_{\mu^{+}\mu^{-}}$ can be calculated 
once the number of events $N_{e^{+}e^{-}}$ is known. 
In turn, determination of the $N_{e^{+}e^{-}}$ becomes possible
with the application of the effective separation of the $e^{+}e^{-}(\gamma)$ and 
$\pi^+\pi^-(\gamma)$ final states.
Currently at CMD-3 we use two independent approaches for the $e^{+}e^{-}(\gamma)$ and $\pi^+\pi^-(\gamma)$ separation: 
1) using the particle momenta; 
2) using the full energy depositions of particles in the calorimeter.
The LXe-based PID provides us another method of $e^{+}e^{-}(\gamma)$ and $\pi^+\pi^-(\gamma)$ separation.

As an example we consider the $e^+e^-(\gamma)$ and $\pi^+\pi^-(\gamma)$ separation at the energies $E_{\rm beam}<500$~MeV 
in the experimental runs of 2018.
We select events having exactly two oppositely charged tracks,
satisfying the following conditions:   
1) the momenta of tracks are larger than 100~${\rm MeV}/c$;
2) the $|\rho|$ and $|z|$ of the
track point of the closest approach to the beam axis should be less than
0.6 and 12~cm, respectively;
3) the polar angles of tracks should be in the range from 1.0 to $\pi-1.0$~rad;
4) the track collinearity conditions: $|\theta_{1}+\theta_{2}-\pi|<0.25$~rad
and $||\varphi_{1}-\varphi_{2}|-\pi|<0.15$~rad.

\begin{figure}
  \begin{center}
    \includegraphics[width=0.6\textwidth]{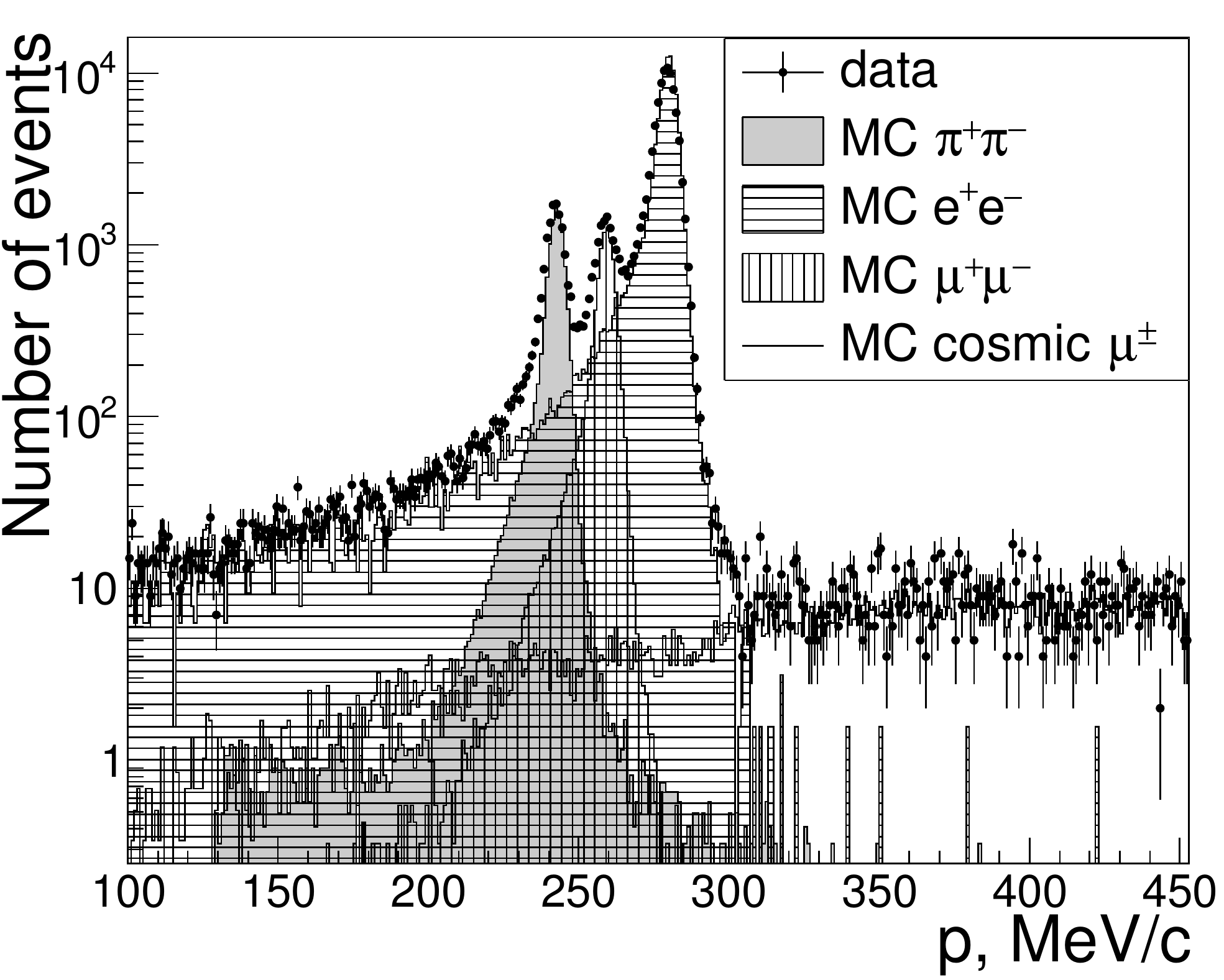}
    \caption{The momentum spectra of the particles selected at 
$E_{\rm beam}=280$~MeV
in the experiment (markers), 
simulation of $\pi^{+}\pi^{-}(\gamma)$ (gray histogram),
$e^{+}e^{-}(\gamma)$ (horizontal hatching), 
$\mu^{+}\mu^{-}(\gamma)$ (vertical hatching) and cosmic muons (open histograsm). 
The black line shows the total MC of the signal and background processes.
 \label{fig:PipPim_mom}}
  \end{center}
\end{figure}

\begin{figure}[hbtp]
  \begin{minipage}[t]{0.49\textwidth}
   \centerline{\includegraphics[width=\textwidth]{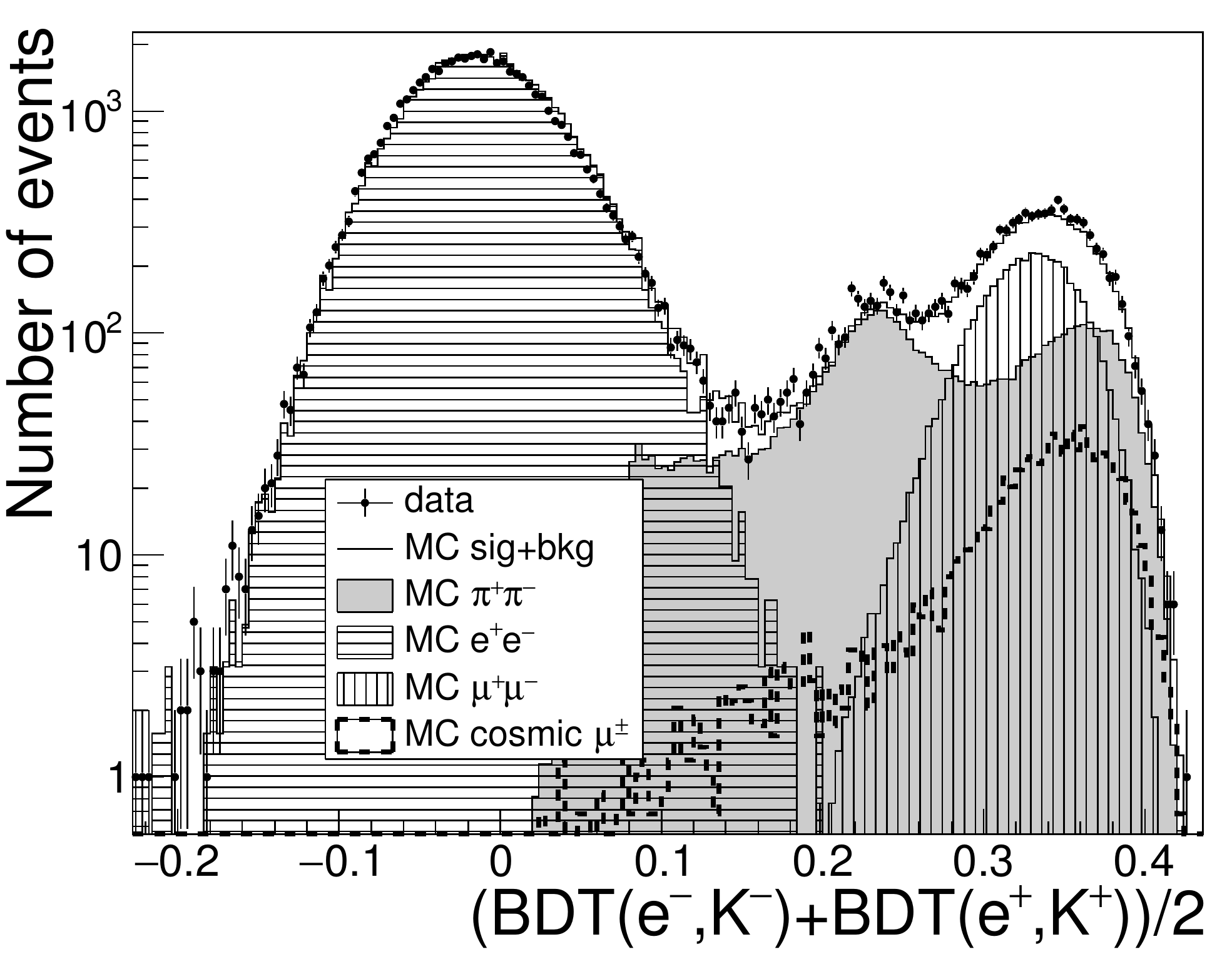}}
  \end{minipage}\hfill\hfill
   \begin{minipage}[t]{0.49\textwidth}
   \centerline{\includegraphics[width=\textwidth]{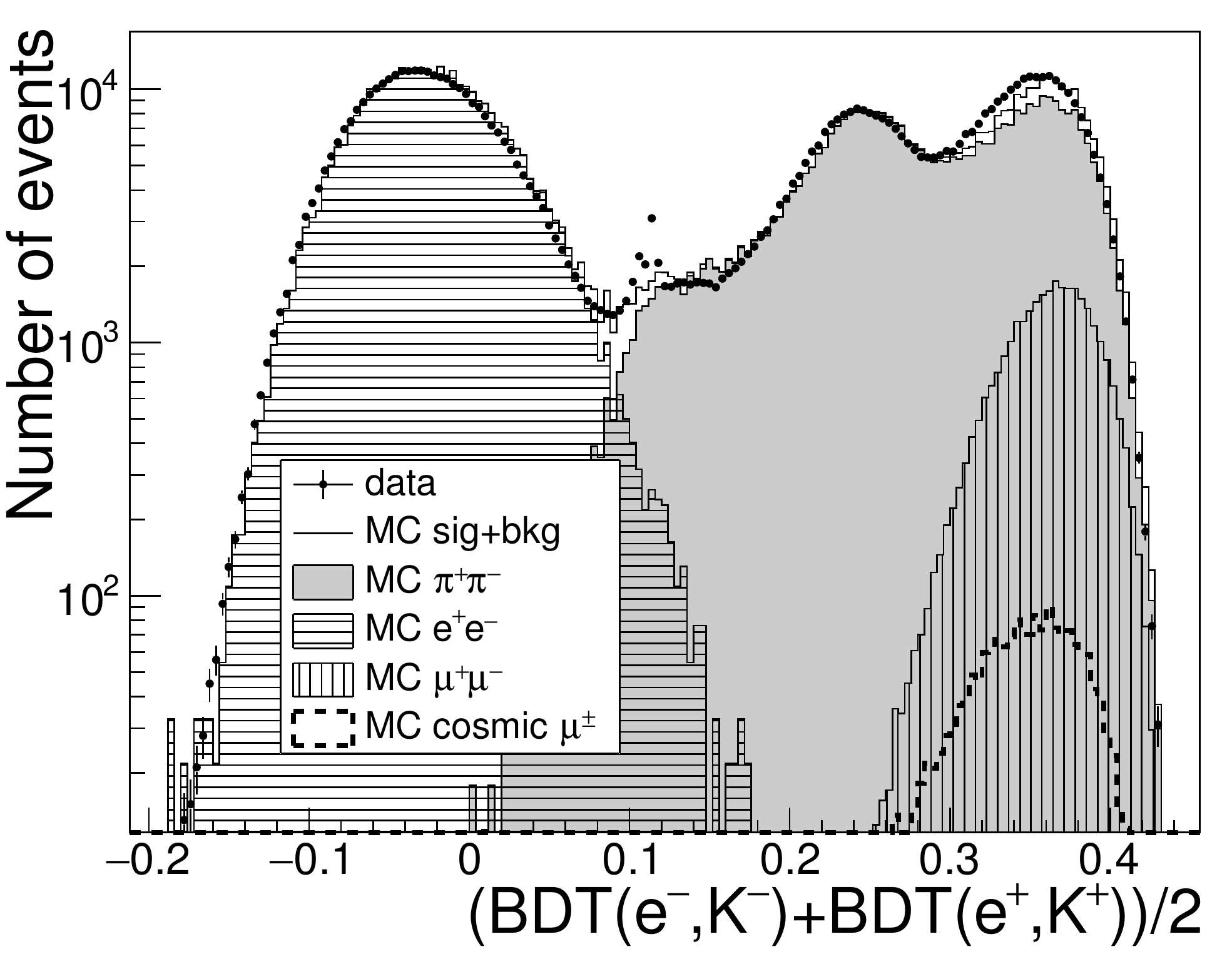}}
  \end{minipage}\hfill\hfill
  \caption{The distribution of the $({\rm BDT}(e^-,\pi^-)+{\rm BDT}(e^+,\pi^+))/2$ 
(left --- for $E_{\rm beam}=280$~MeV, right --- $E_{\rm beam}=380$~MeV)
in the experiment (markers), simulation of 
$\pi^{+}\pi^{-}(\gamma)$ (gray histogram),
$e^{+}e^{-}(\gamma)$ (horizontal hatching), $\mu^{+}\mu^{-}(\gamma)$ 
(vertical hatching) and cosmic muons (dashed line).
The open histogram shows the total MC of the signal and background processes.
          \label{fig:PipPim_bdt_e_pi}}
\end{figure}

\begin{figure}
  \begin{center}
    \includegraphics[width=0.5\textwidth]{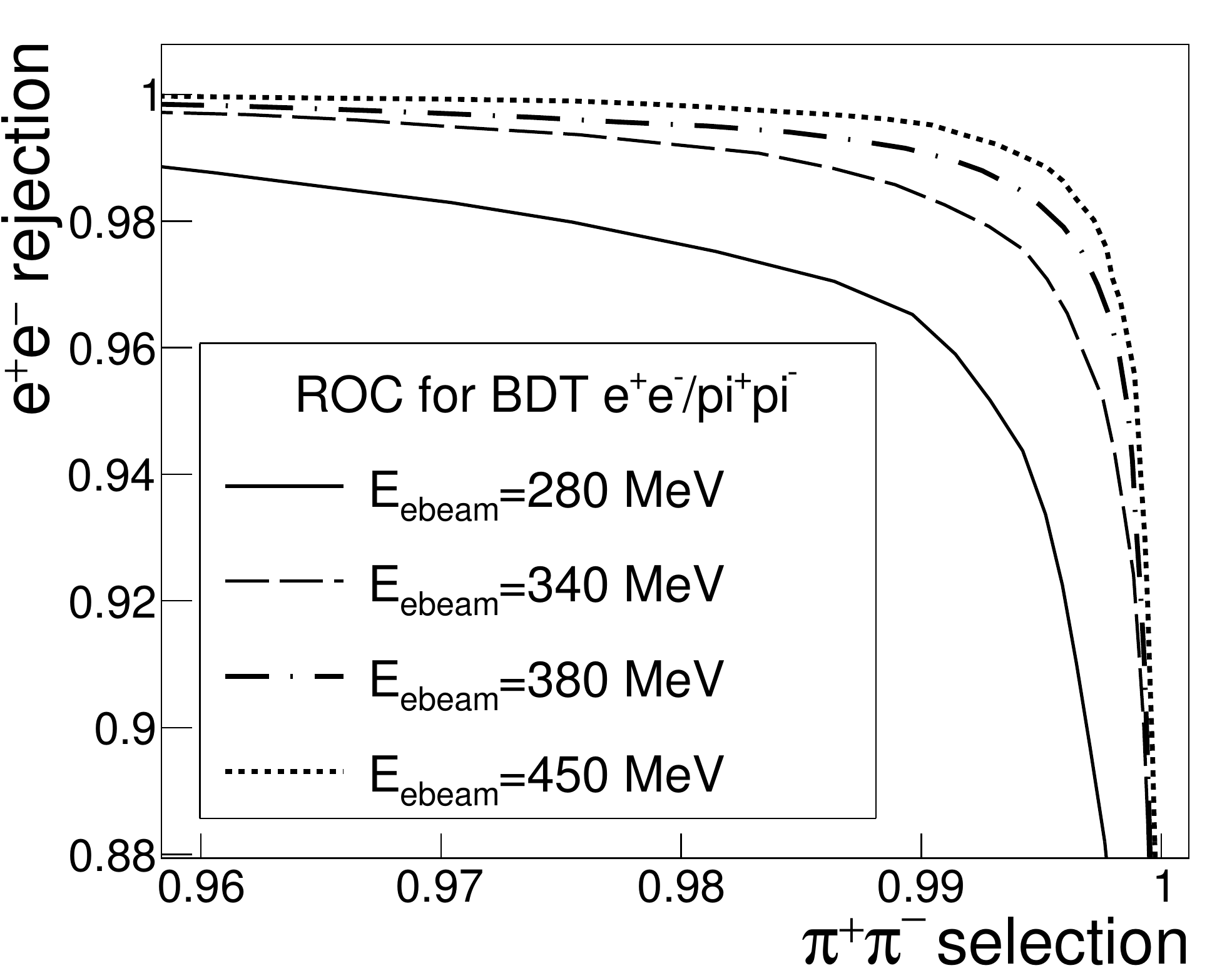}
    \caption{ROC-curves for separation of the $e^+e^-(\gamma)$ and $\pi^+\pi^-(\gamma)$
final states using $({\rm BDT}(e^-,\pi^-)+{\rm BDT}(e^+,\pi^+))/2$ 
at different $E_{\rm beam}$ (see legend) according to simulation. \label{fig:roc_bdt_EpEm_PipPim}}
  \end{center}
\end{figure}

Figure~\ref{fig:PipPim_mom} shows the momentum spectrum for the particles,
selected in the experiment and simulation at $E_{\rm beam}=280$~MeV.  
The contribution of the collinear final states is estimated
according to the known cross sections of the processes and luminosity,
while the contribution of the cosmic muons is estimated using the events 
with the momenta larger than $1.25{\cdot}E_{\rm beam}$.
Further, Fig.~\ref{fig:PipPim_bdt_e_pi} shows the 
distribution of the average ${\rm BDT}(e,\pi)$ response for two tracks,
{\it i.e.} $({\rm BDT}(e^-,\pi^-)+{\rm BDT}(e^+,\pi^+))/2$,
for $E_{\rm beam}=280$~MeV (left tail of the $\rho(770)$) 
and 380~MeV (near the peak of $\rho(770)$).
It is seen that the $({\rm BDT}(e^-,\pi^-)+{\rm BDT}(e^+,\pi^+))/2$ parameter
provides a powerful classifier for $e^{+}e^{-}(\gamma)$ and  $\pi^+\pi^-(\gamma)$ separation,
see corresponding ROC-curves in Fig.~\ref{fig:roc_bdt_EpEm_PipPim}.
At $E_{\rm beam}=380$~MeV the classifier allows to select 99.5\% 
of $\pi^+\pi^-(\gamma)$ events by the 98\% rejection of the $e^{+}e^{-}(\gamma)$ background. 


\subsection{Selection of the $K^{+}K^{-}$ final state at high energies}

\begin{figure}[hbtp]
  \begin{minipage}[t]{0.49\textwidth}
   \centerline{\includegraphics[width=0.98\textwidth]{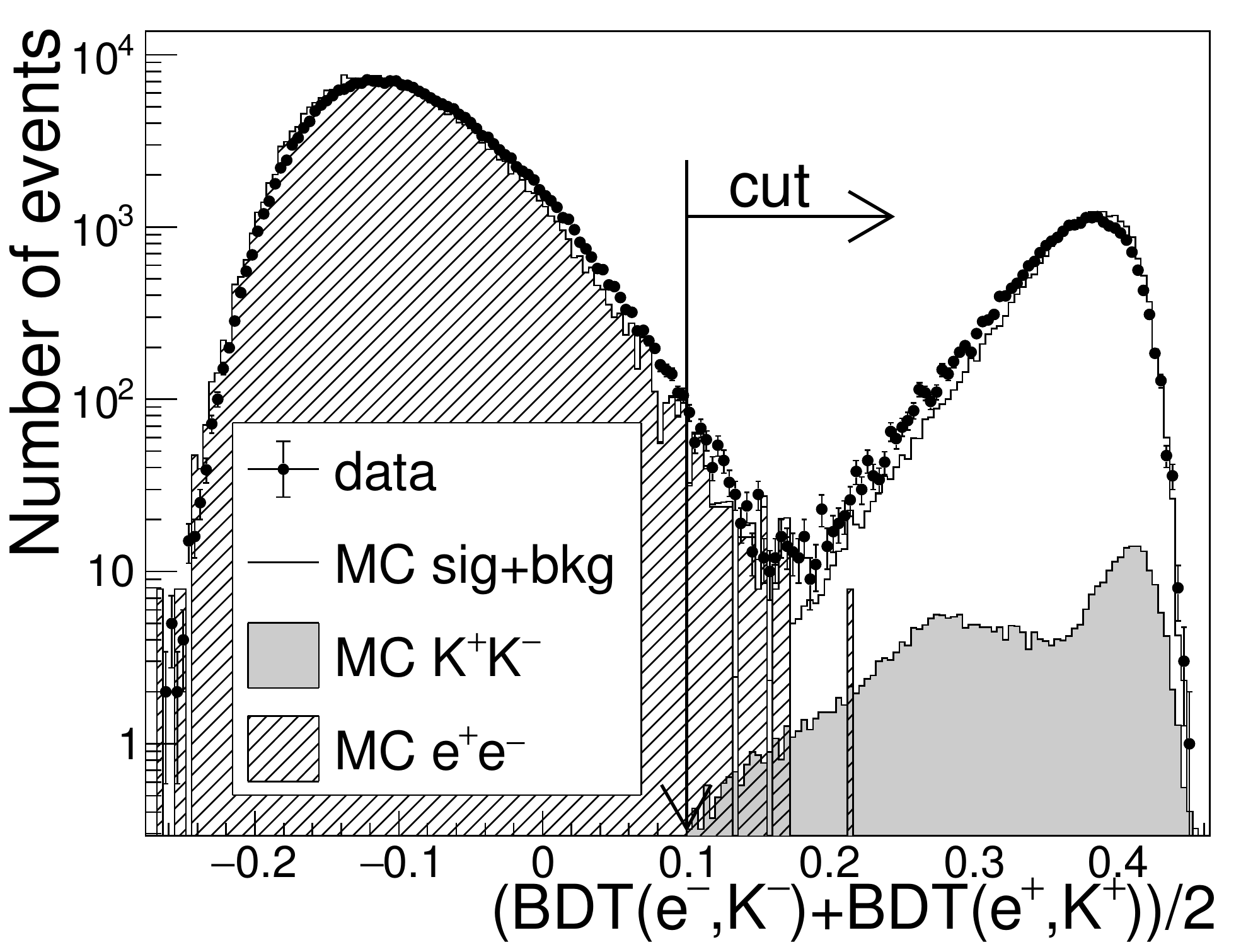}}
   \caption{The $({\rm BDT}(e^{-},K^{-})+{\rm BDT}(e^{+},K^{+}))/2$ 
   spectra in the experiment (markers), MC of the 
   $K^{+}K^{-}(\gamma)$ (gray histogram), 
   $e^{+}e^{-}(\gamma)$ (hatched histogram).   
   The open histogram shows the total MC of the signal and background processes.
   \label{fig:bdt_e_k_KpKm_987_5}}   
  \end{minipage}\hfill\hfill
  \begin{minipage}[t]{0.49\textwidth}
    \centerline{\includegraphics[width=0.98\textwidth]{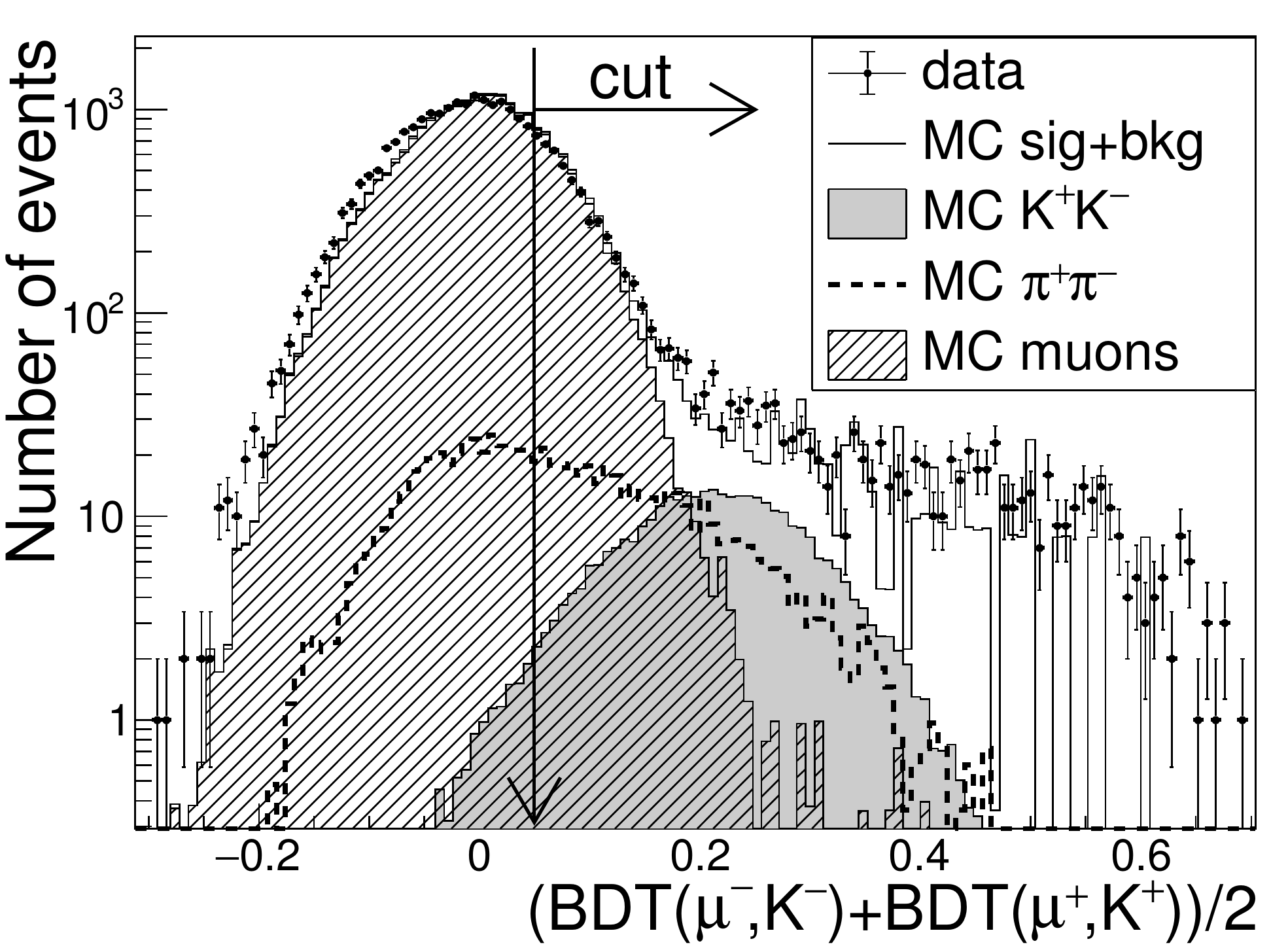}}    
    \caption{The $({\rm BDT}(\mu^{-},K^{-})+{\rm BDT}(\mu^{+},K^{+}))/2$ 
  spectra in the experiment (markers), 
  MC of the $K^{+}K^{-}(\gamma)$ (gray histogram), 
  $\mu^{+}\mu^{-}(\gamma)$ and cosmic muons (hatched histogram), 
  $\pi^{+}\pi^{-}(\gamma)$ (dotted line). 
   The open histogram shows the total MC of the signal and background processes.
    \label{fig:bdt_mu_k_KpKm_987_5}} 
    \end{minipage}\hfill\hfill   
\end{figure}

Another application of the LXe-based PID is the task of the selection of the 
$K^+K^-$ final state at high energies. 
As an example, we perform such selection on the base of 2.2~${\rm pb^{-1}}$ of 
data collected at $E_{\rm c.m.}=1.975$~GeV in the 2019 runs.
To select the two-track collinear events, we apply the selections
listed earlier in Section~\ref{sec:PipPim}.
The main background sources are the 
$e^{+}e^{-}(\gamma)$, $\mu^{+}\mu^{-}(\gamma)$, $\pi^{+}\pi^{-}(\gamma)$ 
final states and the events with cosmic muons,
their contributions in simulation are estimated 
in a way described earlier in Section~\ref{sec:PipPim}. 
The background suppression is done by the cuts, imposed on the
values of the average BDT responses 
$({\rm BDT}(e^{-},K^{-})+{\rm BDT}(e^{+},K^{+}))/2$ and
$({\rm BDT}(\mu^{-},K^{-})+{\rm BDT}(\mu^{+},K^{+}))/2$ for the 2 
tracks, see Figs.~\ref{fig:bdt_e_k_KpKm_987_5}--\ref{fig:bdt_mu_k_KpKm_987_5}.
The cut on $({\rm BDT}(\mu^{-},K^{-})+{\rm BDT}(\mu^{+},K^{+}))/2$ 
leads to the loss of ${\sim}5$\% of signal events
and also provides significant suppression of the 
$e^{+}e^{-}{\to}\pi^{+}\pi^{-}(\gamma)$ process. 
Since the cross section of the latter is relatively low at 
$E_{\rm c.m.}{\sim}2$~GeV and $\pi^{+}\pi^{-}$ events are 
kinematically separated from $K^{+}K^{-}$, 
we do not impose any cuts on ${\rm BDT}(\pi^{\pm},K^{\pm})$.

\begin{figure}[hbtp]
  \begin{minipage}[t]{0.49\textwidth}
   \centerline{\includegraphics[width=0.98\textwidth]{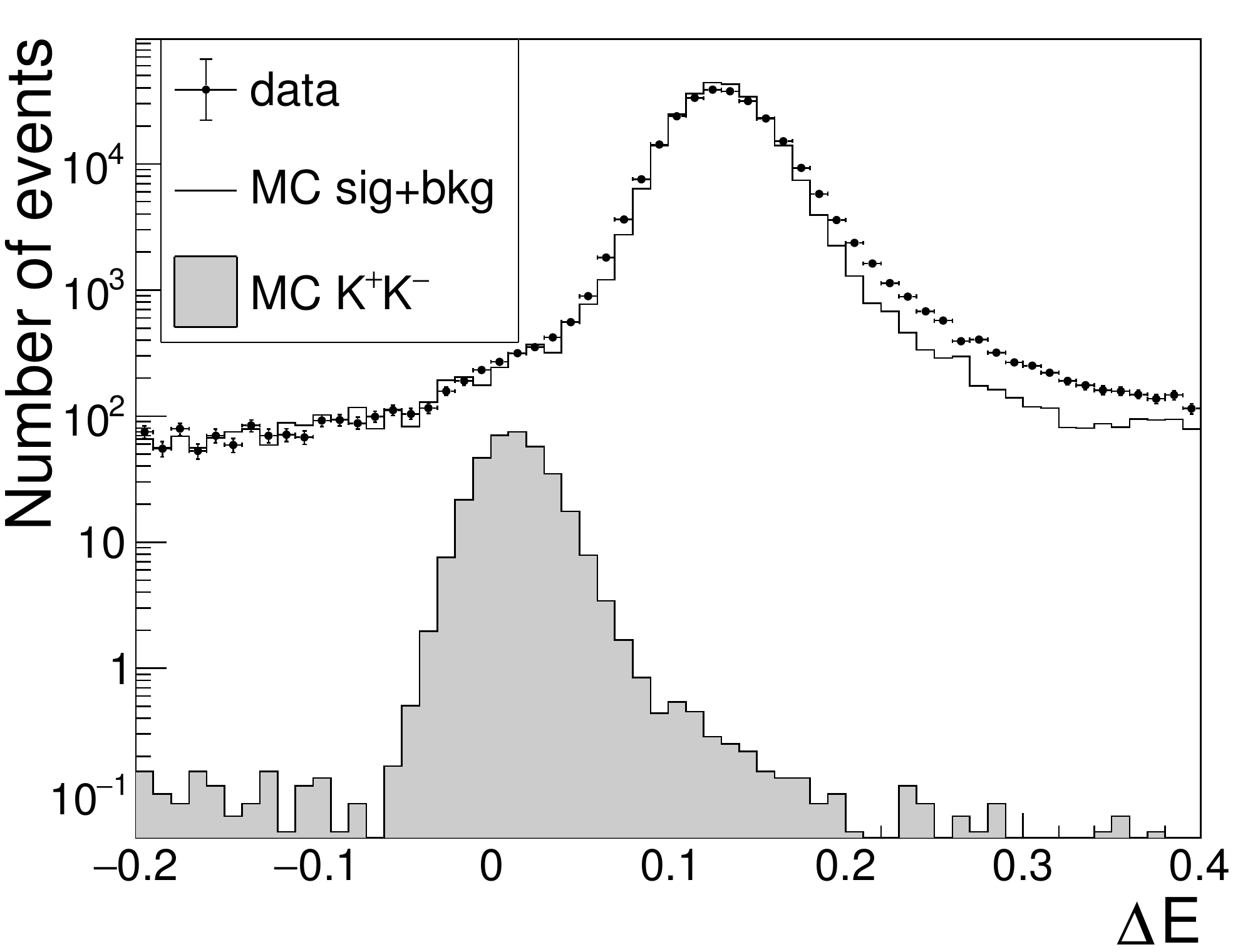}}
  \end{minipage}\hfill\hfill
  \begin{minipage}[t]{0.49\textwidth}
    \centerline{\includegraphics[width=0.98\textwidth]{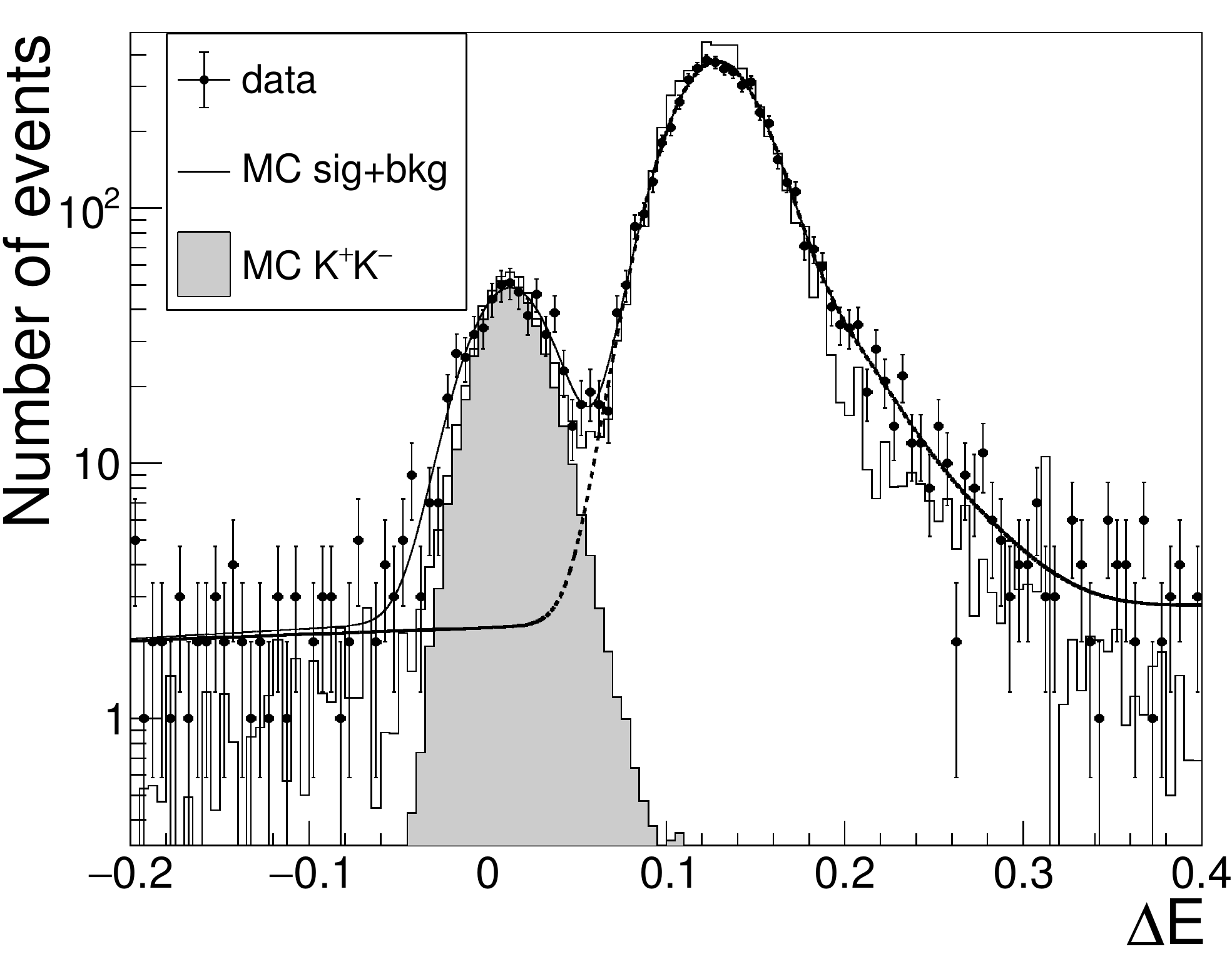}}
    \end{minipage}\hfill\hfill
  \caption{The ${\Delta}E$ spectra before (left) and after (right) 
   background suppression in the experiment (markers) and MC of the 
   $K^{+}K^{-}(\gamma)$ (gray histogram). The open histogram shows 
   the total MC of the signal and background processes. 
   The solid curve on the right picture shows the fit of the 
   distribution in the experiment, 
   dotted curve --- the background part of the fit.
    \label{fig:dE_987_5}}  
\end{figure}

Next, the signal/background separation is performed by approximation 
of the distribution 
of ``energy disbalance'' ${\Delta}E$, defined as

\begin{equation}
  {\Delta}E=
  \frac{\sqrt{{\vec{p}_{+}}^{\, 2}c^{2}+m^{2}_{K^{+}}c^{4}}+\sqrt{{\vec{p}_{-}}^{\, 2}c^{2}+m^{2}_{K^{-}}c^{4}}+|\vec{p}_{+}+\vec{p}_{-}|c}{2E_{\rm beam}}-1,
\end{equation}
where $\vec{p}_{\pm}$ are the particle momenta.
The additional term $|\vec{p}_{+}+\vec{p}_{-}|$, corresponding to the total momentum of two tracks, 
allows to get rid of the superimposition of the signal peak with the $e^{+}e^{-}(\gamma)$ radiative tail.
Figure~\ref{fig:dE_987_5} shows the ${\Delta}E$ spectra before 
and after the application of cuts on BDT.
It is seen that after the background suppression the signal/background separation 
in the ${\Delta}E$ spectrum becomes possible.
To perform the separation, we approximate the experimental ${\Delta}E$ spectra using the sum of three Gaussians 
to approximate the peaking background and the linear function to approximate the contribution of cosmic muons.
The shape of the signal peak is fixed from the approximation of the simulated ${\Delta}E$ spectra for the $e^{+}e^{-}{\to}K^{+}K^{-}$ process,
except for the shift of the signal peak as a whole and its additional broadening, which are added as the floating parameters.
Thus, we obtain $548{\pm}27$ of signal events at $E_{\rm c.m.}=1.975$~GeV (Fig.~\ref{fig:dE_987_5}, right).

It should be noted that for the c.m. energies larger than 1.5~GeV 
usage of the LXe-based PID is the only way to measure the 
$e^{+}e^{-}{\to}K^{+}K^{-}$ process cross section at \mbox{CMD-3}.

\section{\boldmath Conclusions}
\hspace*{\parindent}

The procedure of the charged PID using the LXe
 calorimeter of the \mbox{CMD-3} detector was developed.
The procedure uses the energy depositions, measured in 12 layers of 
the LXe calorimeter, as the input for the set of boosted decision trees 
classifiers, trained for the separation of the electrons, muons, pions and kaons in 
the momentum range from 100 to $1200~{\rm MeV}/c$.
Since the event samples for the classifier training are taken from 
the MC, special attention was paid to the tuning of the 
simulated detector response. From the experimental side the procedure 
of the calibration of strip channels of LXe calorimeter 
with the precision of about 1\% was 
developed. These efforts resulted in good
data/MC agreement for the classifier responses for all particle types.
Finally, the application of the method was demonstrated by the examples of
separation of the $e^+e^-(\gamma)$ and $\pi^+\pi^-(\gamma)$ final states 
at $E_{\rm beam}<500$~MeV 
and of the selection of the $K^+K^-$ final state at high energies.


\section{\boldmath Acknowledgments}
\hspace*{\parindent}

We thank the VEPP-2000 personnel for excellent machine operation. 
Part of work related to the method of the selection 
of $e^{+}e^{-}{\to}K^{+}K^{-}\pi^{+}\pi^{-}$ process events 
mentioned in the Section~\ref{kaons} is partially supported by the grant of the Russian 
Foundation for Basic Research 20-02-00496 A 2020.


\input{Ivanov.bbl}
\end{document}

%% file: authors.tex
\author[adr1,adr2]{V.L.~Ivanov\fnref{tnot}}
\author[adr1,adr2]{G.V.~Fedotovich}
\author[adr1,adr2]{R.R.~Akhmetshin}
\author[adr1,adr2]{A.N.~Amirkhanov}
\author[adr1,adr2]{A.V.~Anisenkov}
\author[adr1,adr2]{V.M.~Aulchenko}
\author[adr1]{N.S.~Bashtovoy}
\author[adr1,adr2]{A.E.~Bondar}
\author[adr1]{A.V.~Bragin}
\author[adr1,adr2,adr3]{S.I.~Eidelman}
\author[adr1,adr2]{D.A.~Epifanov}
\author[adr1,adr2,adr4]{L.B.~Epshteyn}
\author[adr1,adr2]{A.L.~Erofeev}
\author[adr1,adr2]{S.E.~Gayazov}
\author[adr1,adr2]{A.A.~Grebenuk}
\author[adr1,adr2]{S.S.~Gribanov}
\author[adr1,adr2,adr4]{D.N.~Grigoriev}
\author[adr1,adr2]{F.V.~Ignatov}
\author[adr1]{S.V.~Karpov}
\author[adr1,adr2]{V.F.~Kazanin}
\author[adr1,adr2]{A.A.~Korobov}
\author[adr1,adr4]{A.N.~Kozyrev}
\author[adr1,adr2]{E.A.~Kozyrev}
\author[adr1,adr2]{P.P.~Krokovny}
\author[adr1,adr2]{A.E.~Kuzmenko}
\author[adr1,adr2]{A.S.~Kuzmin}
\author[adr1,adr2]{I.B.~Logashenko}
\author[adr1,adr2]{P.A.~Lukin}
\author[adr1]{K.Yu.~Mikhailov}
\author[adr1]{V.S.~Okhapkin}
\author[adr1]{Yu.N.~Pestov}
\author[adr1,adr2]{A.S.~Popov}
\author[adr1,adr2]{G.P.~Razuvaev}
\author[adr1]{A.A.~Ruban}
\author[adr1]{N.M.~Ryskulov}
\author[adr1,adr2]{A.E.~Ryzhenenkov}
\author[adr1,adr2]{A.V.~Semenov}
\author[adr1,adr2,adr5]{V.E.~Shebalin}
\author[adr1,adr2]{D.N.~Shemyakin}
\author[adr1,adr2]{B.A.~Shwartz}
\author[adr1,adr2]{E.P.~Solodov}
\author[adr1]{V.M.~Titov}
\author[adr1,adr2]{A.A.~Talyshev}
\author[adr1]{S.S.~Tolmachev}
\author[adr1]{A.I.~Vorobiov}
\author[adr1,adr2]{Yu.V.~Yudin}

\address[adr1]{Budker Institute of Nuclear Physics, SB RAS, Novosibirsk, 630090, Russia}
\address[adr2]{Novosibirsk State University, Novosibirsk, 630090, Russia}
\address[adr3]{Lebedev Physical Institute RAS, Moscow, 119333, Russia}
\address[adr4]{Novosibirsk State Technical University, Novosibirsk, 630092, Russia}
\address[adr5]{University of Hawaii, Honolulu, Hawaii 96822, USA}
\fntext[tnot]{Corresponding author: V.L.Ivanov@inp.nsk.su}

%% file: Ivanov.bbl
\begin{thebibliography}{00}

\bibitem{vepp1} V.V. Danilov et al., Proceedings EPAC96, Barcelona, p.1593 (1996).

\bibitem{vepp2} I.A. Koop, Nucl. Phys. B (Proc. Suppl.) 181-182, 371 (2008).

\bibitem{vepp3} P.Yu. Shatunov et al., Phys. Part. Nucl. Lett. 13, 995 (2016).

\bibitem{vepp4} D. Shwartz et al., PoS ICHEP2016, 054 (2016).

\bibitem{cmd3} B.I. Khazin et al. (CMD-3 Collaboration), Nucl. Phys. B (Proc. Suppl.) 181-182, 376 (2008).

\bibitem{jegerlehner} F. Jegerlehner, Springer Tracks Mod. Phys. 274, 1 (2017).

\bibitem{davier} M. Davier, A. Hoecker, B. Malaescu, and Z. Zhang, Eur. Phys. J. C 77, 827 (2017).

\bibitem{teubner} A. Keshavarzi, D. Nomura, T. Teubner, Phys. Rev. D 97, 114025 (2018).

\bibitem{hagiwara} K. Hagiwara et al., J. Phys. G 38, 085003 (2011).

\bibitem{cal} D. Epifanov (CMD-3 Collaboration), J. Phys. Conf. Ser. 293, 012009 (2011).

\bibitem{TOF} A.~Amirkhanov et al., Nucl. Instrum. Meth. A936, 598 (2019).

\bibitem{dc} F. Grancagnolo et al., Nucl. Instr. Meth. A623, 114 (2010).

\bibitem{lxe} A.V. Anisyonkov et al., Nucl. Instr. Meth. A598, 266 (2009).

\bibitem{lxe_csi_bgo} V.E. Shebalin et al., JINST 9 (10), C10013 (2014).

\bibitem{tmva} H. Voss, A. Hcker, J. Stelzer, F. Tegenfeldt, PoS(ACAT) 040.

\bibitem{LXe_electroincs} K.I. Kakhuta and Yu.V. Yudin,  Nucl. Instr. and Meth. A {\bf 598}, 342 (2009).

\bibitem{shemyakin_kkpipi} D.N. Shemyakin et al. (CMD-3 Collaboration), Phys.Lett. B 756, 153 (2016).

\bibitem{kkpipi_at_nbarn} R.R. Akhmetshin et al. (CMD-3 Collaboration), Phys. Lett. B 794, 64 (2019).

\bibitem{ignatov} F.V.~Ignatov et al. (CMD-3 Collaboration), EPJ Web Conf. {\bf 218}, 02006 (2019).

\end{thebibliography}
